\documentclass{svjour3}                     
\smartqed  

\newcommand{\kms}{km\,s$^{-1}$}
\newcommand{\ergs}{erg\ s$^{-1}$}

\newcommand{\msol}{M$_{\odot}$}
\newcommand{\xmm}{{\sc XMM}\emph{-Newton}}
\newcommand{\ros}{\emph{{\sc ROSAT}}}
\newcommand{\ch}{\emph{{\sc Chandra}}}
\newcommand{\lxlb}{$L_{\rm X}/L_{\rm bol}$}

\usepackage{graphicx}
\usepackage{natbib}
\usepackage{amssymb}
\usepackage{rotating}
\def\ga{\;\rlap{\lower 2.5pt\hbox{$\sim$}}\raise 1.5pt\hbox{$>$}\;}
\def\la{\;\rlap{\lower 2.5pt\hbox{$\sim$}}\raise 1.5pt\hbox{$<$}\;}
\usepackage{mathptmx}      
\journalname{Astronomy and Astrophysics Review} 
\begin{document}

\title{X-Ray Spectroscopy of Stars
}



\author{Manuel G\"udel         \and
        Ya\"el Naz\'e  
}


\institute{Manuel G\"udel \at
              Institute of Astronomy, ETH Zurich, 8093 Zurich, Switzerland\\
	      Tel.: +41-44-6327129\\
              Fax:  +41-44-6321205\\
              \email{guedel@astro.phys.ethz.ch}          
           \and
           Yael Naz\'e \at
	      Postdoctoral Researcher FNRS, Belgium\\
              Institut d'Astrophysique et de G\'eophysique, Universit\'e de Li\`ege, All\'ee du 6 Ao\^ut 17, Bat B5C, B4000-Li\`ege, Belgium \\
              Tel.: +32-4-3669720 \\
              Fax: +32-4-3669746\\
              \email{naze@astro.ulg.ac.be}
}

\date{Received: 19 February 2009 / Accepted: 3 April 2009}

\maketitle

\begin{abstract}
Non-degenerate stars of essentially all spectral classes are soft X-ray sources. Their X-ray spectra have been important
in constraining physical processes that heat plasma in stellar environments to temperatures exceeding one million degree.
Low-mass stars on the cooler part of the main sequence and their pre-main sequence predecessors define the dominant
stellar population in the galaxy by number. Their X-ray spectra are reminiscent, in the broadest sense, of X-ray spectra 
from the solar corona. The Sun itself as a typical example of a main-sequence cool star has been a pivotal testbed for 
physical models to be applied to cool stars.  X-ray emission from cool stars is indeed ascribed to magnetically trapped hot 
gas analogous to the solar coronal plasma, although plasma parameters such as temperature, density, and element 
abundances vary widely. Coronal structure, its thermal stratification and geometric extent can also be interpreted based on 
various spectral diagnostics. New features have been identified in pre-main sequence stars;
some of these may be related to accretion shocks on the stellar surface, fluorescence on circumstellar
disks due to X-ray irradiation, or shock heating in stellar outflows.
Massive, hot stars clearly dominate the interaction with the galactic interstellar medium: they are the main sources of 
ionizing radiation, mechanical energy and chemical enrichment in galaxies. High-energy emission permits to probe some 
of the most important processes at work in  these stars, and put constraints on their most peculiar feature: the 
stellar wind. Medium and high-resolution spectroscopy have 
shed new light on these objects as well. Here, we review recent advances in our understanding of cool and hot stars through 
the study of X-ray spectra, in particular high-resolution spectra now available from {\xmm} and {\ch}.  
We address issues related to coronal structure, flares, the composition of coronal plasma, X-ray production in 
accretion streams and outflows, X-rays from single OB-type stars,  massive binaries, magnetic hot objects and evolved WR stars.
\end{abstract}

\setcounter{tocdepth}{3}
\tableofcontents

\section{Introduction}
\label{sec:mainintro}
Stars are among the most prominent sources accessible to modern X-ray telescopes. In fact, stars
located across almost all regions of a Hertzsprung-Russell diagram have been identified as X-ray
sources, with only a few exceptions, most notably A-type stars and the coolest giants
of spectral type M. But even for those two classes, important exceptions exist. X-rays have
been identified from the most massive and hottest stars, i.e., O-type and Wolf-Rayet stars,
for which shocks forming in unstable winds are held responsible for the production of 
million-degree plasma and the ensuing X-rays. X-rays are therefore tracers of stellar mass loss
and sensitive diagnostics of stellar-wind physics. In hot star binaries, winds may collide,
thus forming very hot plasma in the wind collision region. The X-ray emission and its
modulation with orbital phase then provide precise constraints on the
colliding wind region, and hence the wind properties of each star. 

In cooler stars of spectral classes F to M, magnetic
coronae, overall analogous to the solar corona, are at the origin of X-rays, enriched by
flares in which unstable magnetic fields reconnect and release enormous amounts of energy in 
a matter of minutes to hours. The presence of coronae in these stars testifies to the
operation of an internal dynamo that generates the  magnetic fields. Although X-rays
provide easy diagnostics for such fields in the corona, the coronal phenomenon is by no means
restricted to X-ray sources but should rather be understood as the ensemble of closed
{\it magnetic field structures} anchored in the stellar photosphere. Some of these magnetic loops
may be in the process of being heated, thus filling up with hot plasma, while others are
not. 

X-rays have also been detected from brown dwarfs; again, the emission mechanism is supposedly
coronal. Similarly, in low-mass pre-main sequence stars, i.e., T Tauri stars (TTS) or (low-mass) protostars,
intense steady and flaring X-ray radiation is also thought to originate predominantly
in hot coronal plasma although the complex environment of such stars allows, in principle,
additional X-ray sources to be present. Shocks forming at the photospheric footpoints 
of disk-to-star accretion flows have been proposed to generate detectable X-ray emission, and
high-resolution X-ray spectra indeed seem to provide the corresponding diagnostics.
Further, outflows and jets form X-rays in internal shocks and shocks with the interstellar medium
(Herbig-Haro objects). 

X-rays not only provide invaluable diagnostics for winds, magnetic fields, accretion and outflow physics.
They can become key players in their own right. In young stellar environments, for example,
they act as ionizing and heating agents in circumstellar disks which then grow unstable
to accretion instabilities in the presence of magnetic fields. Also, X-rays are well known to 
drive a complex chemistry in molecular environments such as circumstellar disks and envelopes.
Once planets have formed, X-rays and the lower-energy extreme ultraviolet (EUV) radiation
may heat and evaporate significant fractions of their outer atmospheres,
contributing to the loss of water and therefore playing a key role in determining the ``habitability''
of a planet.

Many of these topics have been addressed during the past three to four decades of stellar X-ray astronomy.
A decisive boost came, however, with the introduction of high-resolution 
X-ray spectroscopy. While familiar to solar coronal physics for many years, high-resolution X-ray spectroscopy
was the poor cousin of X-ray photometry (possibly with some low energy resolution) until recently, even though a 
few notable experiments like crystal spectrometers or gratings were carried on early satellites 
(e.g., Einstein, EXOSAT). These, however, required exceptionally bright X-ray sources. The Extreme-Ultraviolet
Explorer (EUVE) gave a  first taste of routine high-resolution spectroscopy in the high-energy domain, 
concentrating mostly on the 90--300~\AA\  spectral region that contains many spectral lines of highly ionized 
species formed in million-degree plasmas. Given the large attenuation of EUV photons by the interstellar 
medium, only EUV bright, mostly nearby sources (predominantly cool stars) were the subject of spectroscopic 
observations.  A summary is given by \citet{bowyer00}.

High-resolution X-ray spectroscopy has been provided by grating instruments on the \xmm\ 
and \ch\  X-ray observatories which both were launched in 1999. Their broad wavelength coverage (0.07-15~keV),
effective areas (up to $\approx 200$~cm$^2$) and their impressive spectral resolving power (up to $\approx
1000$) have for the first time allowed many spectral lines to be separated, line multiplets to be resolved, 
and in some cases line profiles and line shifts to be measured. Such {\it spectroscopic} measurements 
have opened the window to stellar coronal and stellar wind physics, including pinpointing the location of 
X-ray emitting sources, determining densities of hot plasmas, measuring their composition 
or assessing X-ray ionization physics in stellar environments. 

These topics define the main focus of the present review. We aim at summarizing our understanding of 
cool and hot star physics contributed by X-ray spectroscopy. Naturally, therefore, we will focus on
observations and interpretations that the \xmm\  and \ch\  high-resolution 
spectrometers have made possible for now almost a decade. Older, complementary results from EUVE will occasionally
be mentioned. However, understanding cosmic sources cannot and should not be restricted to the use
of spectroscopic data alone. Although spectroscopy provides unprecedented richness of physical information,
complementary results from, e.g., X-ray photometric variability  studies or thermal
source characterization based on low-resolution spectroscopy are invaluable in many cases. We highlight,
in particular, the much higher effective areas of present-day CCD detectors that permit a rough
characterization of large samples of stars inaccessible to current grating spectroscopy. A particularly
important example is the study of the 6.4-7 keV line complex  due to highly ionized iron in extremely 
hot gas on the one hand and to fluorescent emission from ``cold'' iron at low ionization stages on the other 
hand. This feature is presently best studied using CCD spectra with resolving powers of 
$\approx 50$, in full analogy to line features at lower energies preferentially investigated with gratings.

We will, however, not concentrate on issues predominantly derived from light curve monitoring,
potentially important for the localization of X-ray sources in the stellar environments or the study of flares; we will
also not focus on the thermal characterization of X-ray sources based on low-resolution spectroscopy available
before the advent of \xmm\  and \ch; further, the many specific subclasses forming a zoo of variations of our 
themes, such as rapidly rotating giants, giants beyond the corona-wind dividing line, Ap stars, ``Luminous 
Blue Variables'' and others will not be treated individually as we wish to emphasize common physics related
to coronae, winds, and accretion/outflow systems. Finally, this review is not concerned with evolutionary and
population studies, for example in star-forming regions, in stellar clusters, or in stellar associations. These topics, 
equally important for a comprehensive picture of stellar physics and stellar evolution, have been reviewed 
elsewhere (e.g., \citealt{favata03b, guedel04}).

We have structured our article as follows. The first  chapter (Sect.~\ref{sec:cool})
addresses X-ray spectroscopy of cool stars, focusing on the thermal and geometric coronal structure, coronal composition, 
and flare physics. We then turn to new features (Sect.~\ref{sec:yso}) found in pre-main sequence stars with accretion 
and outflows. Finally, Sect.~\ref{sec:intro} reviews results from X-ray spectroscopy of massive, hot stars.

\section{X-rays from cool stars}
\label{sec:cool}

\subsection{Coronal X-ray luminosities and temperatures}
\label{sec:coolstars}
The Hertzsprung-Russell diagram rightward of spectral class A is dominated by stars with outer convection zones,
and many of these also have inner radiative zones. In these stars, an interaction of convection with rotation produces
a magnetic dynamo at the base of the convection zone, responsible for a plethora of magnetic phenomena
in or above the stellar photosphere (e.g., magnetic spots, a thin chromosphere, magnetically confined 
coronal plasma occasionally undergoing flares, etc.).

The study of our nearby example of a cool star, the Sun, has provided a solid framework within which to interpret
X-ray emission from cool stars. Indeed, essentially every type of cool star except late-type giants has 
meanwhile been identified in X-rays with characteristics grossly similar to what we know from the 
Sun. This includes objects as diverse as G- and K-type ``solar-like'' main-sequence stars, late-M dwarf
``flare stars'' and brown dwarfs (which are, however, fully convective and for which different dynamos may 
operate), protostars and accreting or non-accreting TTS (many of which are again fully convective),
intermediate-mass A-type and pre-main-sequence Herbig Ae/Be stars, post-main sequence active
binaries, and F--K-type giants. Most of these stars show variable X-ray emission at temperatures of at 
least 1--2~MK and occasional flaring.  

As we expect from a rotation-induced internal dynamo, the luminosity level is fundamentally correlated 
with rotation, as was shown in the early seminal studies based on the {\it Einstein} stellar survey
\citep{pal81, walter81b}; the X-ray luminosity ($L_{\rm X}$), the projected  
rotation period ($v\sin i$; in a statistical sense), the rotation period ($P$), the rotation rate 
($\Omega$), and the Rossby number ($Ro = P/\tau_c$, $\tau_c$ being the convective turnover time)
are related by the following equations,
\begin{eqnarray}\label{lxprot}
L_{\rm X} &\approx& 10^{27}(v\sin i)^2 \quad{\rm [erg~s^{-1}] }, \\
L_{\rm X} &\propto&   \Omega^2 \propto P^{-2}, \\
F_{\rm X}, {L_{\rm X} \over L_{\rm bol}} &\propto& Ro^{-2}   
\end{eqnarray}
(see \citealt{pizzolato03} for further details).
These trends, however, saturate at levels of $L_{\rm X}/L_{\rm bol} \approx 10^{-3}$. At this point, a further
increase of the rotation rate does not change $L_{\rm X}$ anymore. This break occurs at somewhat
different rotation periods depending on spectral type but is typically located at $P\approx 1.5-4$~d, 
increasing toward later main-sequence spectral types (\citealt{pizzolato03}; a description in terms of Rossby 
number provides a more unified picture). The Sun with 
$L_{\rm X} \approx 10^{27}$~erg~s$^{-1}$ is located at the lower end of the activity distribution, with
$L_{\rm X}/L_{\rm bol} \approx 10^{-7...-6}$ and therefore far from any saturation effects. It is presently 
unclear why X-ray emission saturates. Possibilities include a physical saturation of the dynamo, 
or a complete filling of the surface with X-ray strong active regions. For the fastest rotators
($v$ greater than about 100~km~s$^{-1}$), $L_{\rm X}$ tends to slightly decrease again, a regime known as 
``supersaturation'' \citep{randich96}. 

There are some variations of the theme also for pre-main-sequence stars. Given their much deeper convective
zones and ensuing longer convective turnover times, the saturation regime for typical TTS reaches
up to about 30~d; almost all pre-main sequence stars may therefore be X-ray saturated \citep{flaccomio03, preibisch05}. 
For TTS samples, $L_{\rm X}/L_{\rm bol} \approx 10^{-3.5}$, although there is a large scatter around
this saturation value, and the subsample of accreting TTS (``classical T Tauri'' stars or CTTS) reveals
ratios on average about a factor of 2 lower than non-accreting TTS (``weak-lined T Tauri'' stars or WTTS; 
\citealt{preibisch05, telleschi07a}).

The second fundamental parameter of cool-star X-ray sources is the characteristic or ``average'' coronal
temperature. Again, the Sun is located at the lower end of the range of coronal temperatures, showing
$T_{\rm av}\approx 1.5-3$~MK, somewhat depending on the activity level (of course, individual features 
will reveal considerable variation around such averages at any instance). An important although unexpected finding
from early surveys was a correlation between $T_{\rm av}$ and $L_{\rm X}$ (e.g., \citealt{schrijver84, schmitt90a}; see
\citealt{telleschi05} for a recent investigation of solar analogs with similar spectral types and masses).
Roughly, one finds
\begin{equation}\label{TLx}
L_X \propto T^{4.5\pm 0.3}
\end{equation}
although the origin of this relation is unclear. It may involve more frequent magnetic interactions between
the more numerous and more densely packed active regions on more active stars, leading to higher
rates of magnetic energy release (flares) that heat the corona to higher temperatures (we will return
to this point in Sect.~\ref{sec:dem} and \ref{sec:flares}). 

\subsection{The thermal structure of coronae, and the coronal heating problem}
\label{sec:dem}
\begin{figure}[t]
\centerline{\includegraphics[angle=0, width=9.8cm]{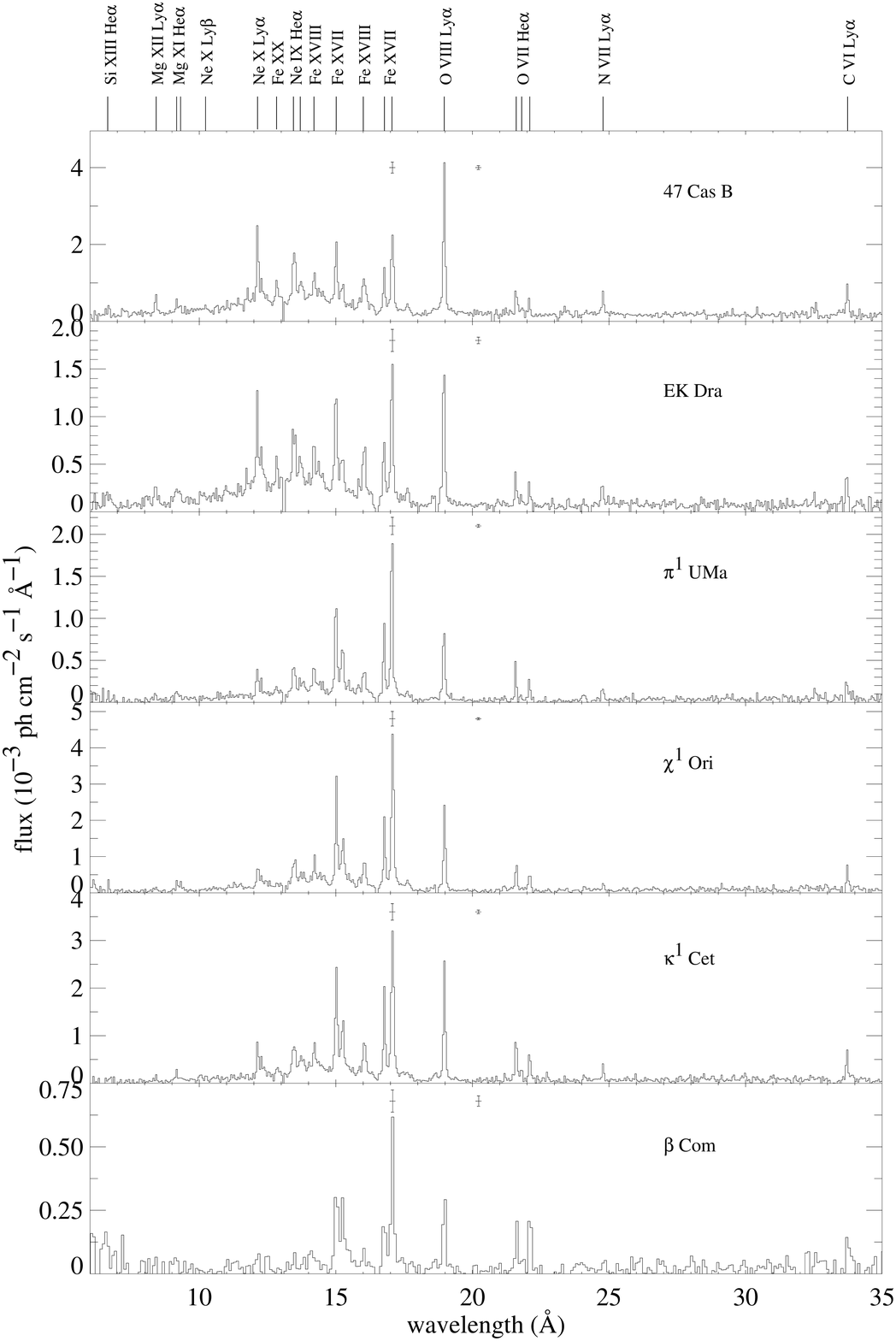}}
\vskip -0.8truecm
\caption{\xmm\ X-ray grating spectra of solar analogs at different activity levels. From top to bottom, age increases, while the 
overall activity level and the characteristic coronal temperatures decrease. (From \citealt{telleschi05}, reproduced by 
permission of the AAS.)}
\label{fig:solaranalogs} 
\end{figure}

Understanding the thermal coronal structure requires spectroscopic data of a number of lines forming 
at different temperatures. The advent of high-resolution X-ray spectroscopy provided by \xmm\ 
and \ch\ has opened a new window to coronal structure, as summarized below. Examples of solar analog stars
with different ages and therefore activity levels are shown in Fig.~\ref{fig:solaranalogs} (see also Fig.~\ref{fig:ttau} 
for further examples of very active and inactive stars). As judged from the pattern of emission lines and the
strength of the continuum, the spectra  show that the dominant temperatures in
the coronal plasma decrease with increasing age (from top to bottom).

Observed X-ray line fluxes as well as the continuum scale with the emission measure, EM $= n_{\rm e}n_{\rm H}V$ at a given 
temperature, where $n_{\rm e}$ and $n_{\rm H}$ ($\approx 0.85n_{\rm e}$ for a fully ionized plasma) are the number densities of electrons 
and hydrogen ions (protons), respectively, and $V$ is the emitting volume. For a distribution of emission measure in temperature, 
the relevant quantity is the differential emission measure (DEM), $Q(T) = n_{\rm e}n_{\rm H} {\rm d}V(T)/{\rm d}{\rm ln}T$ that gives the
emission measure per unit interval in  ${\rm ln}T$. Clearly, the DEM provides only a degenerate description of a complex
corona, but it nevertheless contains statistical information on the average distribution of volumes in temperature and therefore
- indirectly through modeling - also on the operation of heating and cooling mechanisms.

Deriving the thermal structure of an optically thin stellar corona is a problem of {\it spectral inversion},
using temperature-sensitive information available in spectral lines and the continuum.  Many inversion
methods have been designed, but regardless of the methodology, spectral inversion is an ill-conditioned problem and
allows for a large number of significantly different but nevertheless ``numerically correct'' solutions (in the 
sense of reproducing a specified portion of the spectrum sufficiently well). The non-uniqueness of the 
inversion problem is  rooted in the broad temperature range over which a given line forms, combined with  
the discretization of the problem (e.g., on a temperature grid) and uncertainties in the calibration, the tabulated 
atomic physics parameters, and even counting statistics \citep{craig76}. This limitation of spectral inversion is 
fundamentally mathematical and cannot be removed even  if data with perfect spectral resolution and perfect
precision are at hand.
In this sense, any inversion of a spectrum - {\it regardless of the inversion methodology} - is as good as any other if it
reproduces the observed spectrum similarly well.\footnote{Coronal research has developed several classes of inversion
techniques, best separated into {\it discrete inversion techniques} that reconstruct the emission measure distribution
from measured and extracted flux values of selected emission lines, and {\it continuous inversion techniques}
that treat the spectrum as a function to be fitted with a superposition of synthetic template spectra. The often used expressions
``line-based analysis'' and ``global fit'', respectively, miss the essence of these techniques, as all methods
ideally use a large fraction or the entirety of the available spectrum in which the relevant information is
provided mostly by line flux ratios. Basically, equivalent solutions can be found
from both classes of inversion techniques \citep{telleschi05}.} The discrimination between ``physically acceptable'' 
and ``physically unacceptable'' solutions requires that {\it additional conditions} be imposed on the emission measure 
distribution, which are in fact constraining the ``physics'' of the problem  (e.g., smoothness of the emission measure 
distribution as required by thermal conduction for typical coronal features). These physical constraints obviously 
{\it cannot} be identified by any mathematical inversion technique. For further comments on this problem, see 
\citet{guedel04}. 

Understanding emission measure distributions, but also reasonably constraining the spectral inversion process, 
requires theoretical models that link the {\it coronal thermal structure} with the DEM.
We first summarize model predictions for non-flaring magnetic loops, and then address flaring sources. 

Under the conditions of negligible gravity, i.e., constant 
pressure, and negligible thermal conduction at the footpoints, a hydrostatic loop
\citep{rosner78, vesecky79, antiochos86b} reveals a DEM given by
\begin{equation}\label{staticem}
Q(T) \propto pT^{3/4-\gamma/2+\delta}  {1 \over \left( 1 - \left[T/T_{\rm a}\right]^{2-\gamma+\beta}\right)^{1/2}}
\end{equation}
\citep{bray91}. Here, $T_{\rm a}$ is the loop apex temperature,  and $\delta$ and $\beta$ are power-law 
indices of, respectively,  the loop cross section area $S$ and the heating power $q$ as a function of $T$:
$S(T) = S_0T^{\delta}$,  $q(T) = q_0T^{\beta}$, and 
$\gamma$ is the exponent in the cooling function over the relevant temperature range: 
$\Lambda(T) \propto T^{\gamma}$. If $T$ is not close to $T_{\rm a}$, then  constant cross-section loops
($\delta = 0$) have $Q(T) \propto T^{3/4 -\gamma/2}$,
i.e., under typical coronal conditions for non-flaring loops ($T < 10$~MK, $\gamma \approx -0.5$), the DEM 
slope is near unity \citep{antiochos86b}. If strong thermal conduction is included at the footpoints, then
the slope changes to +3/2 if not too close to $T_{\rm a}$ \citep{vdoord97}.
For a distribution of loops with different temperatures, the descending, high-$T$ slope of the DEM is obviously
related to the statistical distribution of the loops in $T_{\rm a}$; a sharp decrease of the DEM then indicates 
that only few loops are present with a temperature exceeding the temperature of the DEM peak \citep{peres01}.

\citet{antiochos80} (see also references therein) discussed DEMs of flaring loops 
that cool by i) static conduction (without flows), or ii) evaporative conduction (including flows), 
or iii) radiation. The DEMs for these three cases scale like (in the above order)
\begin{equation}
Q_{\rm cond} \propto  T^{1.5}, \quad\quad  Q_{\rm evap} \propto T^{0.5}, 
     \quad\quad Q_{\rm rad} \propto T^{-\gamma + 1}. 
\end{equation}
Note that $\gamma \approx 0\pm 0.5$ in the range typically of interest for stellar flares
($5-50$~MK). All above DEMs  are therefore relatively flat (slope $1\pm 0.5$). 

For stellar flares that are too faint for time-resolved spectroscopy, the time-integrated DEM for a 
``quasi-statically'' decaying flare is
\begin{equation}\label{demflareint}
Q \propto T^{19/8}
\end{equation}
up to a maximum $T$ that is equal to the temperature at the start of the decay phase \citep{mewe97}.

In the case of episodic flare heating (i.e., a corona that is heated by a large number of stochastically 
occurring flares), the average, time-integrated DEM of coronal X-ray emission is determined not by the 
internal thermal structure of magnetic loops but by the time evolution of the emission measure and the
dominant temperature of the flaring region. In the case of dominant radiative cooling,
the DEM at a given temperature is roughly inversely proportional to the radiative decay time, which implies
\begin{equation}
Q(T) \propto T^{-\gamma+1}
\end{equation}
up to a maximum $T_{\rm m}$, and a factor of $T^{1/2}$ less if subsonic draining of the cooling loop is 
allowed \citep{cargill94}. Because the cooling function drops rapidly between 1~MK and $\la$10~MK, the DEM in this 
region should be steep, $Q(T) \propto T^4$.

An analytic model of a stochastically flaring corona powered by simple flares rising instantly to a peak 
and then cooling exponentially was presented by \citet{guedel03a}. Assuming a flare distribution in energy
that follows a power law (d$N$/d$E \propto E^{-\alpha}$, see Sect.~\ref{sec:flareover}), and making use of a relation
between flare peak temperature, $T_{\rm p}$, and peak emission measure EM$_{\rm p}$ (Sect.~\ref{sec:flareover}), the DEM is
\begin{equation} 
Q(T) \propto \left\{ 
   \begin{array}{ll}\label{flareheat}
        T^{2/\zeta}\quad\quad\quad\quad\quad\quad\quad\quad\quad\quad\quad  &, afT^{b+\gamma} \le  L_{\mathrm{min}} \\
        T^{-(b+\gamma)(\alpha-2\beta)/(1-\beta) +2b + \gamma} \quad             &, afT^{b+\gamma} >    L_{\mathrm{min}} 
   \end{array} 
   \right. 
\end{equation}
where $\zeta$ relates temperature and density during the flare decay, $T \propto n^{\zeta}$; from the run of
flare temperature and emission measure in stellar flares, one usually finds $0.5 \la \zeta\la 1$. The parameter 
$b$ follows from the relation between flare peak temperature and emission measure, EM$_{\rm p} = aT_{\rm p}^b$, where 
$b \approx 4.3$ (see Sect.~\ref{sec:flareover}; \citealt{guedel04}); $\gamma$ determines the plasma cooling function $\Lambda$ as
above,  $\Lambda(T) = fT^{\gamma}$ (for typical stellar coronal abundances, $\gamma \approx -0.3$ below 10~MK, 
$\gamma \approx 0$ in the vicinity of 10~MK, and 
$\gamma \approx 0.25$ above 20~MK). Further, $\beta$ describes a possible relation between flare decay time $\tau$
and the integrated radiative energy of the flare, $E$, $\tau \propto E^{\beta}$, with  $0 \le \beta \la 0.25$
(see \citealt{guedel04} for further details); $L_{\mathrm{min}}$ is the peak luminosity of the smallest flares
contributing to the DEM. For a flare-heated corona, this DEM model can in principle be used to assess the cooling behavior 
of flares (i.e., through $\zeta$ from the low-temperature DEM slope) and to derive the stochastic occurrence rate of flares 
(i.e., through $\alpha$ from the high-temperature DEM slope).
 
\begin{figure}[t]
\centerline{\includegraphics[angle=0,width=8.7cm]{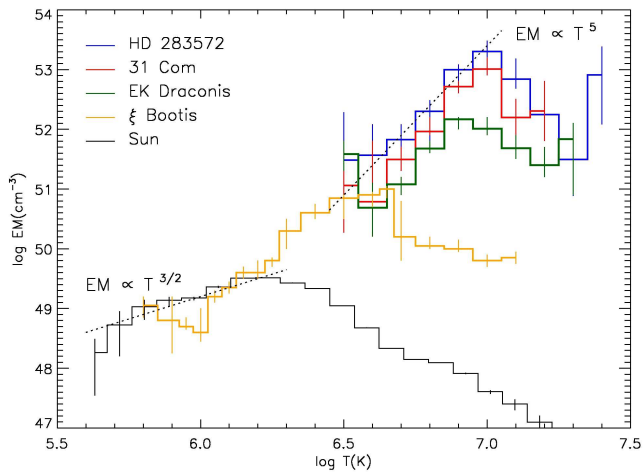}}
\caption{DEMs derived from spectra of stars at different activity levels. The black
histogram refers to the solar corona. Note the steeply increasing distributions for stars at higher activity
levels. (Adapted from \citealt{scelsi05}.)}
\label{fig:vardem} 
\end{figure}

The above relations can easily be applied to DEMs reconstructed from observed spectra provided that DEM slopes have
not been imposed as constraints for the inversion process. In fact, DEMs derived
from stellar coronal spectra almost invariably show a relatively simple shape, namely an increasing power-law on 
the low-temperature side up to a peak, followed by a decreasing power-law up to a maximum temperature (Fig.~\ref{fig:vardem}). 
The DEM peak may itself be a function of activity in the sense that it shifts to higher temperatures in more active
stars, often leaving very little EM at modest temperatures and correspondingly weak
spectral lines of C, N, and O (see, e.g., \citealt{drake00, telleschi05, scelsi05}; Fig.~\ref{fig:vardem}). 
This behavior of course reflects the by now classic result that ``stars at higher activity levels show higher average 
coronal temperatures'' (e.g., \citealt{schmitt90a}). Given the ill-conditioned inversion problem, many additional 
features turn up in reconstructed DEMs, such as local maxima or minima, but their reality is often difficult to assess.

Despite the models now available for a description of  DEMs, we do not clearly understand which stellar parameters 
shape an emission measure distribution. The trend mentioned above, a correlation between 
average coronal temperature and ``activity level'', is reminiscent of a similar relation for individual stellar
flares (``more-energetic flares are hotter''; Sect.~\ref{sec:flareover}), perhaps suggesting a connection between continuous flaring
and overall coronal heating. Other parameters are less relevant. For example, active G stars at very different
evolutionary stages (e.g., giants, main-sequence stars, pre-main sequence stars), with different coronal abundances 
and surface gravities may reveal very similar DEMs \citep{scelsi05}.

A principal finding of major interest are the unusually steep low-$T$ sides of DEMs of active stars, with slopes
in the range of 2--5 \citep{drake00, behar01, mewe01, argiroffi03, guedel03a, telleschi05, scelsi05}. There is evidence that
the slopes increase with increasing activity, as again illustrated in Fig.~\ref{fig:vardem}. Such slopes clearly exceed values expected 
from hydrostatic loops (1--1.5). Numerical simulations of loops undergoing repeated pulse heating at their footpoints do produce
DEMs much steeper than the canonical values \citep{testa05}. An alternative solution may be loops with an expanding 
cross section from the base to the apex. In that case, there is comparatively more hot plasma, namely the plasma located 
around the loop apex, than cooler plasma. The DEM would consequently steepen.

Steep DEMs have also been interpreted in terms of continual flaring using Eq.~\ref{flareheat} \citep{guedel03a, audard04, maggio04, telleschi05}. 
For solar analog stars at different activity levels, \citet{telleschi05} used the high-$T$ slope of the DEMs
to infer $\alpha = 2.1-2.8$ for the flare energy distribution (Sect.~\ref{sec:flareover}), suggesting that the ensemble of ``weak'',
unresolved
flares may generate the observed X-ray emission. Radiative flare energies in the range of $10^{25} - 10^{30}$~erg~s$^{-1}$ would be
required. From the DEM of the rapidly rotating giant FK Com, \citet{audard04} inferred a steep flare energy distribution 
with $\alpha = 2.6-2.7$; such distributions produce relatively flat light curves in which individual, strong flares
are rare, compatible with the observations.

\subsection{Coronal structure from X-ray spectroscopy}
\label{sec:structure_spec}
Understanding stellar coronal structure goes a long way in understanding stellar magnetism and 
dynamos. The distribution of magnetic fields near the stellar surface diagnoses 
type and operation of the internal magnetic dynamo; the structure of coronal magnetic fields
is relevant for the energy release mechanism and determines the thermal structure of 
trapped hot plasma. Further, extended magnetic fields may connect to the companion star in close 
binaries, or the inner border of circumstellar disks in TTS or protostars;
additional physical processes such as angular momentum transfer or mass flows are important consequences.

Despite a suite of methods to infer {\it some} properties of coronal structure, all available
methods are strongly affected by observational bias. This is principally due to the fact
that the defining constituent of a corona, the {\it magnetic field} itself, is extremely
difficult to measure; coronal structure is mostly inferred from observable signatures of magnetically confined, hot 
plasma (as seen in the EUV or X-ray range) or trapped energetic particles (as seen in the
radio range). Considering the complexity of the solar magnetic corona, its large range of size scales,
and the important role that coronal {\it fine-structure} plays, we should not be surprised
that presently available methods provide some limited qualitative sketches of what is a much more complex, 
highly dynamic system driven by continuous release and transformation of energy, coupled with mass motions
and cooling processes. Before concentrating on high-resolution spectroscopic techniques, we briefly summarize alternative
methods and results derived from them (see \citealt{guedel04}  for more details).

\subsubsection{Summary of low-resolution spectroscopic and non-spectroscopic X-ray studies}
\label{sec:structure}
{\it Hydrostatic loop models} \citep{rosner78, vesecky79, serio81} have been extensively
used in solar and stellar X-ray astronomy to relate pressure, apex (peak) temperature, heating
rate, and length of simple, static coronal magnetic loops anchored in the photosphere. In its
simplest form, a half-circular loop of semi-length $L$ (footpoint to apex), pressure $p$,
apex temperature $T_{\rm a}$, and heating rate $\epsilon$ follows the two relations
\citep{rosner78}.
\begin{equation}\label{loop}
T_{\rm a} = 1400(pL)^{1/3} \quad\quad\quad \epsilon = 9.8\times 10^4p^{7/6}L^{-5/6}.
\end{equation}
Measuring, e.g., $T_{\rm a}$ and relating the observed $L_{\rm X}$ to $\epsilon$, the loop length
can be inferred, depending on a surface filling factor $f$ for the loops filled with the
observed plasma. Judged from such assessments, magnetically active stars require very large,
moderate-pressure loops with a large surface filling factor, or alternatively  more solar-sized high-pressure
compact loops with very small filling factors ($<$1\%, e.g., \citealt{schrijver89b, giampapa96}).

Synthetic emission spectra from {\it loop-structure models} of this kind have been applied 
to observed spectra of active stars. One typically finds mixtures
of magnetic loops ranging from cool (1.5--5~MK), moderate-pressure (2--100~dyn~cm$^{-2}$)
loops to hot (10--30~MK) extreme-pressure ($10^2-10^4$~dyn~cm$^{-2}$) loops reminiscent of flaring loops
(\citealt{ventura98}). We need to keep in mind, however, that model solutions are
degenerate in the product $pL$ (see Eq.~\ref{loop}), potentially resulting in multiple
solutions. Caution should therefore be applied when interpreting ``best-fit'' results based
on the assumption of one family of identical magnetic loops. 

Coronal imaging by {\it light-curve inversion} makes use of the fortuitous arrangement of 
components in binary stars producing {\it coronal eclipses}, or a large inclination of the 
rotation axis of a single star resulting in {\it self-eclipses} (``rotational modulation''). Image 
reconstruction from light curves is generally non-unique but can, in many cases, be constrained 
to reasonable and representative solutions. In the 
simplest case, {\it active region modeling}, similar to surface spot modeling, provides 
information on the location and size of the dominant coronal features (e.g., \citealt{white90}). 
More advanced image reconstruction methods (maximum-entropy-based methods, backprojection/clean methods, etc.) 
provide entire maps of coronal emission (e.g., \citealt{siarkowski96}).
Again, a mixture of compact, high-pressure active regions and much more extended ($\approx R_*$),
lower-pressure magnetic features have been suggested for RS CVn-type binaries \citep{walter83b, 
white90}. Light curves also provide important information on inhomogeneities and the global 
distribution of emitting material; X-ray bright
features have been located on the leading hemispheres in binaries \citep{walter83b, ottmann93b}, 
or on hemispheres that face the companion (e.g., \citealt{white90, siarkowski96}), perhaps
suggesting some role for {\it intrabinary magnetic fields}. Such results indicate that even 
the most active stars are not entirely filled by X-ray or radio emitting active regions,
as surface filling factors reach values of sometimes no more than 5--25\% (e.g., \citealt{white90,
ottmann93b}). In a few special cases, flares have been mapped using the fortuitous eclipsing
by a close companion star. Such observations have located the flaring structure either at the pole
or near the equator and have constrained the size of the flaring magnetic fields  to typically 
a few tenths of $R_*$ (for examples, see \citealt{choi98, schmitt99, schmitt03, sanz07}). As a 
by-product, characteristic electron densities of the flaring plasma can be inferred  to be of order 
$10^{11}$~cm$^{-3}$,  exceeding, as expected, typical non-flaring coronal densities (Sect.~\ref{sec:densities}).

{\it Magnetic field extrapolation} using surface Doppler (or Zeeman Doppler) imaging of magnetic 
spots has been used in conjunction with X-ray rotational modulation observations to study distribution
and radial extent of coronal magnetic fields (e.g., \citealt{jardine02a, jardine02b, hussain02, hussain07}). 
Although permitting a 3-D view of a stellar corona, the method has its limitations as part of 
the surface is usually not accessible to Doppler imaging, and small-scale magnetic structure on 
the surface is not resolved. The type of field extrapolation (potential, force-free, etc) must 
be assumed, but on the other hand, the 3-D coronal model can be verified if suitable 
rotational modulation data are available \citep{gregory06}.
   
\subsubsection{Coronal structure from spectroscopic line shifts and broadening}
\label{sec:lineshifts}
Doppler information in X-ray spectral lines may open up new ways of imaging 
coronae of stars as they rotate, or as they orbit around the center of gravity in binaries.
In principle, this method can be used to pinpoint the surface locations, heights, and sizes of 
coronal features. Applications are very limited at the present time given 
the available X-ray spectral resolving power of $\la 1000$. Shifts corresponding to 100~km~s$^{-1}$ 
correspond to less than the instrumental resolution, but such surface (or orbital) velocities are 
attained only in exceptional cases of young, very rapidly rotating stars or rotationally locked, 
very close binaries. We summarize a few exemplary studies.

Amplitudes  of $\approx 50$~km~s$^{-1}$ and phases of Doppler shifts measured in the RS CVn binary HR~1099
agree well with the line-of-sight orbital velocity of the subgiant K star, thus locating the bulk 
of the X-ray emitting plasma  on this star, rather than in the intrabinary region \citep{ayres01b}. 
In contrast, periodic 
line {\it broadening} in the dMe binary YY Gem, consisting of two nearly identical M dwarfs, 
indicates, as expected, that both components are similarly X-ray luminous \citep{guedel01a}.
Doppler shifts in the RS CVn binary AR Lac \citep{huenemoerder03}  are compatible with coronae on both 
companions if the plasma is close to the photospheric level. For the contact binary 44i Boo, 
\citet{brickhouse01} reported periodic line shifts corresponding to a total net velocity change over the 
full orbit of 180 km~s$^{-1}$. From the amplitudes and the phase of the rotational modulation, the authors
concluded that two dominant X-ray sources were present, one being very compact 
and the other being extended, but both being located close to the stellar pole of the larger companion. Similar
results have been obtained for another contact binary, VW~Cep \citep{huenemoerder06}, revealing that almost all X-rays
are emitted by a relatively compact corona (height 0.06-0.2$R_*$) almost entirely located on the primary star.

Applications are more challenging for rapidly rotating single stars. \citet{hussain05} did find periodic line shifts
in the spectrum of the young AB Dor which, together with light curve modulation, suggested a coronal model
consisting of a relatively low-lying distributed corona (height $\la 0.5~R_*$) and several more compact
(height $< 0.3~R_*$) active regions. This result, when combined with coronal extrapolations from surface
Doppler imaging and spectroscopic coronal density measurements, further constrains the corona to heights
of $0.3-0.4~R_*$, and reasonable 3-D models of the coronal structure can be recovered \citep{hussain07}.

A comprehensive study of line shifts and line broadening has been presented for the Algol binary 
\citep{chung04}. Periodic line shifts corresponding to a quadrature radial velocity of 150~km~s$^{-1}$
clearly prove that the X-rays are related to the K subgiant. However, the amplitude of the shifts
indicates that the source is slightly  displaced toward the B star, which may be the result of tidal 
distortions by the latter. Excess line broadening (above thermal and rotational broadening) 
can be ascribed to a radially extended corona, with a coronal scale height of order one stellar 
radius, consistent with expected scale heights of hot coronal plasma on this star.

A rather predestined group of stars for this type of study are the rapidly rotating single giants of the
FK Com class, thought to have resulted from a binary merger \citep{bopp81}. With surface velocities of order
100~km~s$^{-1}$,  shifts or  broadening  of bright lines can be measured. \citet{audard04} found 
significant line broadening corresponding to velocities of about 100--200~km~s$^{-1}$ in the K1~III giant YY Men;
the broadening could in principle 
be attributed to rotational broadening of a coronal source above the equator, confined to a height of 
about a pressure scale height ($\approx 3R_*$). YY Men's extremely hot corona ($T\approx 40$~MK), however,
makes Doppler thermal broadening of the lines a more plausible alternative \citep{audard04}. The prototype of the
class, FK Com, also shows indications of line shifts (of order 50--150~km~s$^{-1}$) and marginal suggestions for
line broadening. The X-ray evidence combined with contemporaneous surface Doppler imaging suggests the presence
of near-polar active regions with a height of $\approx 1R_*$ \citep{drake08a}.
Taken together with other observations, there is now tentative evidence for X-ray coronae around active giants being
more extended (relative to the stellar radius) than main-sequence coronae, which are predominantly compact
(height $\la 0.4R_*$, \citealt{drake08a}).

\subsubsection{Inferences from coronal densities}
\label{sec:densities}
High-resolution X-ray spectroscopy has opened a window to coronal densities because the X-ray range
contains a series of density-sensitive lines; they happen to be sensitive to expected coronal and 
flare plasma densities. Electron densities have mostly been inferred from 
line triplets of He-like ions on the one hand and lines of Fe on the other hand. We briefly 
review results from these in  turn, and then summarize implications for coronal structure.

\begin{figure}[t]
\includegraphics[angle=0,width=12cm]{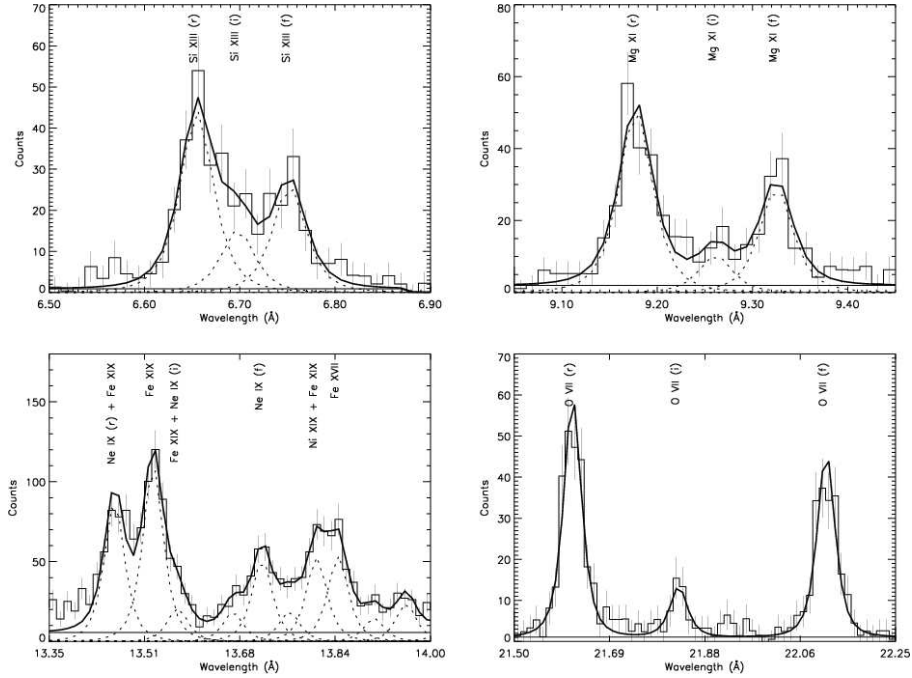}
\caption{He-like triplets of  Si\,{\sc xiii} (upper left), Mg\,{\sc xi} (upper right),
Ne\,{\sc ix} (lower left), and O\,{\sc vii} (lower right) extracted from the \ch\ LETGS spectrum of Capella.
Multi-line fits are also shown. The horizontal lines indicate the continuum level.
(From \citealt{argiroffi03}.)}
\label{fig:hetriplets} 
\end{figure}

\begin{figure}[t!]
  \centerline{\includegraphics[width=6cm,height=5.5cm,bb=30 200 570 580, clip]{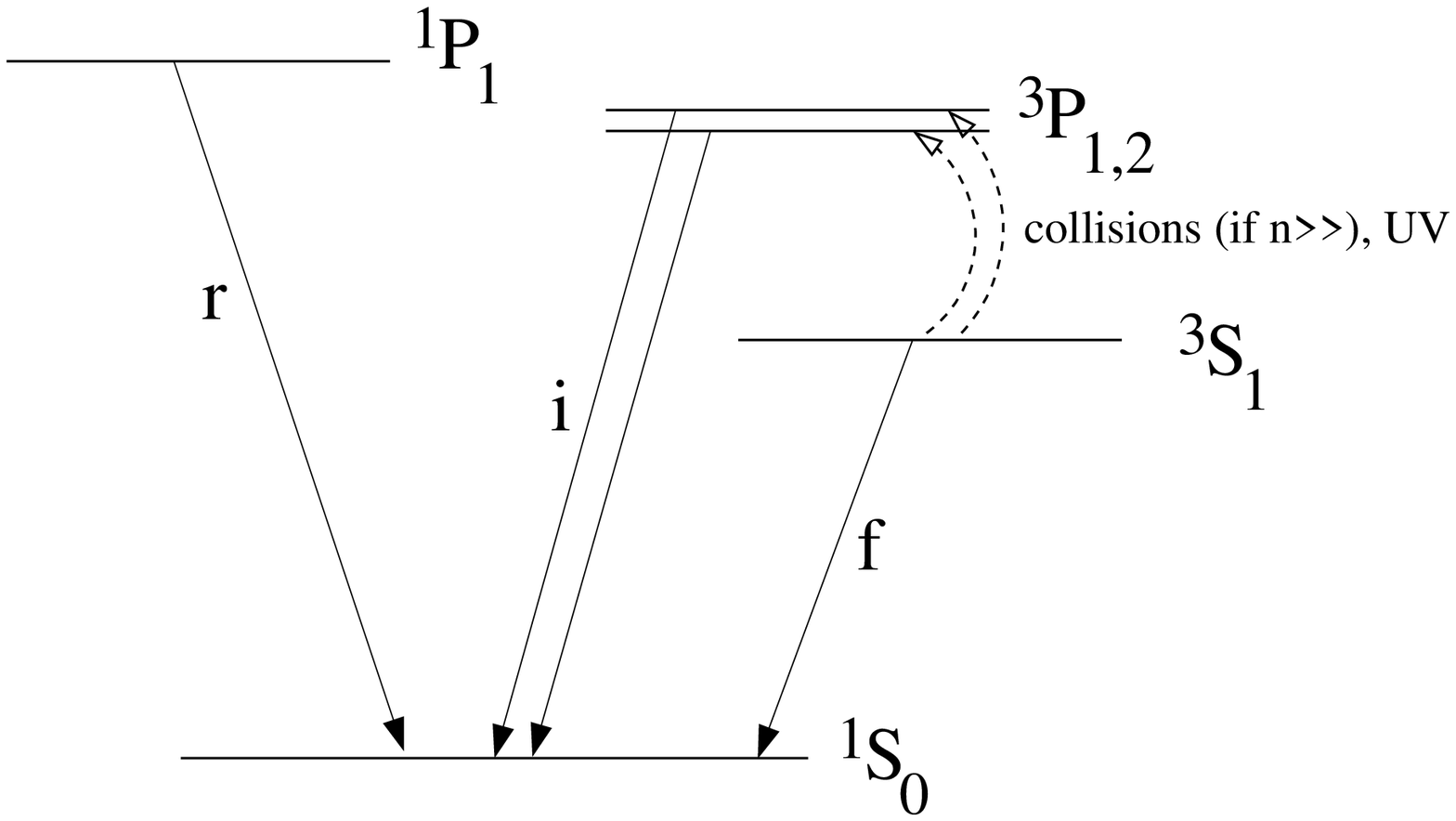}}
  \caption{Schematic diagram showing the origin of the {\it fir} lines in a He-like ion.}
  \label{fir}
  \end{figure}

\paragraph{Coronal densities from He-like triplets}
He-like triplets of C\,{\sc v}, N\,{\sc vi}, O\,{\sc vii}, Ne\,{\sc ix}, Mg\,{\sc xi}, and Si\,{\sc xiii}  
show, in order of increasing wavelength, a resonance ($r$, $1s^2\ ^1S_0 - 1s2p\ ^1P_1$), an intercombination 
($i$, $1s^2\ ^1S_0 - 1s2p\ ^3P_{1,2}$), 
and a forbidden ($f$, $1s^2\ ^1S_0 - 1s2s\ ^3S_1$) line (Fig.~\ref{fig:hetriplets}). 
The ratio between the $f$ and $i$ fluxes is sensitive to density \citep{gabriel69} 
for the following reason: if the electron collision rate is sufficiently 
high, ions in the upper level of the forbidden transition, $1s2s\ ^3S_1$, do not return to the ground level, $1s^2\ ^1S_0$, 
instead the ions are collisionally excited to the upper level of the intercombination transitions, $1s2p\ ^3P_{1,2}$, 
from where  they decay radiatively to the ground state (for a graphical presentation, see Fig.~\ref{fir}). 

The measured ratio ${\cal R} = f/i$ of the forbidden to the intercombination 
line fluxes can be approximated by
\begin{equation}
{\cal R} = {{\cal R}_0 \over 1 + n_{\rm e}/N_c} = {f\over i}
\end{equation}
where ${\cal R}_0$ is the limiting flux ratio at low densities and $N_c$ is the critical
density at which ${\cal R}$ drops to ${\cal R}_0/2$ (we ignore the influence of the photospheric ultraviolet radiation 
field for the time being; see Sect.~\ref{sec:herbigs} and \ref{sec:oldmodel} below). Table~\ref{lineratios} contains 
relevant parameters for triplets interesting for coronal studies; they refer to the case of 
a plasma that is at the maximum formation temperature of the respective ion (for detailed tabulations, see \citealt{porquet01}). 
A systematic problem with He-like triplets is that the critical density $N_c$ increases with the formation temperature of 
the ion, i.e., higher-$Z$ ions measure only high densities at high $T$, while
the lower-density analysis based on C\,{\sc v}, N\,{\sc vi}, O\,{\sc vii}, and Ne\,{\sc ix} is 
applicable only to cool plasma.

\begin{table}
\caption{Density-sensitive He-like triplets$^a$}
\label{lineratios}     
\begin{tabular}{llllll}
\hline\noalign{\smallskip}
Ion       &  $\lambda(r,i,f)$ (\AA)&   ${\cal R}_0$          & $N_c$              & log\,$n_{\rm e}$ range$^b$  &   $T$ range$^c$ (MK) \\
\noalign{\smallskip}\hline\noalign{\smallskip}
C\,{\sc v}       & 40.28/40.71/41.46      &  11.4            & $6\times 10^8$     &  7.7--10         &   0.5--2      \\
N\,{\sc vi}      & 28.79/29.07/29.53      &   5.3            & $5.3\times 10^9$   &  8.7--10.7       &   0.7--3      \\
O\,{\sc vii}     & 21.60/21.80/22.10      &   3.74           & $3.5\times 10^{10}$&  9.5--11.5       &   1.0--4.0    \\ 
Ne\,{\sc ix}     & 13.45/13.55/13.70      &   3.08           & $8.3\times 10^{11}$&  11.0--13.0      &   2.0--8.0    \\ 
Mg\,{\sc xi}     & 9.17/9.23/9.31         &   2.66$^d$       & $1.0\times 10^{13}$&  12.0--14.0      &   3.3--13     \\
Si\,{\sc xiii}   & 6.65/6.68/6.74         &   2.33$^d$       & $8.6\times 10^{13}$&  13.0--15.0      &   5.0--20     \\
\noalign{\smallskip}\hline
\multicolumn{6}{l}{$^a$data derived from \citet{porquet01} at maximum formation temperature of ion} \\
\multicolumn{6}{l}{$^b$range where ${\cal R}$ is within approximately [0.1,0.9] times ${\cal R}_0$ } \\
\multicolumn{6}{l}{$^c$range of 0.5--2 times maximum formation temperature of ion} \\
\multicolumn{6}{l}{$^d$for  measurement with \ch\  HETGS-MEG spectral resolution} \\
\end{tabular}
\end{table}

He-like triplets are usually bright in coronal spectra and have therefore been used extensively for density estimates
(e.g., \citealt{mewe01, ness01}). Large samples of coronal stars were surveyed by \citet{ness04} and \citet{testa04a}.
Although density estimates are roughly in line with experience from the solar corona (at least as far as
analysis of the O\,{\sc vii} triplet forming at temperatures of $\approx$2~MK is concerned), several systematic features 
have become apparent. The following trends predominantly apply to  O\,{\sc vii} derived densities:
\begin{itemize}
\item Low-activity stars tend to show low densities (often represented by upper limits), i.e., $\log n_{\rm e} < 10$ 
\citep{ness01, raassen03a}.

\item Higher densities significantly measured by the O\,{\sc vii} triplet, i.e., $\log n_{\rm e} = 10-11$, are only
reported from magnetically active stars, many of which are located on the main sequence. Examples include very active
solar analogs, very young K dwarfs such as AB Dor, or active M dwarfs (even higher densities have been reported for
accreting TTS; see Sect.~\ref{sec:accretion}).

\item For evolved active binaries (RS CVn binaries, contact binaries), both density ranges are relevant.
\end{itemize}
Higher-$Z$ triplets are more difficult to interpret, in particular because the range of sensitivity shifts to higher
density values that may exceed coronal values. These triplets are also subject to more problematic blending,
which is in particular true for the Ne\,{\sc ix} triplet that suffers from extensive blending by line of Fe,
specifically Fe\,{\sc xix} \citep{ness03b}. \citet{mewe01} found $n_{\rm e} > 3\times 10^{12}$~cm$^{-3}$ for Capella from
an analysis of Mg\,{\sc xi} and Si\,{\sc xiii}, but the results disagree with measurements using lines of Fe\,{\sc xx}-{\sc xxii} 
\citep{mewe01}. Similarly, \citet{osten03}, \citet{argiroffi03}, and \citet{maggio04} found sharply increasing densities 
moving from cooler to  hotter plasma, with densities reaching up to order $10^{12}$~cm$^{-3}$. But the trend reported by 
\citet{osten03} is contradicted by the analysis of 
Si\,{\sc xiii} that indicates $n_{\rm e} < 10^{11}$~cm$^{-3}$ despite its similar formation temperature as Mg\,{\sc xi}. 
The high-density results have been questioned altogether from detailed analyses of the Capella and II Peg spectra, for 
which upper limits to electron densities have been derived from Ne, Mg, and Si triplets 
\citep{canizares00, ayres01b, huenemoerder01, phillips01}.   

The most detailed analyses of the higher-$Z$ triplets are those by \citet{ness04} and \citet{testa04a}. Ne\,{\sc ix} density
measurements are typically higher than those using O\,{\sc vii}, covering the range of $\log n_{\rm e} = 11-12$ despite
the significant temperature overlap between the two ions \citep{ness04}. Unrecognized blends in the Ne\,{\sc ix} triplet 
may still be problematic. Mg\,{\sc xi} and Si\,{\sc xiii} systematically yield even higher densities for various types
of active stars. \citet{testa04a} report
Mg-derived densities of a few times $10^{12}$~cm$^{-3}$, with a trend for stars with higher $L_{\rm X}/L_{\rm bol}$
to show higher densities, a trend paralleling suggestions from O\,{\sc vii} (see above) albeit for much higher 
densities. The situation is less clear for Si\,{\sc xiii} as most measured $f/i$ flux ratios {\it exceed} the 
theoretical upper limit.

\paragraph{Coronal densities from Fe lines}
Many transitions of Fe ions are sensitive to density as well \citep{brickhouse95}. Line ratios of Fe\,{\sc xix}-{\sc xxii}
 in the EUV range have frequently been used for density estimates \citep{dupree93, schrijver95, sanz02}, 
resulting in relatively high densities in magnetically active stars. Given the formation temperatures of the respective Fe ions, 
reported densities in the range of $10^{12}$~cm$^{-3}$-$10^{13}$~cm$^{-3}$ are, however, in agreement with densities inferred 
from Mg\,{\sc xi} (see above).

For inactive and intermediately active stars such as Procyon, $\alpha$ Cen, $\epsilon$ Eri, or $\xi$ Boo A, much lower 
densities, $n_{\rm e} < 10^{10}$~cm$^{-3}$, are inferred from lower ionization stages of Fe (e.g., Fe\,{\sc x}-{\sc xiv}; 
\citealt{mewe95, schmitt94a, schmitt96c, schrijver96, laming96, laming99}).

A number of conflicting measurements are worthwhile to mention, however. Measurements using \ch\  spectroscopy
have shown systematic deviations from earlier EUVE measurements, perhaps due to blending affecting EUVE spectroscopy \citep{mewe01}.
For the active Algol, \citet{ness02b} report rather low densities of $\log n_{\rm e} \la 11.5$ from Fe\,{\sc xxi}. Similarly,
\citet{phillips01} concluded that Fe\,{\sc xxi} line ratios indicate densities below the low-density limits for the respective
ratios ($\log n_{\rm e} < 12$). \citet{ayres01a} found contradicting results from various line ratios for the giant
$\beta$ Cet, suggesting that densities are in fact low. Further conflicting measurements of this kind have been 
summarized by \citet{testa04a}, and  a systematic consideration of Fe-based density measurements 
was presented by \citet{ness04}. The latter authors found that {\it all} Fe line ratios are above the low-density limit, but by an approximately
constant factor, suggesting that all densities are compatible with the low-density limit after potential correction
for systematic but unrecognized  blends or inaccuracies in the atomic databases. 

The present situation is certainly unsatisfactory. Contradictory measurements based on different density diagnostics
or extremely (perhaps implausibly) high densities inferred from some line ratios make a reconsideration of blending
and the atomic databases necessary. Bias is also introduced by high low-density limits; any deviation of flux ratios into
the density-sensitive range, perhaps by slight blending, by necessity results in ``high densities'' while lower densities 
are, by definition, inaccessible.

\paragraph{Coronal structure from density measurements}
Density measurements in conjunction with emission measure analysis provide an order-of-magnitude estimate of coronal 
volumes $V$ (because EM $= n_{\rm e}n_{\rm H}V$ for a plasma with uniform density). Taken at face value, the very high 
densities sometimes inferred for hot plasma require compact sources and imply small surface
filling factors. For example, \citet{mewe01} estimated that the hotter plasma component in Capella 
is confined in magnetic loops with a semi-length of only $L \la 5\times 10^7$~cm, covering  
a fraction of $f \approx 10^{-6} - 10^{-4}$ of the total surface area. Confinement of such high-pressure
plasma would then require  coronal magnetic field strengths of order 1~kG \citep{brickhouse98}. In that case, the 
typical magnetic dissipation time is only a few seconds for $n_{\rm e} \approx 10^{13}$~cm$^{-3}$ if the energy is 
derived from the same magnetic fields, suggesting that the small, bright loops light up only briefly. In other words, 
the stellar corona would be made up of numerous ephemeral loop sources that  cannot be treated as being in 
quasi-static equilibrium \citep{vdoord97}.

Both \citet{ness04} and \citet{testa04a}
calculated coronal {\it filling factors} $f$ for plasma emitting various He-like triplets. The total ``available'' volume,
$V_{\rm avail}$, for coronal loops of a given temperature depends on a corresponding ``characteristic height'' for which the 
height of a hydrostatic loop  (Eq.~\ref{loop}) can be assumed.
The volume filling factor thus derived, $V/V_{\rm avail}$, is surprisingly small for cool plasma detected in O\,{\sc vii} and
Ne\,{\sc ix}, namely of order a few percent and increasing with increasing activity level. The emission supposedly originates in
solar-like active regions that cover part of the surface, but never entirely fill the available
volume. With increasing magnetic activity, a hotter component appears (recall the general correlation between
average coronal temperature and activity level, Sect.~\ref{sec:coolstars}). This component seems to fill in the volume between the
cooler active regions and contributes the bulk part of the emission measure in very active stars although
the high densities suggest very small loop structures and filling factors ($L \la 10^{-2}~R_*$ resp. $f \ll 1\%$ for AD Leo; 
\citealt{maggio04}). Hotter plasma could thus be a natural result of increased flaring in the interaction zones between cooler regions. 
This provides support for {\it flare-induced} coronal heating in particular in magnetically active stars. The higher densities
in the hotter plasma components are then also naturally explained as a consequence of flaring
(see Sect.~\ref{sec:flaredensities}). Explicit support for this picture comes from $f/i$ ratios in the active M dwarf 
AD Leo that vary between observations separated by more than a year, higher densities being inferred for the
more active states; the overall flaring contribution may have changed between the two epochs, although a different
configuration of active regions with different average electron densities cannot be excluded \citep{maggio05}.
 
At this point, a word of caution is in order. There is no doubt (cf. the solar corona!) that coronal
plasma comes in various structures covering a wide range of densities. Because the emissivity of a coronal
plasma scales with $n_{\rm e}^2$, any X-ray observation is biased toward detections of structures at high densities 
(and sufficiently large volumes). The observed $f/i$ line flux ratios may therefore be a consequence of the
{\it density distribution} and may not represent any existing, let alone dominant, electron density in the corona.
Rather, they are dependent on the steepness of the density distribution, but because of the $n_{\rm e}^2$ dependence, 
they do not even correspond to a linear average  of the $f/i$ ratio across all coronal volume elements. A calculated 
example is given in \citet{guedel04}.

\subsubsection{Inferences from coronal opacities}
\label{sec:opacities}
Emission lines in coronal spectra may be suppressed by optical depth effects due to
resonance scattering in the corona. This effect was discussed in the context of ``anomalously faint'' EUV lines  \citep{schrijver94,
schrijver95}, now mostly recognized to be a consequence of sub-solar element abundances. Resonance scattering requires 
optical depths in the line centers of $\tau \ga 1$, and $\tau$ is essentially proportional
to $n_{\rm e}\ell/T^{1/2}$ \citep{mewe95} where $\ell$ is the path length. For static coronal loops, this implies
$\tau \propto T^{3/2}$ (\citealt{schrijver94}; e.g., along a loop or for a
sample of nested loops in a coronal volume). 

Optical depth effects due to scattering are marginal in stellar coronae; initial attempts to identify 
scattering losses in  the Fe\,{\sc xvii}  $\lambda$15.01 were negative \citep{ness01, mewe01, phillips01, huenemoerder01,
ness02b, audard03a}, regardless of the magnetic activity level of the considered stars. Larger survey work by \citet{ness03a}
again reported no significant optical depth for strong Fe\,{\sc xvii} lines and for $f/r$ ratios in He-like triplets,
after carefully correcting for effects due to line blending. Although Fe\,{\sc xvii}
line ratios, taken at face value, do suggest the presence of line opacities, the deviations turned out to be similar for 
{\it all} stars, suggesting absence of line scattering while the deviations should be ascribed
to systematic problems in the atomic physics databases. Similar conclusions were reached in work by 
\citet{huenemoerder01} and \citet{osten03} using the Lyman series for O\,{\sc viii}, Ne\,{\sc x}, or 
Si\,{\sc xiv}.

More recent survey work by \citet{testa04b} and \citet{testa07} supports the above overall findings although significant 
(at the 4--5$\sigma$ level) optical depths were reported for two RS CVn-type binaries and one active M dwarf 
based on Ly$\alpha$/Ly$\beta$ line-flux ratios of O\,{\sc viii} and Ne\,{\sc x}. Moreover, the optical depth was 
found to be variable in one of the binaries. The path length was estimated
at $\ell \approx 2\times 10^{-4}R_* - 4\times 10^{-2}R_*$, with very small  corresponding  surface filling factors,
$f \approx 3\times 10^{-4} - 2\times 10^{-2}$. Although rather small, these scattering source sizes exceed
the heights of corresponding hydrostatic loops estimated from measured densities and temperatures. Alternatively, 
active regions predominantly located close to the stellar limb may produce an effective, non-zero optical depth as well.

We note here that upper limits to optical depth by resonance scattering were also used to assess upper limits to source 
sizes based on simple escape probability estimates (e.g., \citealt{phillips01, mewe01}), but caution that, as detailed 
by \citet{testa07}, due to potential scattering of photons {\it into} the line of sight these estimates in fact provide 
upper limits to lower limits, i.e., no constraint. We should also emphasize that the absence of optical depth 
effects due to resonance scattering does not imply the absence of scattering in individual stellar active regions. 
The question is simply whether there is a net loss or gain of scattered photons along the line of sight, and for most 
coronal sources such an effect is not present for the (disk-integrated) emission.

\subsubsection{Summary: Trends and limitations}
\label{sec:trends}
Despite a panoply of methods and numerous observed examples, it appears difficult to conclude
on how stellar coronae are structured. There is mixed evidence for compact coronae, coronae predominantly
located at the pole but also distributed coronae covering lower latitudes. Filling factors appear
to be surprisingly small even in saturated stars, as derived from rotational modulation but also
from spectroscopic modeling, in particular based on measurements that indicate very high densities.
Larger structures may be inferred from X-ray flares (see \citealt{guedel04} for a summary).

We should however keep in mind that strong bias is expected from X-ray observations. There is little
doubt - judging from the solar example - that coronae are considerably structured and come in 
packets with largely differing temperatures, size scales, and densities. Because the X-ray
emissivity of a piece of volume scales with $n_{\rm e}^2$, the observed X-ray light is inevitably dominated
by dense regions that occupy sufficiently large volumes. Regions of very low density may remain
undetected despite large volumes (as an example we mention the solar wind!).

Keeping with our definition of coronae as the ensemble of closed stellar magnetic fields containing
heated gas and plasma or accelerated, high-energy particles, X-ray observations miss those
portions of the corona into which hot plasma has not been evaporated, and it is likely to miss 
very extended structures. The latter are favorable places for high-energy electrons,  and
radio interferometry has indeed shown extended radio coronae reaching out to several stellar radii (e.g., 
\citealt{mutel85}).

\subsection{X-ray flares}
\label{sec:flares}

\subsubsection{Introduction}
\label{sec:flareintro}
A flare is a manifestation of a sudden release of magnetic energy in the solar or in a stellar corona. Observationally,
flares reveal themselves across the electromagnetic spectrum, usually sowing a relatively rapid (minutes to hours) increase
of radiation up to the ``flare peak'' that  may occur at somewhat different times in different wavelength bands, 
followed by a more gradual decay (lasting up to several hours). Solar X-ray flare classification schemes 
\citep{pallavicini77} distinguish between
{\it compact} flares in which a small number of magnetic loops lighten up on time scales of minutes, and {\it
long-duration}  (also ``2-Ribbon'') flares evolving on time scales of up to several hours. The latter class involves
complex loop arcades anchored in two roughly parallel chromospheric H$\alpha$ ribbons. These ribbons define the footpoint
regions of the loop arcade. Such flares are energized by continuous reconnection of initially open magnetic 
fields above a neutral line at progressively larger heights, so that nested magnetic ``loops'' lighten up
sequentially, and possibly also at different times along the entire arcade. The largest solar flares are usually 
of this type.

Transferring solar flare classification schemes to the stellar case is problematic; most stellar flares reported in the literature reveal
extreme luminosities and radiative energies, some exceeding even the largest solar flares by several orders of magnitude.
This suggests, together with the often reported time scales in excess of one hour, that most stellar flares interpreted in 
the literature belong to the class of  ``2-Ribbon'' or arcade flares involving considerable magnetic complexity. We caution, however,
that additional flare types not known on the Sun may exist, such as flares in magnetic fields connecting the components
of close binary systems, flares occurring in dense magnetic fields concentrated at the stellar poles, reconnection
events on a global scale in large stellar  ``magnetospheres'', or flares occurring in magnetic fields connecting a
young star and its circumstellar disk. 

In a standard model developed for solar flares, a flare event begins with a magnetic instability that eventually leads
to magnetic reconnection in tangled magnetic fields in the corona. In the reconnection region, heating, particle
acceleration, and some bulk mass acceleration takes place. The energized particles (e.g., electrons) travel along
closed magnetic fields toward the stellar surface; as they reach denser layers in the chromosphere, they deposit
their energy by collisions, heating the ambient plasma explosively to millions of K. The ensuing overpressure drives 
the hot plasma into coronal loops, at which point the ``X-ray flare'' becomes manifest.

Clearly, understanding the physical processes that lead to a flare, and in particular interpreting the microphysics of plasma 
heating, is mostly a task for the solar coronal physics domain. Nevertheless, stellar flare observations have largely extended the
parameter range of flares, have added new features not seen in solar flares, and have helped understand the structure of 
stellar magnetic fields in various systems. Most of the information required for an interpretation of coronal
flares is extracted from ``light curve analysis'', most importantly including the evolution of the characteristic 
temperatures that relate to heating and cooling processes in the plasma. A summary of the methodology has been given in
\citet{guedel04} and will not be  addressed further here. High-resolution spectroscopy, ideally obtained
in a time-resolved manner, adds important information on stellar flares; specifically, it provides information for which
spatially resolved imaging would otherwise be needed  (as in the solar case), namely clues on densities in the flaring 
hot plasma, opacities, and signatures of fluorescence that all provide information on the size of flaring regions. Furthermore,
line shifts may be sufficiently large to measure plasma flows in flares.

\subsubsection{An overview of stellar flares}
\label{sec:flareover}
A vast amount of literature on flares is available from the past three decades of stellar X-ray astronomy. A collection of
results from flare interpretation studies until 2004 is given in tabular form in \citet{guedel04}. Here, we summarize the
basic findings.

\begin{figure}[t]
\centerline{\includegraphics[angle=0,width=12cm]{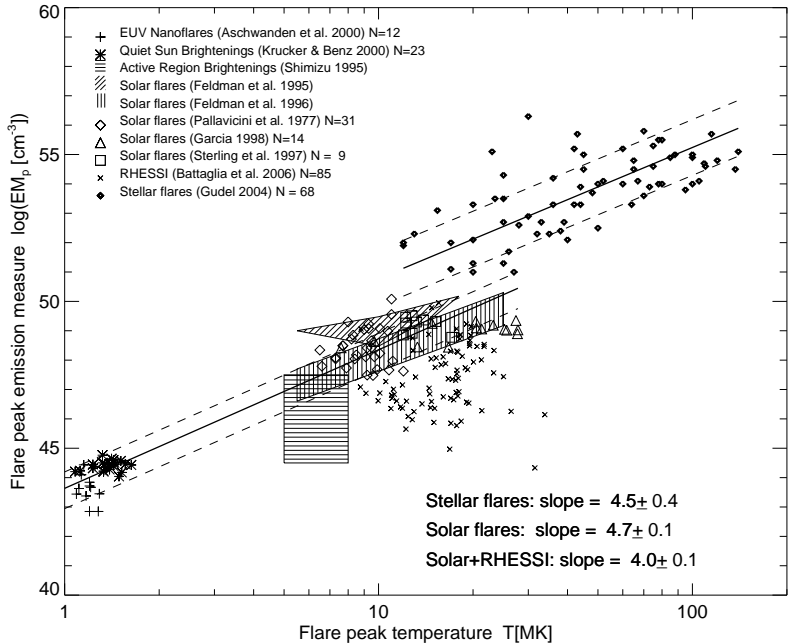}}
\caption{Flare peak emission measure as a function of flare peak temperature for solar and stellar flares. Note the 
similar trends for solar and stellar flares although there is an offset between the regression lines.
(From \citealt{aschwanden08}, reproduced by permission of the AAS.)}
\label{fig:tempEM} 
\end{figure}

Increased temperatures during flares are a consequence of efficient heating mechanisms. Spectral observations of 
large stellar flares have consistently shown electron temperatures up to 100~MK, in some cases even more \citep{tsuboi98,
favata99, osten05, osten07}, much beyond  typical solar-flare peak temperatures (20--30~MK). Somewhat unexpectedly,
the flare peak temperature, $T_{\rm p}$, correlates with the peak emission measure, EM$_{\rm p}$ (or, by implication, 
the peak X-ray luminosity), roughly as \begin{equation}\label{flaretemp}
{\rm EM_{\rm p}} \propto T_{\rm p}^{4.30\pm 0.35}
\end{equation} 
\citep{guedel04} although observational bias may influence the precise power-law 
index. For solar flares, a similar trend holds with a normalization (EM) offset between solar and stellar flares 
 - again perhaps involving observational bias (\citealt{aschwanden08}; Fig.~\ref{fig:tempEM}). The 
EM$_{\rm p}$-$T_{\rm p}$ relation was interpreted based on MHD flare modeling \citep{shibata99, shibata02}, with the
result that larger flares require larger flaring sources (of order $10^{11}$~cm for the most active stars) while magnetic
field strengths ($B \approx 10-150$~G) should be comparable to solar flare values.

Following the standard flare scenario exposed above, we should expect that a flare reveals itself first
in the radio regime through gyrosynchrotron emission from the injected, accelerated electron population, and also
in the optical/UV as a result of prompt emission from the heated chromospheric region at the loop footpoints. 
Further, as the electrons impact in the denser layers, they promptly emit non-thermal hard X-rays (HXR, $>$10~keV)
that have indeed been of prime interest in solar flare research. These initial emission bursts characterize and define
the {\it impulsive phase} of the flare during which the principal energy is released. The more extended phase
of energy conversion, mass motion, and plasma cooling characterizes the {\it gradual phase}.
In particular, soft X-ray emission is a consequence of plasma accumulation in the coronal loops, which roughly 
increases with the integral of the deposited 
energy.\footnote{\citet{schmitt08} report the case of a strong flare on the M dwarf CN Leo in which an initial thermal 
soft-Xray pulse was observed on time scales of a few seconds during the impulsive flare phase, coincident with the 
optical flare peak. This radiation may originate from the initial plasma heating at the bottom of the magnetic loops,
at a time when evaporation only starts to fill the loops.}
A rough correlation is therefore expected between the time behavior of the radio, optical/UV, and hard X-rays on
the one hand and soft X-rays on the other hand, such that the former bursts resemble the time derivative
of the increasing soft X-ray light curve, $L_\mathrm{R, O, UV, HXR}(t) \propto {\rm d}L_{\rm X}(t)/{\rm d}t$, a relation known
as the ``Neupert Effect'' (after \citealt{neupert68}). 

The Neupert effect has been observed in several
stellar flares (e.g., \citealt{hawley95, guedel02a, osten04, osten07, smith05, mitra05}; see
Fig.~\ref{fig:proxcen} in Sect.~\ref{sec:flaredensities} below), 
in cases with emission characteristics very similar to the standard case of solar flares. Such observations
testify to the chromospheric evaporation scenario in many classes of stars. Equally important is the lack of 
correlated behavior - also observed in an appreciable fraction of solar flares - which provides important clues
about ``non-standard'' behavior. Examples of X-ray flares without accompanying radio bursts or radio bursts peaking
at the time of the X-rays or later were presented by \citet{smith05}. In a most outstanding case, described by
\citet{osten05},  a sequence of very strong X-ray, optical, and radio flares occurring on an M dwarf show
a complete breakdown of correlated behavior. While the presence of X-ray flares without accompanying
signatures of high-energy electrons can reasonably be understood (e.g., due to flares occurring in high-density
environments in which most of the energy is channeled into direct heating), the reverse case, radio and correlated
U-band flares  without any indication of coronal heating, is rather puzzling; this is especially true given 
that the non-thermal energy must eventually thermalize, and the thermal plasma is  located close to
or between the non-thermal coronal radio source and the footpoint U-band source. Possible explanations include
an unusually low-density environment, or heating occurring in the lower chromosphere or photosphere after deep
penetration of the accelerated electrons without appreciable evaporation at coronal temperatures \citep{osten05},
but a full understanding of the energy transformation in such flares is still missing.

Lastly, we mention the fundamental role that flares may play in the heating of entire stellar coronae.
The suggestion that stochastically occurring flares may be largely responsible for
coronal heating is known as the ``microflare'' or ``nanoflare'' hypothesis in solar 
physics \citep{parker88}. Observationally, solar flares are distributed in energy following a  
power law,  ${\rm d}N/{\rm d}E = k E^{-\alpha}$ where ${\rm d}N$ is the number of flares per unit time with a total 
energy in the  interval [$E,E+{\rm d}E$], and $k$ is a constant. If $\alpha\ge 2$, then the energy integration 
(for a given  time interval) diverges for $E_{\rm min} \rightarrow 0$, i.e., a lower energy cutoff is required,
and depending on its value, an arbitrarily large amount of energy could be related to flares.
Solar studies indicate $\alpha$ values of $1.6-1.8$ for ordinary
solar flares \citep{crosby93}, but some recent studies of low-level  flaring
suggest $\alpha = 2.0 - 2.6$ \citep{krucker98, parnell00}. Stellar studies have provided interesting evidence
for $\alpha > 2$ as well, for various classes of stars including TTS and G--M-type dwarf stars
\citep{audard00, kashyap02, guedel03a, stelzer07, wargelin08}. There is considerable further evidence that flares 
contribute fundamentally to coronal heating, such as correlations between average X-ray emission on the
one hand and the observed (optical or X-ray) flare rate or UV emission on the other hand. Continuous
flaring activity in light curves, in some cases with little evidence for a residual, truly constant baseline 
level, add to the picture. Evidence reported in the literature has been more comprehensively summarized 
in \citet{guedel04}.

\subsubsection{Non-thermal hard X-ray flares?}
\label{sec:nonthermal}
The X-ray spectrum of solar flares beyond approximately 15--20~keV is dominated by {\it non-thermal} hard X-rays.
These photons are emitted when a non-thermal, high-energy population of electrons initially accelerated in the coronal
reconnection region collides in the denser chromosphere and produce ``thick-target'' bremsstrahlung. The spectrum
is typically a power law, pointing to a power-law distribution of the accelerated electrons \citep{brown71}.
Detection of such emission in magnetically active stars would be of utmost importance as it would provide information
on the energetics of the initial energy release, the particle acceleration process in magnetic field configurations
different from the solar case, the relative importance of particle acceleration and direct coronal heating, 
and possibly travel times and therefore information on the size of flaring structures. The presence of
high-energy electron populations is not in doubt: they are regularly detected from their non-thermal gyrosynchrotron 
radiation at radio wavelengths.

\begin{figure}[t]
\hbox{
\hskip -0.6truecm\includegraphics[angle=-90,width=6.7cm]{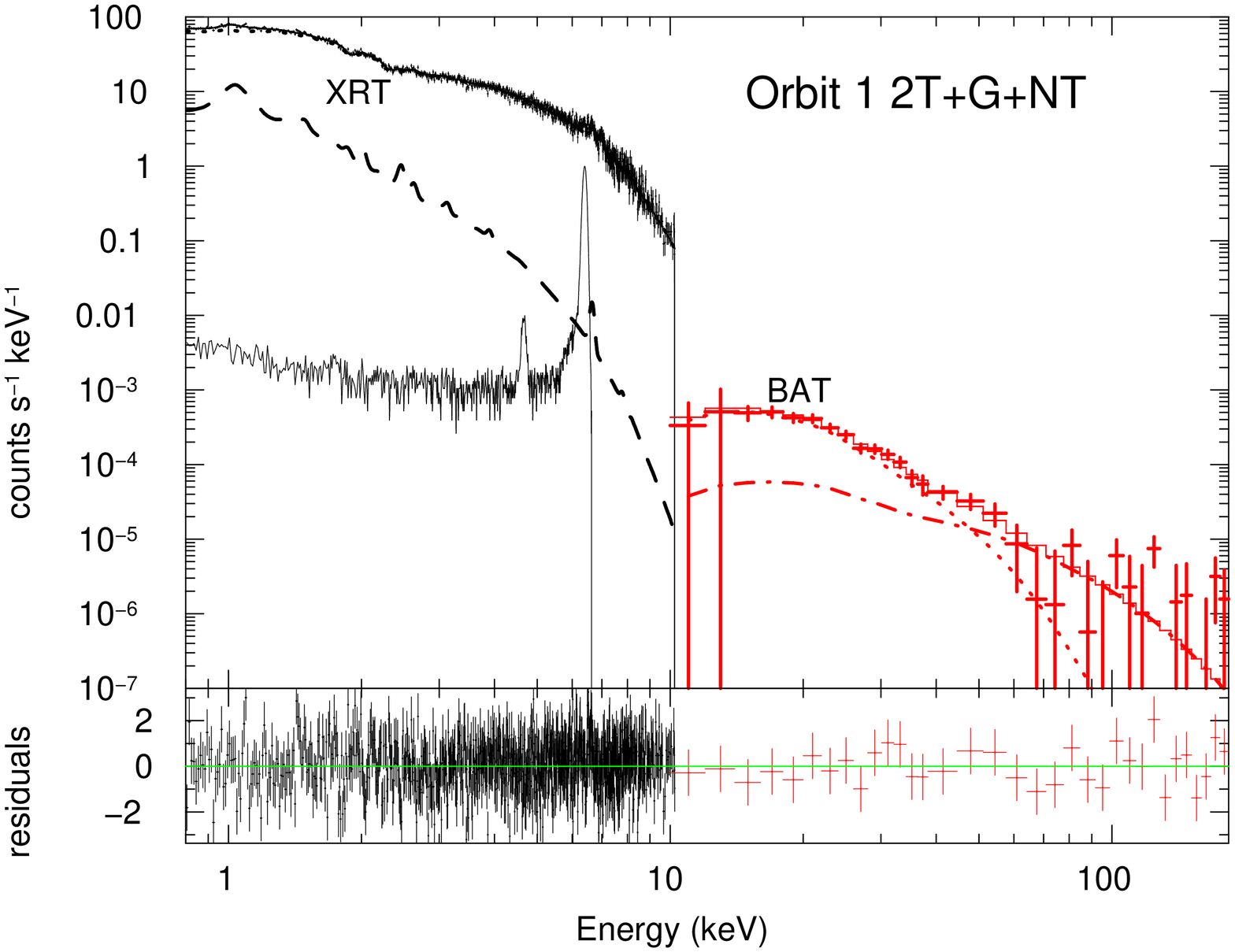}
\hskip -0.6truecm\includegraphics[angle=-90,width=6.7cm]{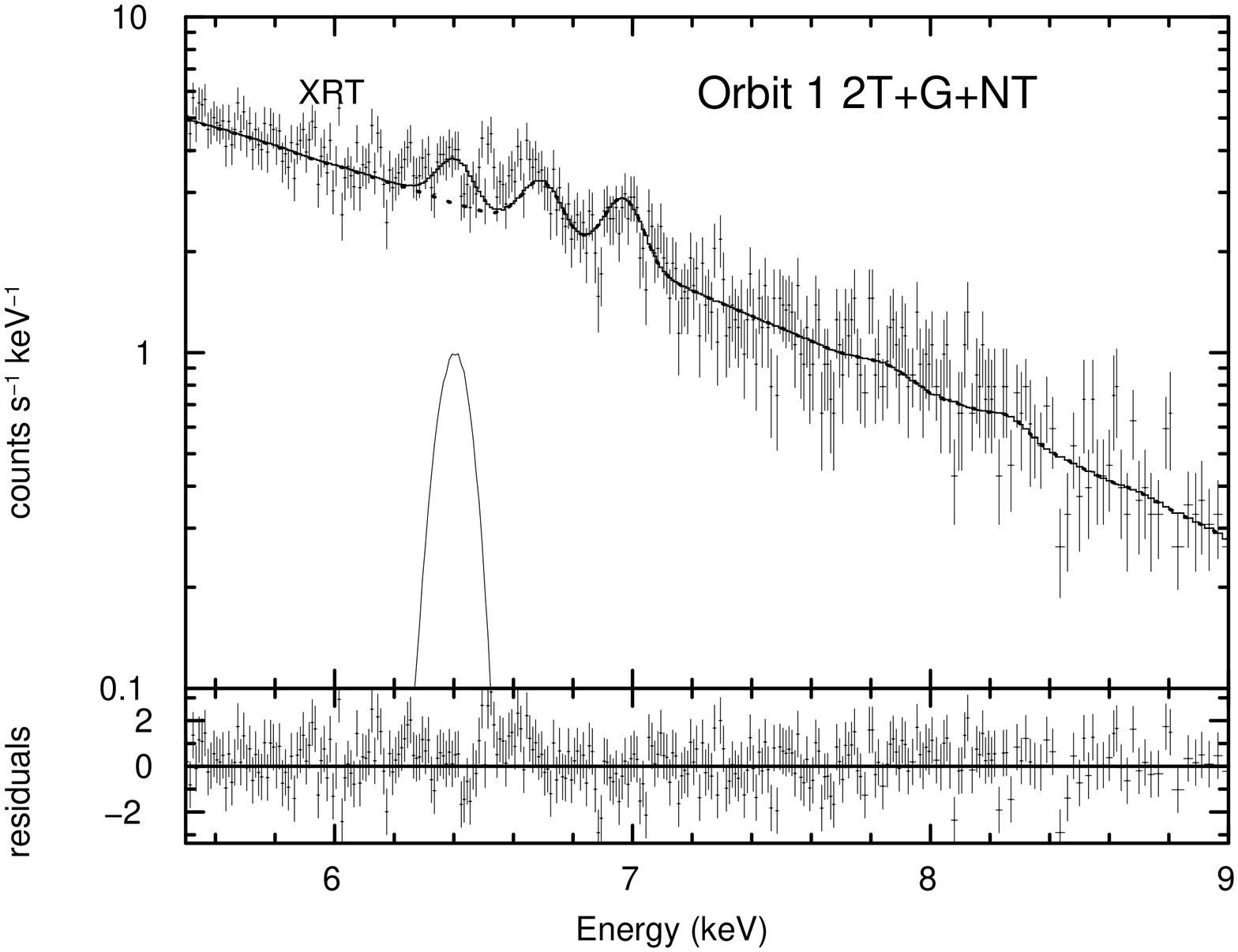}
}
\caption{Left (a): The observed spectrum of a large flare on the RS CVn binary II Peg is shown with a spectral fit (consisting 
         of two thermal components, a Gaussian for the 6.4~keV Fe K$\alpha$ line, and a power law for the highest 
	 energies). The spectrum was obtained by different detectors (above and below $\approx$10~keV). The individual 
	 contributions by the model components are shown dashed and dotted, or by a thin  solid line (6.4~keV feature). 
	 Fit residuals are shown in the bottom panel. -- Right (b): Extract from (a), showing the 
	 region around the 6.4~keV K$\alpha$ feature and the 6.7~keV Fe\,{\sc xxv} complex. (From \citealt{osten07},
	 reproduced by permission of the AAS.)}
\label{fig:nonthermalflare} 
\end{figure}

Detection of non-thermal hard X-rays is hampered by the low expected fluxes, but also by the trend that large 
flares produce very hot plasma components that dominate the bremsstrahlung spectrum up to very high photon energies
(Eq.~\ref{flaretemp}).
This latter effect has clearly been demonstrated in observations successfully recording X-ray photons  up to about
100~keV from large stellar flares \citep{favata99, rodono99, franciosini01}, the extended X-ray spectrum 
being compatible with an extrapolation of the thermal soft X-ray spectrum. The most promising case has been reported 
from a very large flare occurring on the RS CVn binary II Peg in which activity was recorded up to 100~keV during the 
entire flare episode (\citealt{osten07}; Fig.~\ref{fig:nonthermalflare}).  Although the spectrum could be interpreted
with bremsstrahlung from a very hot, $\approx$300~MK component, \citet{osten07} favored a non-thermal interpretation,
arguing that conductive losses would be excessive for a thermal component; also, the concurrent Fe K$\alpha$ 6.4~keV emission
recorded during the flare may be the result from non-thermal electron impact ionization rather than from photoionization
fluorescence (Sect.~\ref{sec:fluorescence}), given the high hydrogen column densities required. However, while suggestive,
these arguments remain somewhat inconclusive; first,
conductive losses across extreme temperature gradients cannot exceed the free-streaming electron limit at which conduction
saturates. Second, the hard component persists into the flare decay phase, at high levels, unlike in solar
flares. And third, recent detailed fluorescence calculations suggest that the observed Fe K$\alpha$ feature can
in fact be explained as arising from photoionization, while the impact ionization mechanism is inefficient 
\citep{drake08b, ercolano08, testa08a}. Unequivocal detection of non-thermal hard X-rays from stellar coronae remains an important 
goal, in particular for future detectors providing high sensitivity and low background up to at least 100~keV.

\subsubsection{Fluorescence and resonance scattering during stellar flares}
\label{sec:fluorescence}
Photoionization of cool material by X-ray photons above the Fe~K edge at 7.11~keV produces a prominent line feature
at 6.4~keV (Fe~K$\alpha$ feature; see Sect.~\ref{sec:fluordisk} for further details). This feature is usually too faint to be detected 
in any stellar X-ray spectrum; exceptions include a few TTS for which fluorescence on the circumstellar disk due 
to irradiation by stellar X-rays has been proposed (Sect.~\ref{sec:fluordisk}). In more evolved stars, 6.4~keV K$\alpha$ emission would originate
from the stellar {\it photosphere}, but has so far been detected in only two cases. Prominent Fe K$\alpha$ emission
was recorded from a giant flare on the RS CVn binary II Peg (\citealt{osten07}; Fig.~\ref{fig:nonthermalflare}b), 
although a model based on electron impact
ionization was put forward, as discussed above (Sect.~\ref{sec:nonthermal}). More recently, \citet{testa08a} presented evidence
for photospheric fluorescence in the single G-type giant HR~9024, using detailed fluorescence calculations to estimate
a source height of $\la 0.3R_*$ ($R_* = 13.6R_{\odot}$).

\begin{figure}[t]
\centerline{\hskip -0.6truecm\includegraphics[angle=90,width=9cm]{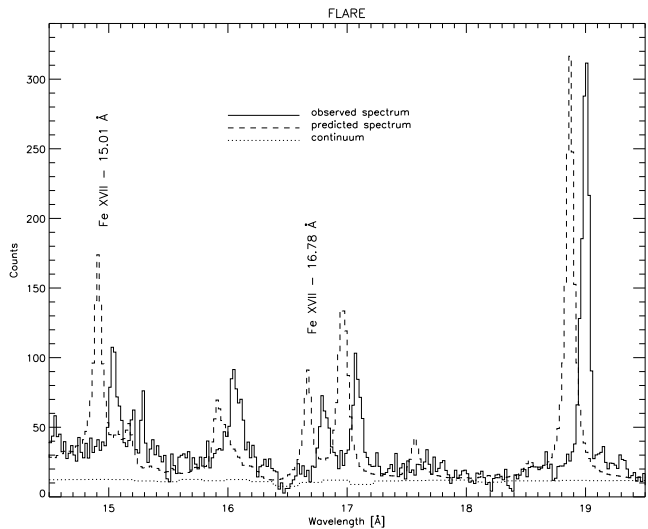}}
\caption{Evidence for optical depth effects due to resonance scattering. The plot shows the observed spectrum
(solid), the predicted spectrum (dashed, slightly shifted in wavelength for clarity), and the continuum (dotted). 
Although there are discrepancies for several lines, the model overprediction of the 15.01~\AA\ line is significantly larger
than during quiescence. (From \citealt{matranga05}, reproduced by permission of the AAS.)}
\label{fig:flareopticaldepth} 
\end{figure}

Resonance scattering (Sect.~\ref{sec:opacities}) is another potential method to measure the size of
flaring coronal structures. A suppression of the strong Fe\,{\sc xvii}~$\lambda$15.01 line compared to the
Fe\,{\sc xvii}~$\lambda$16.78 line was recorded during a flare on AB Dor (Fig.~\ref{fig:flareopticaldepth}) 
and interpreted in terms of an optical depth of 0.4 in the line center,  implying a path length of order 8000~km \citep{matranga05}.

\subsubsection{Flare densities: Evidence for evaporation}
\label{sec:flaredensities}
According to the standard flare scenario, densities in flaring loops should largely increase  as a consequence
of chromospheric evaporation. This is the essential cause of the large emission measure increase in the flaring
corona. Spectroscopic density measurements in {\it solar} flares using He-like triplets (Sect.~\ref{sec:densities}) 
confirm this picture, suggesting density increases to several times $10^{12}$~cm$^{-3}$ \citep{mckenzie80, doschek81, 
phillips96, landi03}.

\begin{figure}[t]
\includegraphics[angle=0,width=12.3cm]{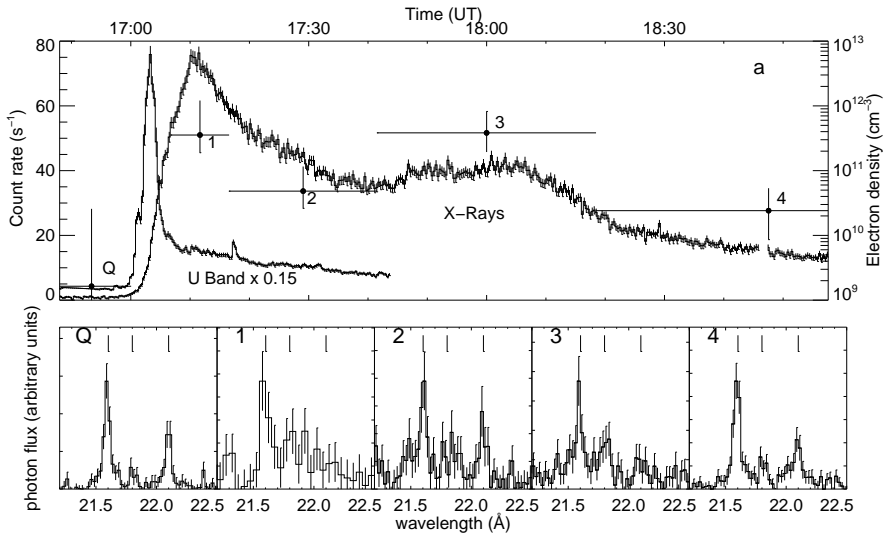}
\caption{Evolution of a large flare on Proxima Centauri. The upper panel shows the X-ray light curve
together with the short pulse in the U band peaking at about 17:05 UT. The large crosses show
electron densities during the intervals defined by the horizontal arms, derived from the O\,{\sc vii} line
triplet fluxes (the density scale is given on the right y axis). The triplets for the five intervals are 
shown in the bottom panel, marked ``Q'' for quiescence and ``1'' -- ``4'' for the four flare episodes.  The three
marks in the upper parts of the figures show the locations of the resonance, intercombination, and forbidden
lines (from left to right). (Adapted from \citealt{guedel02a}.)}
\label{fig:proxcen} 
\end{figure}

Stellar evidence is still limited given the high signal-to-noise ratio required for short observing intervals.
First significant spectral evidence for strong density increases were reported for a large flare on Proxima Centauri
\citep{guedel02a, guedel04a}, both for the O\,{\sc vii} and (more tentatively) for the Ne\,{\sc ix} triplet. The forbidden
line in the O\,{\sc vii} triplet nearly disappeared during the flare peaks, while a strong intercombination
line showed up (Fig.~\ref{fig:proxcen}).  The derived densities rapidly increased from a pre-flare level of 
$n_{\rm e} < 10^{10}$~cm$^{-3}$  to $\approx 4\times 10^{11}$~cm$^{-3}$ at flare peak, then again rapidly decayed to 
$\approx 2\times 10^{10}$~cm$^{-3}$, to increase again during a secondary peak, followed by a gradual decay. 
The instantaneous mass involved in the cool, O\,{\sc vii} emitting source was estimated at $\approx 10^{15}$~g, 
suggesting similar (instantaneous) potential and thermal energies in the cool plasma, both of which are much smaller 
than the total radiated X-ray energy. It is therefore probable that the cool plasma
is continuously replenished by the large amount of material that is initially heated to higher 
temperatures and subsequently cools to O\,{\sc vii} forming temperatures and below.
The measured densities agree well with estimates from hydrodynamic simulations \citep{reale04} and, together with
light curve analysis, provide convincing evidence for the operation of chromospheric evaporation in stellar 
flares.

Marginal signatures of increased densities during flares have been suggested from He-like triplet flux ratios  
for several further stars, in particular for YY Gem \citep{stelzer02}, $\sigma^2$ CrB \citep{osten03},
AD Leo \citep{besselaar03}, AT Mic \citep{raassen03b}, AU Mic \citep{magee03}, and CN Leo \citep{fuhrmeister07}. 

\subsection{The composition of stellar coronae}
\label{sec:composition}

\subsubsection{The FIP and IFIP effects}
\label{sec:fip}
Studies of element abundances in stars are of fundamental interest as they contribute to our understanding  
of galactic chemistry and its evolution as well as to refined models of stellar interiors. The composition
of material available in young stellar environments is of course also relevant for the planet-formation process. 

The composition of stellar material could change as it is driven  from the surface into the corona or
the stellar wind, owing to various fractionation processes. Specifically, elements with a low first
ionization potential (``low-FIP'' elements Mg, Si, Ca, Fe, Ni) are predominantly
ionized at chromospheric levels, while  high-FIP elements (C, N, O, Ne, Ar, and marginally also S) 
are predominantly neutral. Ions and neutrals could then be affected differently by electric and magnetic fields. 

It is well known that the composition of the solar corona and the solar wind is indeed at variance
with the photospheric mixture; low-FIP elements are enhanced in the corona and the wind by factors of a
few, whereas high-FIP elements show photospheric abundances (this is the essence of the ''FIP effect''; \citealt{feldman92}). 
There are considerable variations between different solar coronal features, e.g., active regions, quiet-Sun regions, 
old magnetic loop systems, or flares. A discussion of the physics involved in this fractionation is 
beyond the scope of this review, and in fact a universally accepted model does not yet exist; see, e.g., 
\citet{henoux95} and \citet{drake03b} for a few selected model considerations. 

Although a solar-like FIP effect could be identified in a couple of nearby stars based on EUVE
observations \citep{laming96, laming99, drake97}, early observations of magnetically active stars
started painting a different - and confusing - picture when abundances were compared with standard solar 
photospheric abundances (assumed to be similar to the respective stellar composition). In many low-resolution 
spectra, unusually weak line complexes required significantly subsolar abundances (e.g., 
\citealt{white94, antunes94, gotthelf94} and many others) also confirmed by EUVE observations (e.g., 
\citealt{stern95a}). 

\begin{figure}[t]
\includegraphics[angle=0,width=12.1cm]{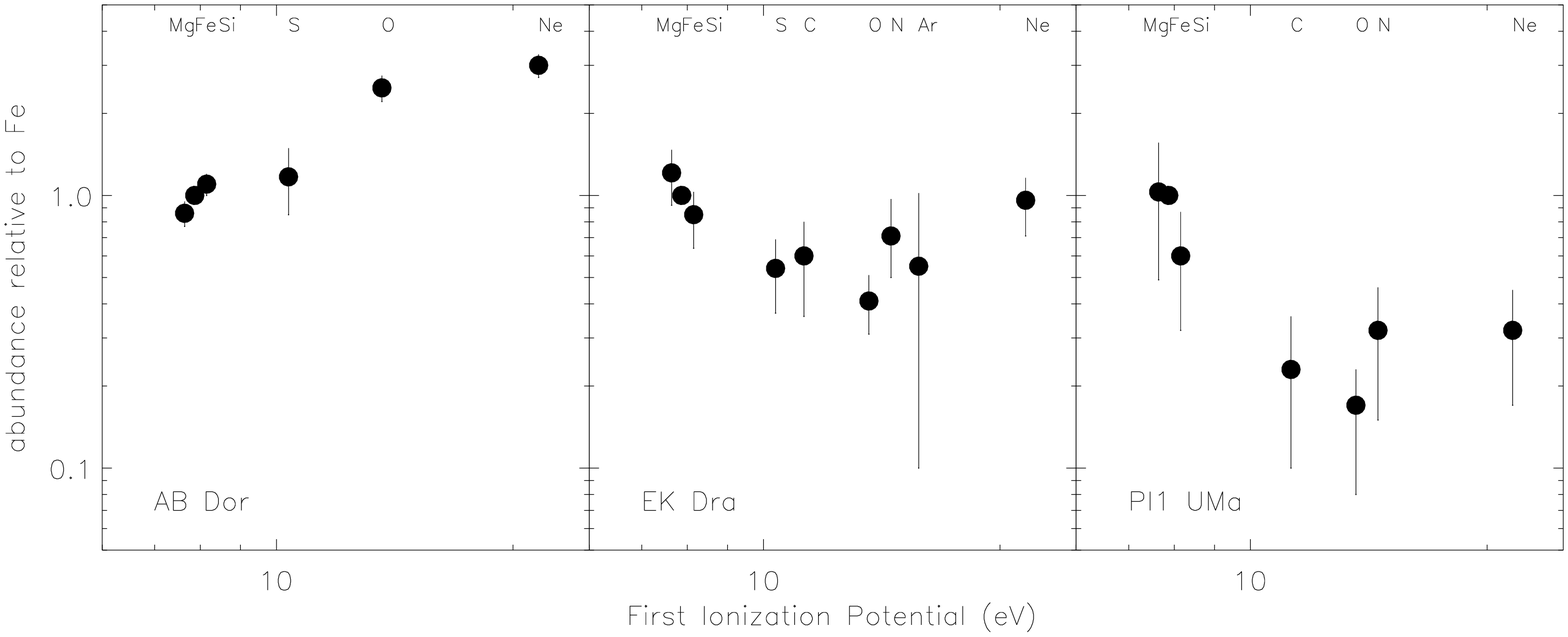}
\caption{Coronal abundances as a function of FIP, normalized to the coronal abundance of Fe. 
Left (a): AB Dor, an example for an inverse-FIP effect (data from \citealt{garcia05}). --
Middle (b): EK Dra, an intermediate case, showing increasing abundances toward the lowest and highest 
          FIP (data from \citealt{telleschi05}). --
Right (c): $\pi^1$~UMa, an intermediately active star showing a solar-like FIP effect (data from 
\citealt{telleschi05}). 
}
\label{fig:abundancedist} 
\end{figure}

A proper derivation of abundances requires well-measured fluxes of lines from different elements,
and this derivation  is obviously tangled with the determination of the emission measure distribution.
It is therefore little surprising that at least partial clarification came only with the introduction of
high-resolution X-ray spectroscopy from \xmm\  and \ch. High-resolution X-ray spectra of magnetically
active stars revealed a new trend that runs opposite to the solar FIP effect and that has become known
as the ``inverse FIP (IFIP) effect''.  Coronae expressing an IFIP effect show low-FIP abundances   
systematically depleted with respect to high-FIP elements (\citealt{brinkman01}; see example in Fig.~\ref{fig:abundancedist}).
As a consequence of this anomaly, the ratio between the abundances of Ne (highest FIP) and Fe (low FIP)
is unusually large, of order 10 in some cases, compared to solar photospheric conditions (see also \citealt{drake01, audard01a}).
The IFIP effect has been widely confirmed for many active stars or binaries (e.g., \citealt{huenemoerder01,
huenemoerder03, garcia05}). With respect to the hydrogen abundance, most elements in active stars remain, 
however, depleted. 

\begin{figure}[t]
\hbox{
\includegraphics[angle=0,width=6.1cm]{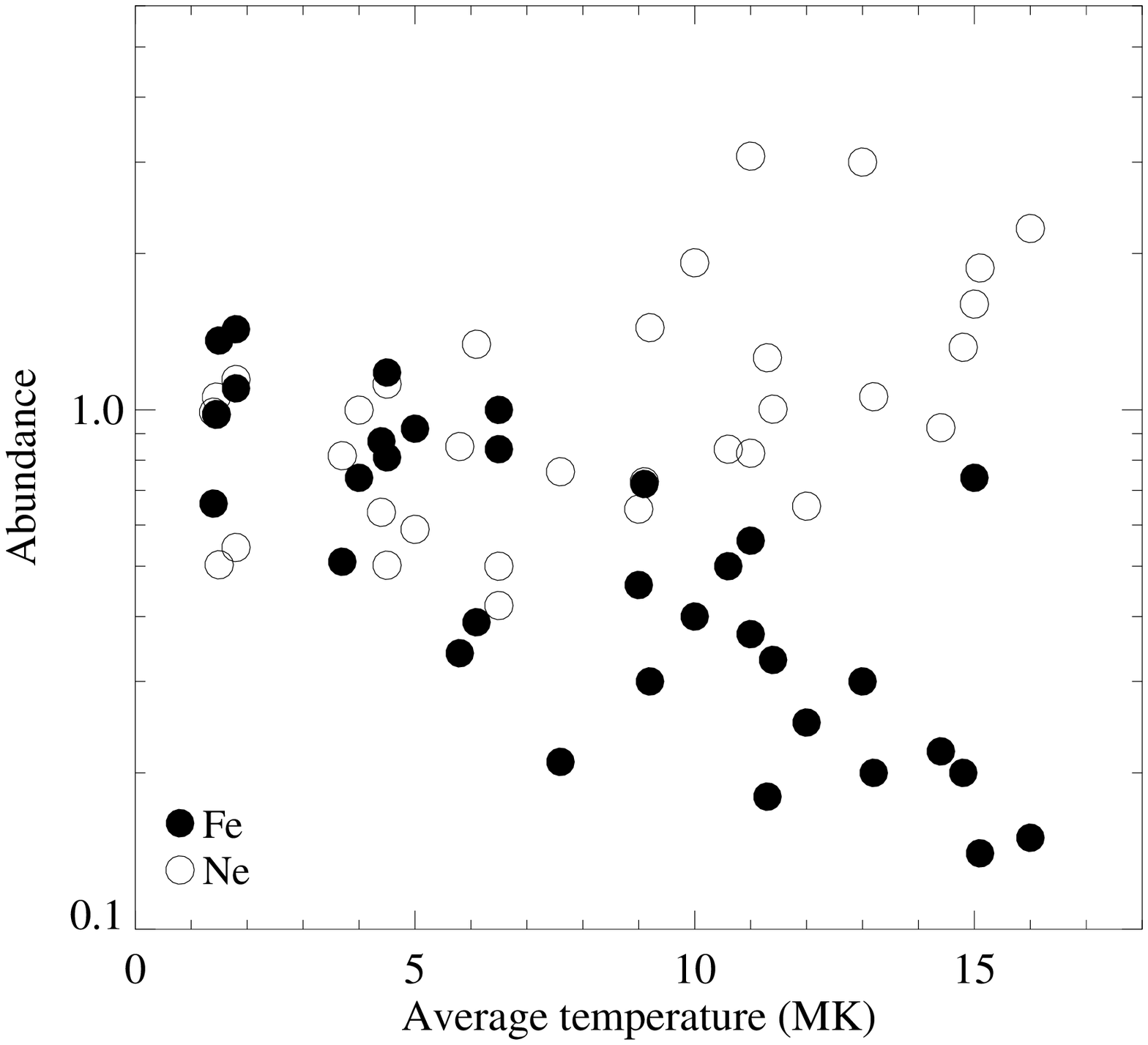}
\includegraphics[angle=0,width=6.1cm]{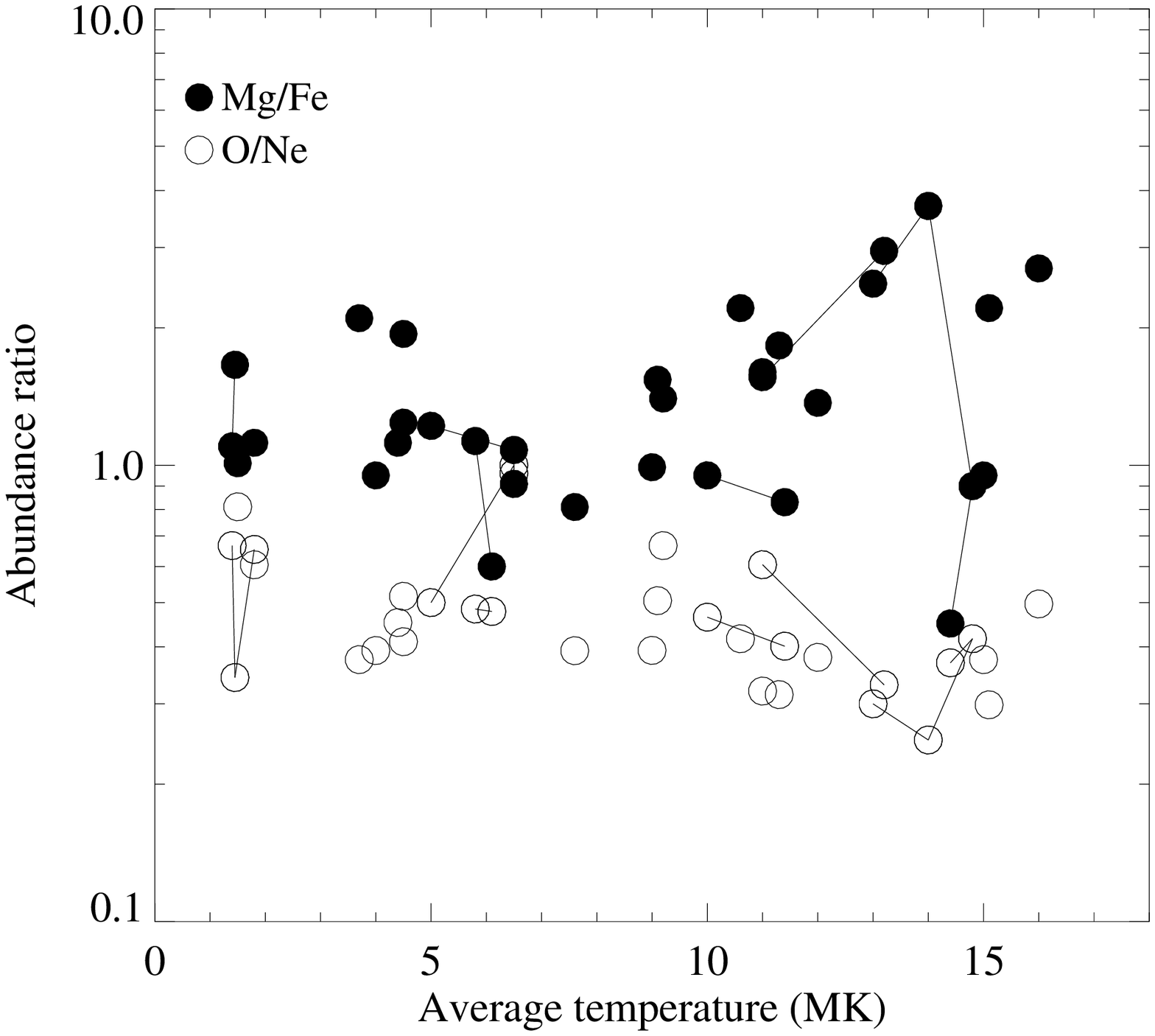}
}
\caption{Left (a): Abundances of Fe and Ne, normalized to the solar photospheric values, for a sample of stars 
at different activity levels as characterized by the average coronal temperature. --
Right (b): Abundance ratios of Me/Fe (both low-FIP elements) and O/Ne (both high-FIP elements), with respect to the
solar photospheric mixture. Lines 
connect different measurements for the same star.
(From \citealt{guedel04}.)}
\label{fig:abunstat} 
\end{figure}

The FIP and IFIP effects in fact represent only two extremes in a bigger picture related
to coronal evolution. With decreasing activity (e.g., as parameterized by the $L_{\rm X}/L_{\rm bol}$ ratio, or
the average coronal temperature), the IFIP effect weakens until the abundance distribution becomes flat
\citep{audard03a, ball05}, and eventually turns into solar-like FIP trend
for less active stars with log\,$L_{\rm X}/L_{\rm bol} \la -4$ or an average coronal temperature
$\la 7-10$~MK (\citealt{telleschi05, guedel04}; Fig.~\ref{fig:abundancedist}). At this critical activity 
level, we also observe that the dominant very hot plasma  component disappears. This trend is best seen in the Ne/Fe 
abundance ratio (high-FIP vs. low-FIP) that decreases by an order of magnitude from the most active to the 
least active coronae (Fig.~\ref{fig:abunstat}a). On the high-activity
side, the trends continue into the ``supersaturation regime''  (Sect.~\ref{sec:coolstars}) of the fastest rotators, where 
$L_{\rm X}/L_{\rm bol}$ converges to  $\approx 10^{-3}$ and becomes unrelated to rotation; for these stars, the 
Fe abundance continues to drop sharply, while the O abundance continues to increase, possibly reaching a saturation level 
for the most active stars \citep{garcia08}. 

While these trends have been well studied in main-sequence stars and subgiant RS CVn binaries, similar 
trends apply to giant stars. A FIP effect is seen in giants 
evolving through the Hertzsprung gap and into the red giant phase \citep{garcia06}, a transformation during which stars
develop a deep convection zone anew, regenerate magnetic activity and become subject to rapid braking. In contrast,
strong enhancements of Ne but low abundances of low-FIP elements characterize the class of extremely active, rapidly 
rotating FK Com-type giants and active giants in tidally locked binaries \citep{gondoin02, gondoin03, audard04}.

Furthermore, the strength of the IFIP effect is also a function of the spectral type or $T_{\rm eff}$ of the star. 
As noted by \citet{telleschi07b} based on studies of high-resolution X-ray spectra, the IFIP trend is much more 
pronounced in active K and M type stars than in G stars, the Fe/O and Fe/Ne abundance ratios indicating 
much stronger depletion of Fe in later-type active stars. This trend is evident for young main-sequence stars 
and also for TTS regardless of whether they are accreting or not. For TTS, the spectral-type
dependent strength of the IFIP effect has been confirmed from low-resolution spectroscopy \citep{scelsi07}.

Variations of these general abundance features have been noted. In some cases, the lowest-FIP elements
(such as Al, Na, and Ca, sometimes others) seem to show abundances in excess of the IFIP trend in active stars, resulting
in an abundance distribution with a minimum around FIP $\approx 10$~eV and increasing trends to both lower and
higher FIP (\citealt{sanz03, argiroffi04, scelsi05, maggio07}; see Fig.~\ref{fig:abundancedist}). Some of these element 
abundances are extremely difficult to assess, and conflicting results have been reported (see, e.g. \citealt{garcia05}).

Caution must be applied when interpreting stellar coronal abundances as derived from high-resolution spectra.
First, as we know from the Sun, coronal abundances vary from  region to region, and they also evolve in
time. A stellar spectrum represents a snapshot of the integrated corona, averaging out all variations 
that might be present. Abundance measurements therefore most likely represent weighted means, and the formal
uncertainties do not include the true variation in location and time. Extreme precision in the abundance
tabulations may indeed be useless - the important information is in overall trends in stellar samples.

Significant differences have also been 
found among stars that share most of their fundamental properties; the primary of the intermediately active
K star binary 70 Oph shows a distinct solar-like FIP effect, while such bias is absent in the otherwise
very similar secondary \citep{wood06}. On the other hand, very similar coronal thermal and abundance properties 
can be found in stars with similar properties but different evolutionary histories; examples include
two rapidly rotating K stars, the younger of which (AB Dor) still keeps its primordial angular momentum
while the older (V471 Tau) is subject to tidal spin-up due to a companion star \citep{garcia05}; and finally, 
the post-main-sequence active binaries Algol and HR~1099 contain very similar K-type subgiants revealing very similar
X-ray spectra, a similar coronal thermal structure, and similar coronal abundance patterns (except for mass-loss
related abundances of C and N in Algol) despite their fundamentally different evolutionary histories \citep{drake03a}.

Caution is in order also because the coronal material is processed gas from the stellar photosphere, and 
appropriate {\it stellar photospheric abundances} should therefore be taken as a reference. For most stars of interest 
such measurements are not available, or are difficult to obtain due to, e.g., rapid rotation. Do the FIP and IFIP
trends survive once the coronal measurements are related to the {\it stellar} photospheric composition rather than
the standard solar mixture? Some reports
suggest they do not. Studies of a few individual examples with well-measured photospheric abundances
\citep{drake95b, raassen02, sanz04} as well as a larger sample of stars in the Orion Nebula \citep{maggio07} or
a sample of M dwarfs (with relatively poorly known photospheric metallicities; \citealt{robrade05}) indicate 
that  the coronal abundances reflect the photospheric mix, or at least the global photospheric metallicity, 
of the respective stars. However, while suggestive, 
these observations must be contrasted with cases for which non-photospheric coronal abundances are
undisputed. Specifically, IFIP and FIP trends have been identified in nearby stars with well-measured
near-solar photospheric abundances such as AB Dor or several solar analogs \citep{sanz03, telleschi05, garcia05};
further, the similarity of the FIP or IFIP effects in stars with similar activity levels, or the trends
depending on activity would be difficult to explain as being due to the photospheric mixture. Younger (and therefore more active)
stars would be expected to be more enhanced in metals, contrary to the observed trends. Also, the FIP and
IFIP effects would reflect a photospheric anomaly of apparently most stars when compared with the
Sun. But it is the Sun itself that proves otherwise: the solar FIP effect is a true
coronal property, suggesting that equivalent trends observed in stars are, by analogy, coronal in origin as well. 
Solar and stellar flares in which abundances change systematically compared to quiescence 
support this picture further, suggesting genuine FIP-related abundance biases in stellar coronae (see Sect.~\ref{sec:flareabun}). 

Given the somewhat contradictory findings, it should then not be surprising that there is also no fully accepted 
physical model explaining all types of FIP bias coherently. For the normal FIP effect, solar physics has provided 
several valuable models. Ideas proposed specifically for the 
stellar case involve stratification of the atmosphere, with different scale heights for ions of different
mass and charge \citep{mewe97}; enrichment of the coronal plasma by some type of ``anomalous flares'' also
seen on the Sun (e.g., Ne-rich flares; \citealt{brinkman01}); electric currents that affect diffusion of
low-FIP elements into the corona \citep{telleschi05}; or a recent model based on a resonance between
chromospheric Alfv\'en waves and a coronal loop, explaining either a FIP or an IFIP effect \citep{laming04}. 
Given the present uncertainties, we do not elaborate further on any of these models in the stellar context. 

\subsubsection{The Ne/O abundance ratio: an indicator of magnetic activity?}
\label{sec:NO}
Not only stellar photospheric abundances remain uncertain for most stars commonly observed
in X-rays, but tabulations  of the solar photospheric composition themselves have seen various revisions 
in the recent past.  The solar composition is a determining factor for the depth of the solar convective zone, 
and the abundances given by \citet{grevesse98}, widely used in the stellar community, have
led to an excellent agreement with the ``standard solar model'' describing the solar interior. 
The latest revision, correcting  solar C, N, and O abundances downward \citep{asplund05a}, however led to 
serious disagreement with the observed helioseismology results \citep{bahcall05}. Regardless of these 
revisions, the absolute solar Ne/O abundance ratio remained at similar levels of 0.15--0.18. 

The helioseismology problem could be solved by adopting a solar photospheric Ne abundance higher by a factor of
about 3--4, or somewhat less if slight re-adjustments are also made for CNO \citep{antia05}. There is
freedom to do so as the photospheric Ne abundance is not really well known, given the absence
of strong photospheric Ne lines in the optical spectrum.

\citet{telleschi05} were the first to point out that the systematically non-solar coronal Ne/O abundance 
ratios measured in several solar-analog stars (offset by similar factors) may call for a revision of the 
adopted solar photospheric Ne abundance, thus at the same time solving the solar helioseismology problem. 
Although X-rays measure {\it coronal} abundances, the derived stellar coronal Ne/O ratios indeed seem to be 
consistently high (by a factor of 2--3) for stars of various activity levels when compared with the 
adopted solar photospheric value; given that both O and Ne are high-FIP elements, the assumption here 
is that the coronal abundance ratio faithfully reflects the photospheric composition.
This was further elaborated in detail by \cite{drake05a} who suggested a factor of 2.7 upward 
revision of the adopted solar Ne abundance.
However, as part of a subsequent debate, analysis of solar active region X-ray spectra and of EUV spectra 
from transition-region levels \citep{schmelz05, young05} both reported robust Ne/O abundance ratios
in agreement with the ``standard'' ratios, rejecting an upward correction of Ne.  

A major problem may in fact be due to selection effects, as initially discussed by \citet{asplund05b}. 
Most of the stars in the sample studied  by \citet{drake05a} are magnetically active stars with 
coronae unlike that of the Sun. The only stars considered with activity indicators $L_{\rm X}/L_{\rm bol}$ in the
solar range were subgiants or giants; in one case (Procyon) different Ne/O abundance ratios, some in
agreement with the solar value, can be found in the published literature, and two further
reports on $\alpha$ Cen and $\beta$ Com suggest values close to the solar mix. \citet{asplund05b} therefore
proposed that the Ne/O ratio is {\it activity dependent}, high values referring to magnetically active
stars while lower, solar-like values refer to inactive, ``solar-like'' stars. This echoes suggestions already
made by \citet{brinkman01} for the high Ne abundance seen in a very active RS CVn-type binary to be possibly due to
flaring activity. The dependence of the Ne/O abundance ratio on activity is in fact
fully recovered in a statistical compilation presented in \citet{guedel04} (Fig.~\ref{fig:abunstat}b). These suggestions 
have meanwhile been confirmed from high-resolution spectroscopy of 
low-activity stars in the solar neighborhood (including the above objects), indicating an abundance ratio of Ne/O 
$\approx 0.2\pm 0.05$ for the Sun and weakly active stars, increasing to Ne/O $\approx 0.4-0.5$ for magnetically active stars
\citep{liefke06, robrade08}. 

The solar Ne/O ratio is thus in line with coronal abundances of inactive stars,
i.e., the ratio is low as adopted by both the older and revised solar abundance lists.
A selective increase of the solar Ne abundance to reconcile measurements of solar composition with 
helioseismology results seems to be ruled out. The discrepancy with the solar model may require other 
modifications yet to be identified.

\subsubsection{Abundance changes in flares}
\label{sec:flareabun}
The chromospheric evaporation scenario for flares predicts that bulk mass motions emerging from the chromosphere
fill coronal loops. One would therefore anticipate that any FIP or IFIP effect becomes suppressed
during large flares, as the corona is replenished with material of chromospheric/photospheric
origin. This is indeed what has been derived from X-ray spectroscopy of stellar flares. Given
the commonly observed IFIP trend outside flares in active stars, flares tend to enhance low-FIP elements more than
high-FIP elements, with abundances eventually approaching the photospheric composition (e.g., 
\citealt{osten00, audard01a}), but such FIP-related abundance changes are not always observed \citep{osten03, 
guedel04a, nordon06}.

\begin{figure}[t]
\centerline{\includegraphics[angle=0,width=12.2cm]{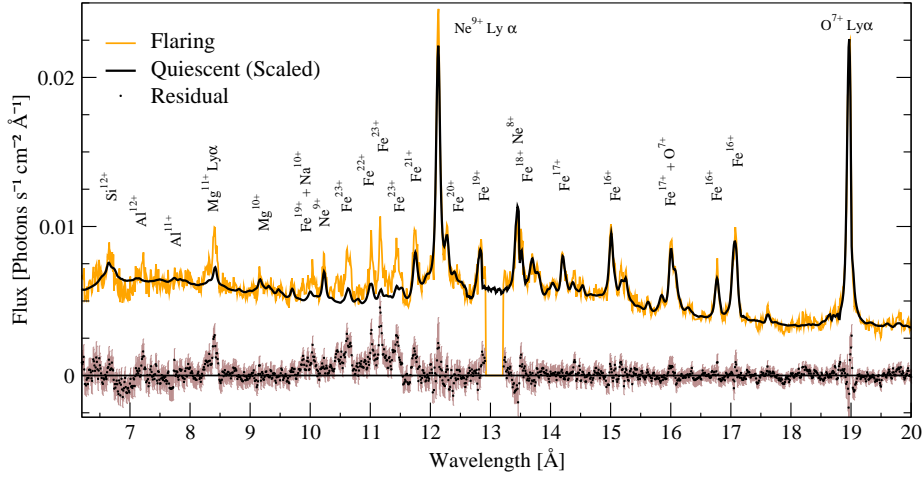}}
\caption{Comparison of a spectrum taken during a large flare on the RS CVn-type binary $\sigma$ Gem (orange) with
a spectrum taken during the quiescent stage (black). An ad-hoc continuum has been added to the quiescent
spectrum to match the flare spectrum (because of the broad-band nature of the bremsstrahlung continuum).
The lowest spectrum shows the residuals ``flare - quiescent''. Note that only short-wavelength lines
forming at high temperatures react to the flare. 
(From \citealt{nordon06}.)}
\label{fig:flarespec} 
\end{figure}

The systematics in FIP-related abundance changes in flares was clarified considerably by \citet{nordon07} and
\citet{nordon08} who spectroscopically studied 14 flares observed with \xmm\ and \ch. The analysis was performed
{\it relative} to the quiescent emission, i.e., trends are {\it independent of the mostly unknown photospheric abundances.}
The majority of the flares indeed showed a ``relative'' solar-like FIP effect (with respect to quiescence),
although relative IFIP effects and absence of relative changes were observed as well. The latter case can
be explained by flares changing the total stellar emission measure only at high temperatures (above 10--20~MK; Fig~\ref{fig:flarespec}), 
i.e., above limit of significant line formation in the 5--20~\AA\ range. While flare-induced abundance changes 
may occur, they thus remain undetected as the observable lines are still generated by quiescent plasma.

\begin{figure}[t]
\vskip 0.7truecm
\centerline{\includegraphics[angle=0,width=8cm]{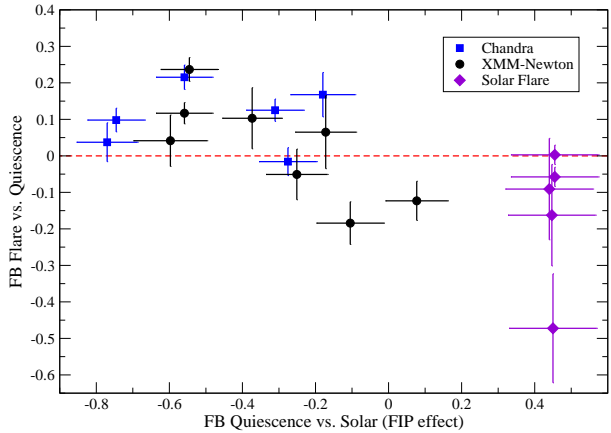}}
\vskip 0.2truecm
\caption{The abscissa gives the logarithm of the {\it quiescent} abundance ratio between low-FIP elements and high-FIP 
elements normalized to solar photospheric values; high values indicate quiescent coronae with a solar-like FIP effect,
low values coronae with an IFIP effect. The ordinate gives a similar ratio for {\it flares} normalized to quiescence.
Here, high values indicate a {\it relative} FIP effect during the flare, and low values a relative IFIP effect. Points to
the right refer to solar flares. Note the anticorrelation, indicating that flares in FIP-effect coronae show a relative
IFIP effect and vice versa. (From \citealt{nordon08}.)}
\label{fig:flarefip} 
\end{figure}

More interestingly, a clear correlation is apparent between quiescent and flaring composition in the
sense that quiescent coronae with a FIP effect typically show a relative IFIP bias during flares, and quiescent IFIP-biased
coronae change to a FIP effect (Fig.~\ref{fig:flarefip}). This relation leads to three significant conclusions:
First, it confirms the chromospheric evaporation scenario that posits that fresh, unfractionated chromospheric/photospheric  
material is brought up into the corona; second, it suggests that the observed coronal (I)FIP biases are genuine
and are not a consequence of the photospheric composition. And third, if active stellar coronae are generated
by a large number of small-scale flares (Sect.~\ref{sec:flareover}), then these should should enrich the corona with
FIP-biased material, different from individual large flares that reveal an opposite trend.

\section{X-rays from young stellar objects and their environments}
\label{sec:yso}

\subsection{From protostars to T Tauri stars: coronal properties}
\label{sec:ysooverview}
X-ray observations of young, forming stars are ideal to address the question on the first appearance of
stellar magnetic fields. It is presently  unknown whether such fields would preferentially be fossil 
or are generated by emerging internal dynamos. 

A rather comprehensive picture of the X-ray characteristics is available for the latest stages of the
star formation process, the T Tauri phase when stars are essentially formed but may still be moderately 
accreting from a circumstellar disk. Significant information has been collected  from large recent 
surveys conducted with \xmm\ and \ch\ \citep{getman05, guedel07a, 
sciortino08}, but also from case studies of individual, close TTS. 
In short, as judged from the temperature structure, flaring behavior, or rotational modulation, X-ray 
emission from TTS mostly originates from magnetic coronae, with characteristics comparable  
to more evolved active main-sequence stars. X-rays from CTTS saturate at a level of  
log($L_{\rm X}/L_{\rm bol}) \approx -3.5$, again similar to main-sequence stars. Because for a 
typical pre-main sequence association $L_{\rm bol}$ roughly correlates with stellar mass $M_*$, one also 
finds a distinct correlation between $L_{\rm X}$ and $M_*$, $L_{\rm X} \propto M_*^{1.7\pm 0.1}$ 
\citep{telleschi07a}. Flaring is common in TTS \citep{stelzer07}, the most energetic examples reaching 
temperatures of $\approx 10^8$~K \citep{imanishi01}.

Our view of X-ray production in younger phases, protostars of Class I (in which most of the final mass has been
accreted) and Class 0 (in which the bulk of the final mass is still to be accreted)  is much less 
complete owing to strong X-ray attenuation by the circumstellar material. Non-thermal radio emission from
Class I protostars suggests the presence not only of magnetic fields but also accelerated
particles in a stellar corona (e.g., \citealt{smith03}).

Class I protostars have amply been detected in X-rays by \xmm\  and \ch\  although 
sample statistics are biased by the X-ray attenuation, favoring detection of the most luminous and 
the hardest sources. Overall, X-ray characteristics are very similar to TTS, confirming that
magnetic coronae are present at these earlier stages. A comprehensive study is available for stars in
the Orion region \citep{prisinzano08}. Here, the X-ray luminosities increase from  protostars to TTS by 
about an order of magnitude, although the situation is unknown below 1--2~keV. No significant trend is found 
for the electron temperatures, which are similarly high ($\approx$1--3~keV) for all detected classes. Again,
flaring is common in Class I protostars \citep{imanishi01}.

For  Class 0 objects, a few promising candidates but no definitive cases have been 
identified (e.g., \citealt{hamaguchi05}). \citet{giardino07a} reported several non-detections of nearby Class 0
objects, with a ``stacked Class 0 data set'' corresponding to 540~ks of Chandra ACIS-I 
exposure time still giving no indication for a detection. In the absence of detailed information on
the absorbing gas column densities or the intrinsic spectral properties, an interpretation within
an evolutionary scenario remains difficult.

\subsection{X-rays from high-density accretion shocks?}
\label{sec:accretion}
Accretion streams falling from circumstellar disks toward the central stars reach a maximum
velocity corresponding to the free-fall velocity,
\begin{equation}
v_{\rm ff} = \left({2GM_*\over R}\right)^{1/2} 
    \approx 620 \left({M\over M_{\odot}}\right)^{1/2}\left({R\over R_{\odot}}\right)^{-1/2}
    \left[{\rm km~s^{-1}}\right].
\end{equation}
This velocity is an upper limit as the material starts only at the inner border of the circumstellar
disk, probably following curved magnetic field lines down toward the star; a more realistic 
terminal speed is $v_{\rm m} \approx 0.8v_{\rm ff}$ \citep{calvet98}. Upon braking at the stellar surface, 
a shock develops which, according to the strong-shock theory, reaches a temperature of
\begin{equation}
T_{\rm s} = {3\over 16k}m_{\rm p}\mu v_{\rm m}^2 \approx 3.5\times 10^6 {M\over M_{\odot}}\left({R\over R_{\odot}}\right)^{-1}~{\rm \left[K\right]}
\end{equation}
(where $m_{\rm p}$ is the proton mass and $\mu$ is the mean molecular weight, i.e., $\mu \approx 0.62$ for ionized gas).
For typical TTS, $M = (0.1-1)M_{\odot}$, $R = (0.5-2)R_{\odot}$, and $M/R \approx (0.1-1)M_{\odot}/R_{\odot}$ 
and therefore $T_{\rm s} \approx  (0.4-4)\times 10^6$~MK. Such electron 
temperatures should therefore produce soft X-ray radiation \citep{ulrich76}.
The bulk of the ensuing X-rays is probably absorbed in the shock, contributing to its heating, although part of
the X-rays may escape and heat the pre-shock gas \citep{calvet98}. 

The pre-shock electron density, $n_1$, can be estimated from the mass accretion rate, $\dot{M}_{\rm acc}$, and the surface 
filling factor of the accretion streams, $f$: using the strong-shock condition for the post-shock density, $n_2 = 4n_1$, together
with $\dot{M}_{\rm acc} \approx 4\pi R_*^2fv_{\rm m}n_1 m_{\rm p}$ one finds
\begin{equation}\label{n2}
n_2 \approx {4\times 10^{11}\over f} \left({\dot{M}_{\rm acc}\over 10^{-8}~M_{\odot}~{\rm yr}^{-1}}\right)\left({R_*\over R_{\odot}}\right)^{-3/2}
            \left({M_*\over M_{\odot}}\right)^{-1/2}~{\rm \left[cm^{-3}\right]}
\end{equation}
which, for typical stellar parameters, accretion rates, and filling factors $f \approx 0.001-0.1$ \citep{calvet98},
predicts densities of order $10^{11}- 10^{13}$~cm$^{-1}$. Given the expected shock temperatures and electron
densities, X-ray grating spectroscopy of the density-sensitive He-like C\,{\sc v}, N\,{\sc vi}, O\,{\sc vii}, and 
Ne\,{\sc ix} line triplets should be ideal to detect accretion shocks. We emphasize that in Eq.~\ref{n2}, the shock
density is proportional to $\dot{M}_{\rm acc}/f$, i.e., strongly dependent on the magnetic structure and the accretion
rate, but it also depends on the fundamental stellar properties via $R_*^{-3/2}M_*^{-1/2}$ \citep{telleschi07b, robrade07}.

The first suggestion for this mechanism to be at work was made by \citet{kastner02}, based on grating spectroscopy of the 
nearby CTTS TW Hya. TW Hya's spectrum shows an unusually high $f/i$ ratio for the Ne\,{\sc ix} triplet, pointing to 
$n_{\rm e} \approx 10^{13}$~cm$^{-3}$. Also, and again at variance with the typical coronal properties of CTTS, the X-ray
emitting plasma of TW Hya is dominated by a cool component, with a temperature of only $T\approx 2-3$~MK. Similarly high
electron densities are suggested from the O\,{\sc vii} triplet \citep{stelzer04}, and supporting albeit tentative evidence 
is also found from ratios of Fe\,{\sc xvii} line fluxes \citep{ness05}. Simple 1-D plane-parallel shock models including 
ionization and recombination physics can successfully explain the observed temperatures and densities, resulting 
in rather compact accretion spots with a filling factor of $\approx 0.3\%$, requiring a mass accretion rate responsible for
the X-ray production of $2\times 10^{-10}~M_{\odot}~{\rm yr}^{-1}$ \citep{guenther07}.

\begin{figure}[t]
\centerline{\includegraphics[angle=0,width=10cm]{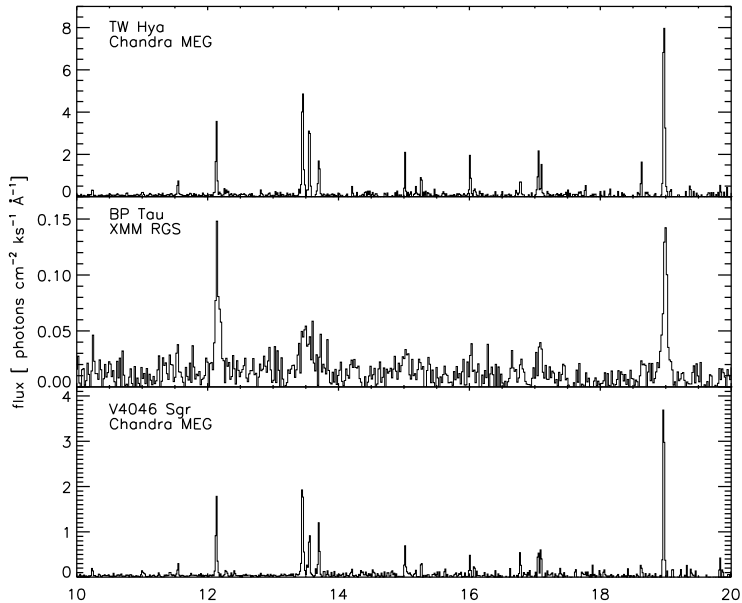}}
\caption{Three spectra from CTTS (from top to bottom, TW Hya, BP Tau, and V4046 Sgr). Note the strong intercombination
lines of Ne\,{\sc ix} (the middle line in the line triplet at 13.4--13.7~\AA).  (From \citealt{guenther06}.)}
\label{fig:cttsspec} 
\end{figure}

Several further CTTS have been observed with X-ray gratings, although the moderate spectral flux of CCTS
and the modest effective areas of detectors presently available have kept the number of well exposed spectra small (see examples
in Fig.~\ref{fig:cttsspec}).
In most cases, high densities are present as judged from the O\,{\sc vii} and Ne\,{\sc ix} triplets, but all spectra
are dominated by a hot, coronal component with the exception of TW Hya. X-ray emission from CTTS may thus have at
least two origins, namely magnetic coronae and accretion flows.
Table~\ref{tab:cttsdens} summarizes published observations of  O\,{\sc vii} triplets in CTTS (and Herbig stars, see
Sect.~\ref{sec:herbigs}) together with derived electron densities.

\begin{table}[t]
\caption{Density measurements in CTTS and Herbig stars based on O\,{\sc vii} triplet line fluxes$^a$}
\label{tab:cttsdens}       
\begin{tabular}{p{1.31cm}p{1.1cm}p{1.0cm}p{1.2cm}p{1.4cm}p{3.7cm}}
\hline\noalign{\smallskip}
Star         & Spectral     & Mass          & O\,{\sc vii}$^b$       & Electron                & Reference \\
             & type         & ($M_{\odot}$) & ${\cal{R}} = f/i$      & density                 &         \\
             &              &               &                        & (cm$^{-3}$)             &         \\
\noalign{\smallskip}\hline\noalign{\smallskip}
Hen~6-300    & M(3+3.5)     &  ...          & 1.0(0.5)               & \hfill 6$\times$10$^{10}$      & \citet{huenemoerder07} \\
TW~Hya       & K8           &  0.7$^c$      & 0.054(0.045)           & \hfill 1.3$\times$10$^{12}$    & \citet{robrade06} \\    
RU~Lup       & K            &  0.8          & 0.26(0.23)             & \hfill 4.4$\times$10$^{11}$    & \citet{robrade07} \\
BP~Tau       & K5-7         &  0.75$^d$     & 0.37(0.16)             & \hfill 3$\times$10$^{11}$      & \citet{schmitt05}  \\
V4046~Sgr    & K5+K7        & 0.86+0.69     & 0.33(0.28)             & \hfill 3$\times$10$^{11}$      & \citet{guenther06}\\     
CR~Cha       & K2           & ...           & 0.64(0.44)             & \hfill 1.6$\times$10$^{11}$    & \citet{robrade06} \\  
MP~Mus       & K1           & 1.2$^c$       & 0.28(0.13)             & \hfill 5$\times$10$^{11}$      & \citet{argiroffi07} \\  
T~Tau        & K0           &  2.4$^d$      & 4.0                    & \hfill$\la$8$\times$10$^{10}$  & \citet{guedel07c} \\ 
HD~104237    & A(7.5-8)     & 2.25          &1.8$_{-0.9}^{+2\ \ b}$  & \hfill 6$\times$10$^{11}$      & \citet{testa08a} \\         
HD~163296    & A1           & 2.3           & 4.6;$>$2.6             & \hfill$\la$2.5$\times$10$^{10}$& \citet{guenther09}\\         
AB~Aur       & A0           & 2.7           & $\approx 4$            & \hfill$\la$1.3$\times$10$^{11}$& \citet{telleschi07c} \\ 
\noalign{\smallskip}\hline
\multicolumn{6}{l}{$^a$values generally from references given in last column and further references therein} \\
\multicolumn{6}{l}{$^b$errors in parentheses;}\\
\multicolumn{6}{l}{\phantom{$^b$}${\cal{R}}$ for HD~104237 refers to the Ne\,{\sc ix} triplet, O\,{\sc vii} being too weak for detection} \\
\multicolumn{6}{l}{$^c$values from \citet{robrade07}} \\
\multicolumn{6}{l}{$^d$values from \citet{guedel07a}} \\
\end{tabular}
\end{table}

\subsection{The ``X-Ray Soft Excess'' of classical T Tauri stars}
\label{sec:softexcess}
\begin{figure}[t]
\includegraphics[angle=0,width=12.3cm]{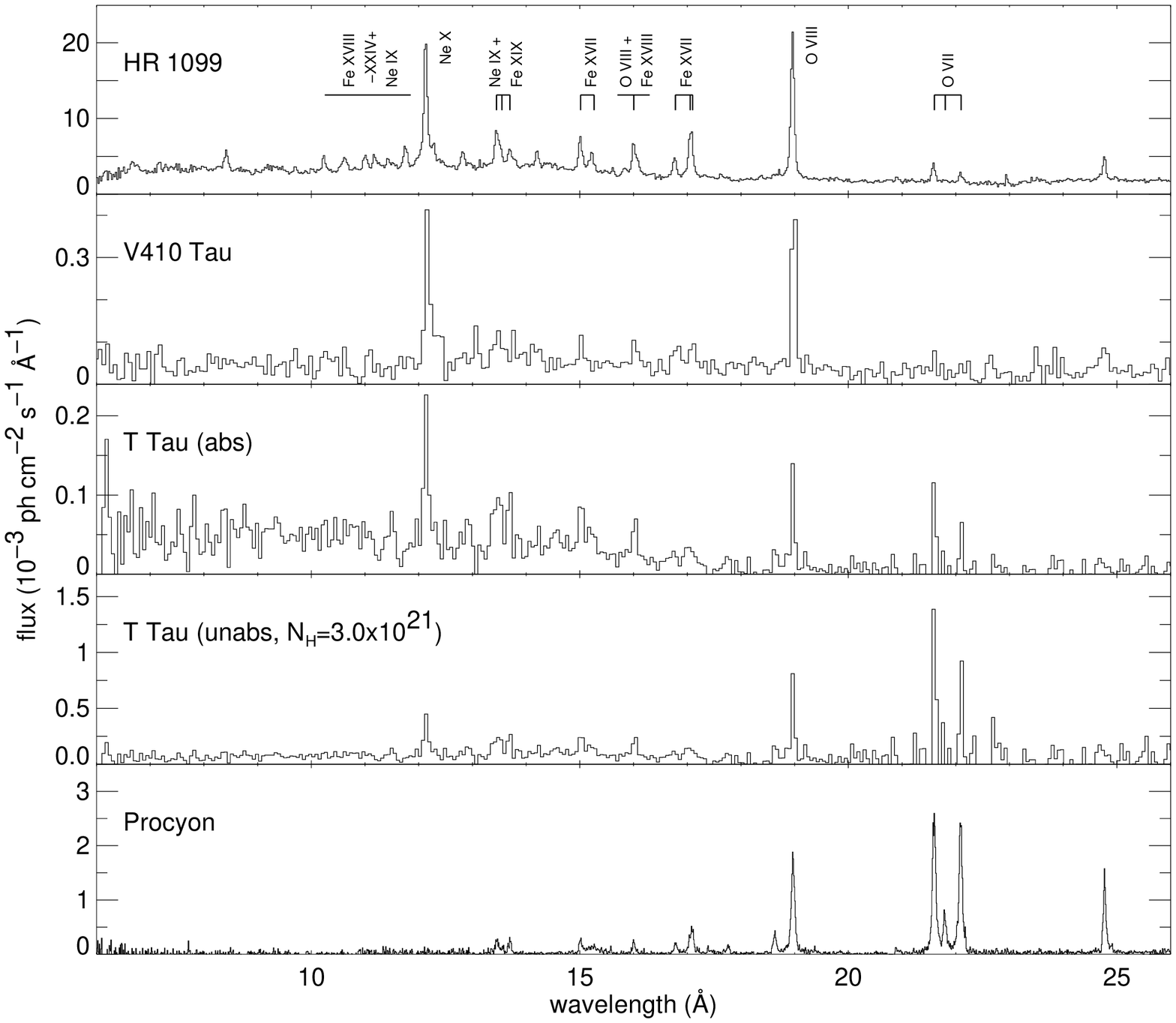}
\caption{Comparison of fluxed \xmm\ RGS X-ray photon spectra of (from top to bottom) the active binary HR~1099, 
         the WTTS V410~Tau, the CTTS T Tau, the T Tau spectrum modeled after removal of absorption  
	 (using $N_{\rm H} = 3\times 10^{21}$~cm$^{-2}$), 
	 and the inactive MS star Procyon. Note the strong O\,{\sc vii} line triplet in T Tau, while the same lines
	 are absent in the spectrum of the WTTS V410~Tau.  (Adapted from \citealt{guedel07d}.)}
\label{fig:ttau} 
\end{figure}

Although TW Hya's soft spectrum remains exceptional among CTTS, lines forming at temperatures of only a few MK 
reveal another anomaly for seemingly {\it all} CTTS, illustrated in
Fig.~\ref{fig:ttau}. This figure compares \xmm\  RGS spectra of the active binary HR~1099 (X-rays mostly 
from a K-type subgiant), the weakly absorbed WTTS V410~Tau, the CTTS T Tau N, and the old F subgiant Procyon.  
HR~1099 and V410~Tau show the typical signatures of a hot, active corona such as a strong continuum, strong lines 
of Ne\,{\sc x} and highly-ionized Fe but little flux in the O\,{\sc vii} line triplet. In contrast, lines of  
C, N, and O dominate the soft spectrum of Procyon, the O\,{\sc vii} triplet exceeding the O\,{\sc viii} Ly$\alpha$ 
line in flux.  T Tau reveals a ``hybrid spectrum'': we see signatures of a very active corona shortward of 19~\AA\ 
but also an unusually strong O\,{\sc vii} triplet. Because T Tau's $N_{\rm H}$ is large (in contrast to  $N_{\rm H}$ of
the other stars), the intrinsic behavior of the O\,{\sc vii} line becomes only evident after correction for
absorption: {\it The {\rm O\,{\sc vii}} lines are in fact the strongest lines in the intrinsic X-ray spectrum,} 
reminiscent of the situation in the inactive Procyon.

While the O\,{\sc vii} line triplet remains undetected in almost all WTTS grating spectra despite 
the usually low absorption column densities toward WTTS, the same triplet is frequently detected as an 
unusually strong  feature in CTTS with their typically larger column density. This reflects quantitatively 
in an anomalously large flux ratio, $S$, between the  intrinsic (unabsorbed) O\,{\sc vii} resonance ($r$) line and 
the O\,{\sc viii} Ly$\alpha$ line, defining the {\it X-ray soft excess} in CTTS
\citep{guedel06, telleschi07b, guedel07b, guedel07d}. 
 
\begin{figure}[t]
\hbox{
\includegraphics[angle=0,width=6cm]{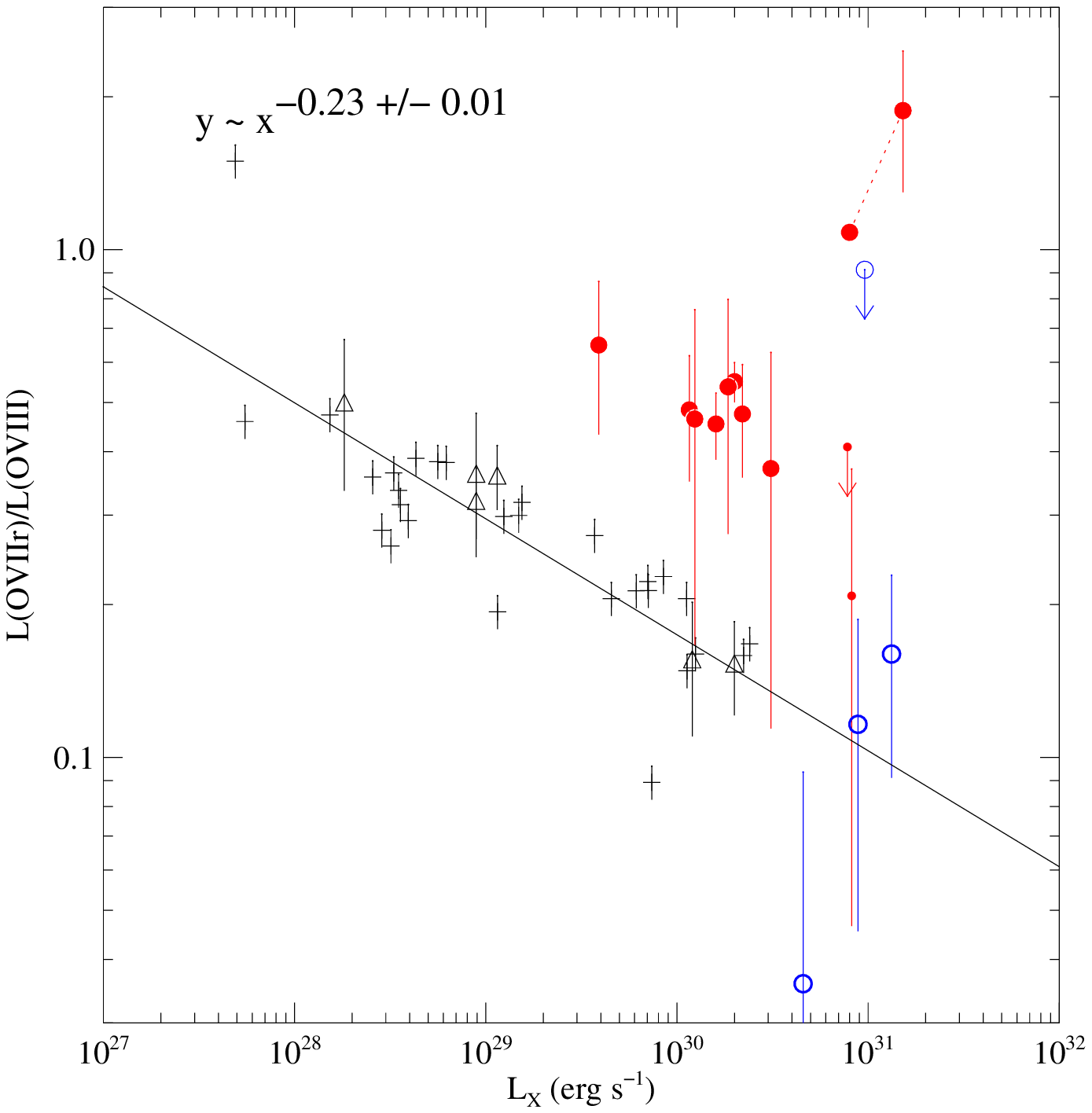}
\includegraphics[angle=0,width=6cm]{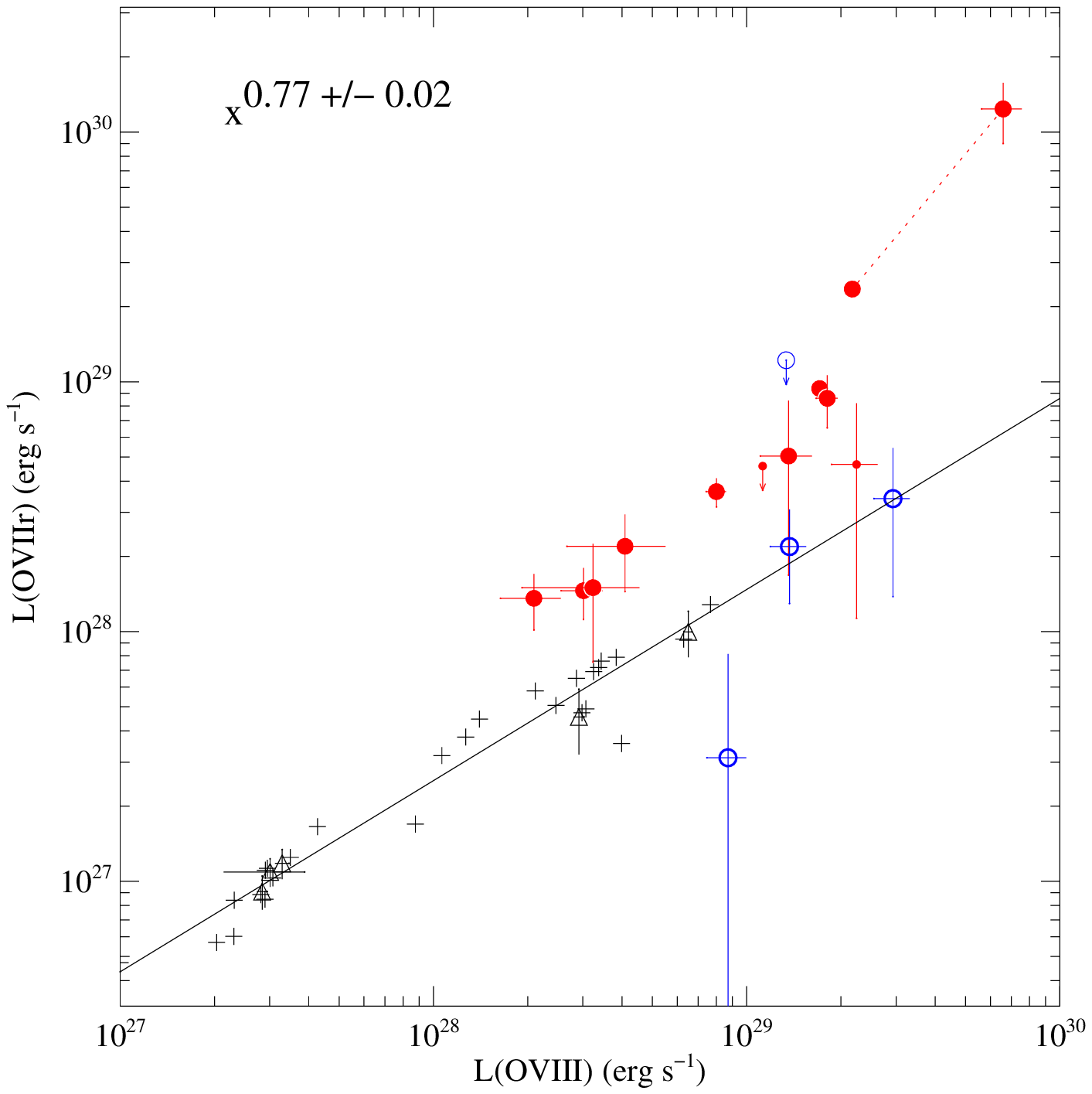}
}
\caption{Left (a): The ratio between O\,{\sc vii} $r$ and O\,{\sc viii} Ly$\alpha$ luminosities  vs. 
$L_{\rm X}$.  Crosses mark MS stars (from \citealt{ness04}), triangles solar analogs (from \citealt{telleschi05}), 
filled (red) circles CTTS, and open (blue) circles WTTS. The solid line is a power-law fit to the 
MS stars with $L_{\rm X} > 10^{27}$~erg~s$^{-1}$ (the relation between the variables is given in the upper left 
corner). --
Right (b): Correlation between $L$(O\,{\sc vii} $r$) and $L$(O\,{\sc viii}). Symbols are as in (a).
(Adapted from \citealt{guedel07d}.)}
\label{fig:softexcess} 
\end{figure}

Figure~\ref{fig:softexcess}a shows the $S$ ratio as a function of $L_{\rm X}$, 
comparing CTTS and WTTS with a larger MS sample (from \citealt{ness04}) and MS solar analogs
(from \citealt{telleschi05}). The trend for MS stars (black crosses and triangles, cf. Eq.~\ref{TLx}, Sect.~\ref{sec:coolstars}) is 
evident: as the coronae get hotter toward higher $L_{\rm X}$, the ratio of O\,{\sc vii} $r$/O\,{\sc viii} 
Ly$\alpha$ line luminosities decreases. This trend is followed by the sample of WTTS,  while CTTS show 
a systematic excess. Interestingly, the excess emission itself seems to correlate with the stellar (coronal)
X-ray luminosity diagnosed by the O\,{\sc viii} Ly$\alpha$ flux (Fig.~\ref{fig:softexcess}b). The soft excess
appears to {\it depend on stellar coronal activity while it requires accretion.} 

The origin of the additional cool plasma is thus likely to be related to the (magnetic) accretion process. 
One possibility is shock-heated plasma at the accretion funnel footpoints as discussed in Sect.~\ref{sec:accretion}.  
Alternatively, the cool, infalling material may partly cool pre-existing heated coronal plasma, or reduce the 
efficiency of coronal heating in the regions of infall \citep{preibisch05, telleschi07b}, which could explain 
the correlation between the soft excess and the coronal luminosity. Such a model would  at the same time explain 
why CTTS are X-ray weaker than WTTS \citep{preibisch05, telleschi07a}. In any case, it seems clear that the 
soft excess described here argues in favor of a substantial influence of accretion on the X-ray production in 
pre-main sequence stars.
 
\subsection{Abundance anomalies as tracers of the circumstellar environment?}
\label{sec:environment}
Initial studies of a few accreting TTS, in particular the old ($\approx 10$~Myr) TW~Hya, showed an abundance pattern in
the X-ray source similar to the IFIP effect although the Ne/Fe abundance ratio is unusually high, of order 10 with respect to the
solar photospheric ratio, and the N/O and N/Fe ratios are enhanced by a factor of $\approx$3.

These anomalous abundance ratios have been suggested \citep{stelzer04, drake05b} to reflect depletion of Fe and O in the accretion disk
where almost all elements condense into grains except for N and Ne  that remain in the gas phase. 
If accretion occurs predominantly from the gas phase in the higher layers of the disk while the grains grow and/or settle at the 
disk midplane, then the observed abundance anomaly may be a consequence.

Enlarging the sample
of surveyed stars has made this picture less clear, however. Several CTTS {\it and non-accreting WTTS} have revealed large Ne/Fe 
ratios ($\approx$4 or higher, \citealt{kastner04, argiroffi05, argiroffi07, telleschi05, telleschi07b, 
guenther06}), suggesting that accretion is not the determining factor for the abundance ratio.
Similar ratios are also found for evolved RS~CVn binaries \citep{audard03a} for which the high Ne/Fe is a consequence 
of the inverse FIP effect (Sect.~\ref{sec:composition}). On the other hand, the {\it accreting} CTTS SU Aur reveals a low Ne/Fe abundance 
ratio of order unity \citep{robrade06, telleschi07b}, similar to several other massive CTTS \citep{telleschi07b}.

Partial clarification of the systematics has been presented by \cite{telleschi07b} who found that
the abundance trends, and in particular the Ne/Fe abundance ratios, do not depend on the accretion status but seem to be a function of
spectral type or surface $T_{\rm eff}$, the later-type magnetically active stars showing a stronger IFIP effect (larger Ne/Fe abundance ratios).
The same trend is also seen in disk-less zero-age main-sequence stars. Further, the same statistical study showed that
the Ne/O abundance ratio is, within statistical uncertainties, indistinguishable between WTTS and CTTS (excluding TW Hya, see below),
arguing against accretion of selectively depleted disk material in these objects.

Anomalously high Ne/O abundance ratios remain, however, for TW~Hya \citep{stelzer04} and V4046~Sgr \citep{guenther06} when compared to
the typical level seen in magnetically active stars or TTS. \citet{drake05b} proposed that the
selective removal of some elements (e.g., O) from the accretion streams should occur only in old accretion disks such as that of TW~Hya where
coagulation of dust to larger bodies is ongoing, whereas younger TTS still accrete the entire gas and dust mix of the inner
disk. However, the old CTTS MP~Mus does not show any anomaly in the Ne/O abundance ratio \citep{argiroffi07}. Larger samples are needed
for clarification.

\subsection{Summary: Accretion-induced X-rays in T Tauri stars}
\label{sec:accretionsummary}
The present status of the search for accretion-induced X-ray emission and current open problems can be summarized as follows:
\begin{itemize}
\item Densities higher than typical for non-flaring coronal plasma, i.e., $n_{\rm e} \ge 10^{11}$~cm$^{-3}$, are seen in the 
      majority of CTTS. A clear exception is T Tau.
      
\item Densities scatter over a wide range; this may be the result of a variety of accretion parameters 
      \citep{guenther07, robrade07}.

\item While TW Hya's X-ray emission is dominated by an unusually soft component, essentially all other CTTS are
      dominated by much hotter plasma, with principal electron temperatures mostly in the 5--30~MK range, similar to coronal 
      temperatures of magnetically active, young main-sequence stars. The presence of flares is also indicative of 
      coronal emission dominating the X-ray spectrum.         

\item A ``soft excess'', i.e., anomalously high fluxes observed in lines forming at low temperatures, e.g., the O\,{\sc vii}
      line triplet, is seen only in accreting TTS and is therefore likely to be related, in some way, to 
      accretion. However, a correlation with the coronal luminosity points toward a relation with coronal heating
      as well. The two relations can be reconciled if the soft excess is due to an interaction of the accretion streams
      with the coronal magnetic field \citep{guedel07d}.         
      
\item While TW Hya's anomalously high Ne abundance \citep{drake05a} and the low abundances of refractory elements 
       (such as Fe or O, \citealt{stelzer04})
       may be suggestive of heated accretion streams from a circumstellar disk, other CTTS do not generally show abundance
       anomalies of this kind. 
        Low abundances of low-FIP elements are a general characteristic of magnetically active stars
        regardless of accretion. And finally, element abundances appear to be a function of spectral type
       or photospheric effective temperature in CTTS, WTTS, and young main-sequence stars while accretion does not matter
       (\citealt{telleschi07b}). 
      
\item If accretion streams are non-steady, correlated X-ray and optical/ultraviolet time variability should be seen
      during ``accretion events'' when the mass accretion rate increases for a short time. Despite detailed searches
      in long time series of many CTTS, no such correlated events have been detected on time scales of minutes to hours
      \citep{audard07} or hours to days \citep{stassun06}.
\end{itemize}         
Accretion-induced X-ray radiation is clearly a promising mechanism deserving detailed further study and theoretical
modeling. Observationally, the sample accessible to grating observations remains very small, however.

\subsection{X-rays from Herbig stars}
\label{sec:herbigs}
Herbig Ae/Be stars are defined after \citet{herbig60} as young intermediate-mass ($\approx 2-10~M_{\odot}$) stars predominantly
located near star-forming regions.  They reveal emission lines in their optical spectra, and their location in the 
Hertzsprung-Russell diagram proves that they are pre-main sequence stars. Herbig stars are therefore considered to 
be the intermediate-mass analogs of TTS. This analogy extends to infrared emission indicative of circumstellar
disks.

Like their main-sequence descendants, B and A-type stars, Herbig stars are supposed to have radiative interiors, lacking the 
convective envelopes responsible for the $\alpha\omega$ magnetic dynamo in late-type stars. Fossil magnetic fields left over
from the star formation process may still be trapped in the stellar interior. Transient convection may, however, be present
during a short phase of the deuterium-burning phase in a shell \citep{palla93}. A dynamo powered by rotational shear
energy may also produce some surface magnetic fields \citep{tout95}. Magnetic fields have indeed been detected on
some Herbig stars with longitudinal field strengths of up to a few 100~G (e.g., \citealt{donati97, hubrig04, wade05}).

X-ray observations provide premium diagnostics for magnetic fields in Herbig stars and their interaction with stellar
winds and circumstellar material. Surveys conducted with low-resolution spectrometers remain, however, ambiguous.
Many Herbig stars are X-ray sources
\citep{damiani94, zinnecker94, hamaguchi05}, with $L_{\rm X}/L_{\rm bol} \approx 10^{-7}-10^{-4}$, i.e., ratios between 
those of TTS ($10^{-4}-10^{-3}$) and O stars ($\approx 10^{-7}$). This finding may suggest that undetected 
low-mass companions are at the origin of the X-rays (e.g., \citealt{skinner04}; the ``companion hypothesis''). The companion 
hypothesis is still favored for statistical samples of Herbig stars including multiple systems studied with {\it Chandra's} 
high spatial resolution  \citep{skinner04, stelzer06, stelzer09}. 

The X-ray spectra of Herbig stars are clearly thermal; intermediate-resolution CCD spectra show the usual indications
of emission lines, partly formed at rather high temperatures \citep{skinner04}.  High electron temperatures ($>10$~MK)
derived from spectral fits as well as flares observed in light curves further support the 
companion hypothesis for Herbig star samples, as both features are common to TTS 
\citep{giardino04, skinner04, hamaguchi05, stelzer06}. On the other hand, $L_{\rm X}$ correlates with 
wind properties such as wind velocity or wind momentum flux (but not with rotation parameters of the Herbig stars), 
perhaps pointing to shocks in unstable winds as a source of X-rays \citep{damiani94, zinnecker94}.

\begin{figure}[t]
\includegraphics[angle=0,width=12.3cm]{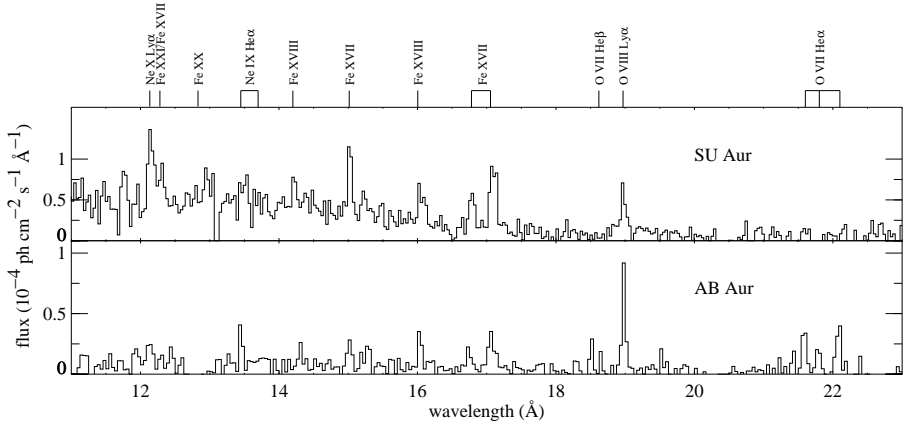}
\caption{\xmm\ grating spectrum of the Herbig star AB Aur (lower panel) compared with the spectrum of the 
CTTS SU Aur (upper panel). Note the softer appearance of the AB Aur spectrum (e.g., absence of a strong continuum)
and also the O\,{\sc vii} triplet
at 22~\AA\ indicating low densities {\it and} the absence of a strong UV radiation field in the X-ray source.
(Adapted from \citealt{telleschi07c}.)}
\label{fig:abaur} 
\end{figure}

High-resolution X-ray spectroscopy has now started revealing new clues about the origin of X-rays from Herbig
stars. The first such spectrum, reported by \citet{telleschi07c} from AB Aur (Fig.~\ref{fig:abaur}), is at variance with X-ray spectra
from CTTS in three respects: First, X-rays from AB Aur are unusually soft (compared to general Herbig X-ray samples or
TTS); electron temperatures are found in the range of 2--7~MK. Second, the X-ray flux 
is modulated in time with a period consistent with periods measured in optical/UV lines originating from the stellar wind.
And third, the O\,{\sc vii} line triplet indicates a high $f/i$ flux ratio (Sect.~\ref{sec:densities}) in contrast to the usually low 
values seen in CTTS.

In the case of sufficiently hot stars ($T_{\rm eff} \ga 10^4$~K), the $f/i$ ratio of the X-ray source
not only depends on the electron density (Sect.~\ref{sec:densities}) but also on the ambient UV radiation field. In that case,
\begin{equation}
{\cal R} = {f\over i} = {{\cal R}_0\over 1+ \phi/\phi_c + n_{\rm e}/N_{\rm c}}
\end{equation}
where ${\cal R}_0$ is the limiting flux ratio at low densities and negligible radiation fields, $N_{\rm c}$ is the
critical density at which ${\cal R}$ drops to ${\cal R}_0/2$ (for negligible radiation fields), and $\phi_c$ is
the critical photoexcitation rate for the $^3S_1 - ^3P_{1,2}$ transition, while $\phi$ characterizes the ambient UV flux. 
The ratio $\phi/\phi_c$ can be expressed using  $T_{\rm eff}$ and fundamental atomic parameters (for applications to
stellar spectra, see \citealt{ness02b} and, specifically for AB Aur, \citealt{telleschi07c}). 
A high $f/i$ ratio thus not only requires the electron density to be low ($\la 10^{11}$~cm$^{-3}$; Table~\ref{tab:cttsdens})
but also the emitting plasma to be located sufficiently far away from the strong photospheric UV source. Considering 
all X-ray properties of AB Aur, \citet{telleschi07c} suggested the presence of magnetically confined winds shocks.
These shocks form in the equatorial plane where stellar winds collide after 
flowing along the closed magnetic field lines from different stellar hemispheres (see also Sect.~\ref{sec:magnetic}). 
As this shock is located a few stellar radii away from the photosphere, a high $f/i$ ratio is easily explained.

HD~104237 is another Herbig star observed with X-ray gratings \citep{testa08b}. This object is of slightly 
later spectral type (A7.5~Ve - A8~Ve) than the other Herbigs observed with gratings, implying that the photospheric radiation field is
irrelevant for the $f/i$ ratio of the He-like O\,{\sc vii}, Ne\,{\sc ix}, Mg\,{\sc xi}, and Si\,{\sc xiii} triplets.
At first sight, both the low-resolution spectrum \citep{skinner04} and the high-resolution grating spectrum
\citep{testa08b} indicate high temperatures similar to TTS, and indeed HD~104237 is accompanied by
a close (0.15~AU), relatively massive K3-type companion (see \citealt{testa08b} and references therein). However, an anomalously cool
component in the spectrum and rotational modulation compatible with the measured rotation period of the Herbig
star point to the latter as the source of the softer emission \citep{testa08b}, while much of the harder emission
may originate from the K-type T Tauri companion. This interpretation fits well into the emerging picture of 
Herbig X-ray emission sketched above although the origin of the soft emission (magnetically confined winds,
accretion shocks, jet shocks [see below], cool coronae induced by shear dynamos) needs further investigation.

The third Herbig star observed at high resolution in X-rays, HD~163296, had previously attracted attention as another
unusually soft X-ray source \citep{swartz05}. Although the latter authors suggested that the soft emission is
due to accretion shocks in analogy to the mechanism proposed for CTTS, grating spectra paint a different
picture: again, the $f/i$ flux ratio is high, requiring the X-ray source to be located at some distance from the 
stellar surface and to be of low density (\citealt{guenther09}; Table~\ref{tab:cttsdens}). Although a similar model as for AB Aur may be applicable,
HD~163296 is distinguished by driving a prominent jet marginally indicated in an image taken by 
\ch\  \citep{swartz05}. A possible model therefore involves shock heating in jets, requiring an initial 
ejection velocity of 750~km~s$^{-1}$, but shock heating at such velocities is required only for a fraction of 
$\approx$1\% of the outflowing mass \citep{guenther09}.

\subsection{Two-Absorber X-ray spectra: evidence for X-ray jets}
\label{sec:tax}
Shocks in outflows and jets may produce X-ray emission. Terminal shocks between jets and the interstellar 
medium, forming ``Herbig-Haro (HH) objects'', are obvious candidates. Sensitive imaging observations have indeed 
detected faint X-ray sources at the shock fronts of HH objects 
(e.g., \citealt{pravdo01}). Shocks may also form internally to the jet, close to the driving star. Such X-ray
sources have been found close to the class I protostar L1551 IRS5 \citep{favata02, bally03} and the CTTS
DG Tau \citep{guedel08}. The faint low-resolution spectra are soft and indicate temperatures of a few million K, 
compatible with shock velocities of a few 100~km~s$^{-1}$.

\begin{figure}[t]
\hbox{
\hskip -3.2truecm\centerline{\includegraphics[angle=-90,width=5.8cm]{TAX.ps}}
\hskip -6truecm\centerline{\includegraphics[angle=0,width=5.6cm]{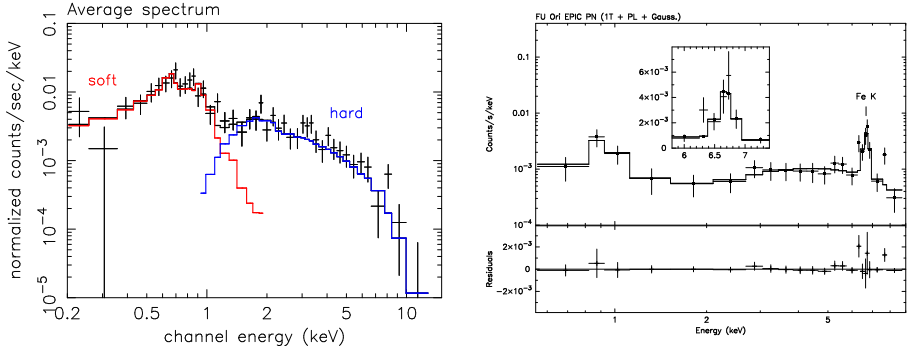}}
}
\vskip -4truecm
\caption{Left (a): Spectrum of the Two-Absorber X-ray source DG Tau. The soft and the hard components are indicated. The solid
histograms show model fits. (Adapted from \citealt{guedel07b}.) --
Right (b): Spectrum of FU Ori. Note the essentially flat appearance  between 0.6~keV and 7~keV, with two broad
peaks related to two different spectral components absorbed by different gas columns. (From \citealt{skinner06},
reproduced by permission of the AAS).}
\label{fig:taxspec} 
\end{figure}

X-ray spectra of several very strongly accreting, jet-driving CTTS exhibit a new spectral phenomenology 
(Fig.~\ref{fig:taxspec},  Table~\ref{tab:tax}): These ``Two-Absorber X-ray'' (TAX) spectra reveal a cool component
subject to very low absorption and a hot component subject to absorption
about one order of magnitude higher. The temperature of the cool component,$T \approx 3$--6~MK, is atypical for TTS.

The hard component of TAX sources  requires an absorbing hydrogen column density
$N_{\rm H} \ga 10^{22}$~cm$^{-2}$, higher by typically an order of magnitude than predicted from the visual 
extinction $A_{\rm V}$ if standard gas-to-dust ratios are assumed (\citealt{guedel07b}
and references therein). Because the hard component requires electron temperatures of tens 
of MK and occasionally shows flares, it is likely to be of coronal origin.
The {\it excess absorption} can be generated by accretion streams falling down along the magnetic 
field lines and absorbing the X-rays from the underlying corona. The 
{\it excess absorption-to-extinction} (or equivalently, $N_{\rm H}/A_{\rm V}$) ratio then is an indicator 
of dust sublimation: {\it the accreting gas streams are dust-depleted}, which is to be expected given that dust
is heated to sublimation temperatures ($\approx 1500$~K) at distances of several stellar radii.

In contrast, $N_{\rm H}$ of the soft component is {\it lower} than suggested from the stellar $A_{\rm V}$.
A likely origin of these very soft X-rays is {\it the base of the  jet} \citep{guedel07b}, suggested by 
i) the unusually soft emission compatible with the jet spectrum, 
ii) the low $N_{\rm H}$ values, and
iii) the explicit evidence of jets in \ch\  imaging \citep{guedel08}. \citet{schneider08}  
supported this scenario by demonstrating that the soft component is very slightly offset from the harder, 
coronal component in the direction of the optical jet. We also note that some CCTS with jets show blueshifts in
O\,{\sc vi} and  C\,{\sc iii} lines (forming at a few $10^5$~K) as observed in {\it FUSE} spectra, again
suggesting shock-heated hot material in jets \citep{guenther08}.

We should note here that one of the sources listed in Table~\ref{tab:tax}, FU Ori, is an eruptive variable presently
still in its outburst stage (which has lasted for several decades). This object is discussed in the context of
eruptive variables below (Sect.~\ref{sec:eruptive}).

\begin{table}[t]
\caption{TAX sources: X-ray parameters and general properties$^a$}
\label{tab:tax}       
%
%
\begin{tabular}{p{1.1cm}p{0.9cm}p{1.3cm}p{1.0cm}p{1.45cm}p{1.3cm}p{0.4cm}p{1.6cm}}
\hline\noalign{\smallskip}
        &               &   \multicolumn{3}{c}{ Soft component}	   &  \multicolumn{3}{c}{Hard component}	   \\
        &               &   \multicolumn{3}{c}{\hrulefill}	   &  \multicolumn{3}{c}{\hrulefill}	   \\
Star    & $A_{\rm V}$   &   N$_{\rm H,s}$	    & $T_{\rm s}$ & $L_{\rm X,s}^b$	   & N$_{\rm H,h}$	   & $T_{\rm h}$ & $L_{\rm X,h}^b$	   \\
        & (mag)         &   ($10^{22}$~cm$^{-2}$)   & (MK)	& ($10^{29}$~erg~s$^{-1}$) & ($10^{22}$~cm$^{-2}$) & (MK)	 & ($10^{29}$~erg~s$^{-1}$)    \\
\noalign{\smallskip}\hline\noalign{\smallskip}
DG Tau  &  1.5-3        &    0.11		    &	3.7   &    0.96 		   &   1.8		     &  69	&   5.1 	       \\
GV Tau  &  3-5          &    0.12		    &	5.8   &    0.54 		   &   4.1		     &  80	&  10.2 	       \\
DP Tau  &  1.2-1.5      &    $\approx 0$	    &	3.2   &    0.04 		   &   3.8		     &  61	&   1.1 	       \\
HN Tau  &               &    0.15		    &2.0, 6.6 &    1.46 		   &   1.1		     &  62	&   3.5 	       \\
CW Tau  &  2-3          &    ...		    &	...   &    ...  		   &	...		     &  ...	&  ...  	       \\
FU Ori  &  1.8-2.4      &    0.42		    &	7.8   &    2.7  		   &   8.4		     &  83	&    53 	       \\
Beehive &               &    0.08		    &	6.6   &    ...  		   &   6.3		     &  41	&  ...  	       \\
\noalign{\smallskip}\hline\noalign{\smallskip}
\end{tabular}
\begin{list}{}{}
\item[$^{\mathrm{a}}$]{X-ray data and $A_{\rm V}$ for Taurus sources from  \citet{guedel07a} and \citet{guedel07b}; 
     for FU Ori, see \citet{skinner06}, 
     and for Beehive, see \citet{kastner05}.
     Index ``s'' for soft component, ``h'' for hard component.}
\item[$^{\mathrm{b}}$]{$L_{\rm X}$ for 0.1--10~keV range (for FU Ori, 0.5-7.0~keV), in units of $10^{29}$~erg~s$^{-1}$}
\end{list}
\end{table}

\subsection{X-rays from eruptive variables: coronae, accretion, and winds}
\label{sec:eruptive}
A small class of eruptive pre-main sequence stars deserves to be mentioned for their peculiar 
X-ray spectral behavior, both during outbursts and during ``quiescence''. They are roughly classified
into FU Orionis objects (FUors henceforth) and EX Lupi objects (EXors). For reviews, we refer to 
\citet{herbig77} and  \citet{hartmann96}.

The classical FUors show optical outbursts during which the systems' brightness increases 
by several magnitudes on time scales of 1--10~yrs, and then decays on time scales of 20--100~yr.
The presently favored model posits that the brightness rises owing to a strongly increasing
accretion rate through the circumstellar disk (from $10^{-7}~M_{\odot}$~yr$^{-1}$ to 
$10^{-4}~M_{\odot}$~yr$^{-1}$). FUor outbursts may represent a recurrent but transient phase 
during the evolution of a TTS. EXors show faster outbursts  with time scales of a few 
months to a few years, smaller amplitudes (2--3 mag) and higher repetition rates.

The first FUor X-ray spectrum, observed from the prototype FU Ori itself, shows a phenomenology
akin to the TAX spectra discussed for the jet sources (Sect.~\ref{sec:tax}; Fig.~\ref{fig:taxspec}b), but 
with notable deviations (\citealt{skinner06}; Table~\ref{tab:tax}). 
First, the cooler and less absorbed spectral component shows a 
temperature ($\approx 8$~MK) characteristic of {\it coronal} emission from TTS, with an absorption 
column density compatible with the optical extinction. An interpretation based on shocks in accretion flows or 
jets is thus unlikely, and jets have not been detected for this object although it does shed a very strong wind, 
with a mass loss rate of $\dot{M}_{\rm w}\approx 10^{-5}~M_{\odot}$~yr$^{-1}$. The hard spectral component shows
very high temperatures ($T\approx 65-83$~MK), but this is reminiscent of TAX sources. Possible models for
the two-absorber phenomenology include patchy absorption in which emission from the cooler coronal component 
suffers less absorption, while the dominant hotter component may be hidden behind additional gas columns, 
e.g., dense accretion streams. Absorption by a strong, neutral wind is an alternative, although this model
requires an inhomogeneous, e.g., non-spherical geometry to allow the soft emission to escape along a path
with lower absorption. 

The above X-ray peculiarity does not seem to be a defining property of FUors; X-ray emission from the FUor
V1735 Cyg does neither show TAX phenomenology nor anomalous absorption but does reveal an extremely
hard spectrum, reminiscent of the hard spectral component of FU Ori \citep{skinner09}. It is also unclear whether
the spectral phenomenology is related with the outburst status of the stars; both FU Ori and V1735 Cyg 
have undergone outbursts, starting about 70 years and 50 years ago (e.g., \citealt{skinner09} and references 
therein), and the former is still in its declining phase.

Two objects, both probably belonging to the EXor group, have been observed from the early outburst phase into
the late decay. The first, V1647 Ori, revealed a rapid rise in the X-ray flux by a factor of $\approx 30$,
closely tracking the optical and near-infrared light curves \citep{kastner04a, kastner06}. The X-ray spectra 
revealed a hardening during the peak emission, followed by a softening during the decay. The absorbing column 
density did not change. Measured temperatures of order $6\times 10^7$~K are much too high for accretion or 
outflow  shocks; explosive magnetic reconnection in star-disk magnetic fields, induced by the strong rise of 
the disk-star accretion rate, is a possibility, and such events may be at the origin of ejected
jet blobs \citep{kastner04a}. The picture of enhanced winds or outflows is supported by a gas-to-dust mass 
ratio enhanced by a factor of two during outburst, compared to standard interstellar ratios \citep{grosso05}.

A very different picture arose from the EXor-type eruption of V1118 Ori \citep{audard05, lorenzetti06}. While 
the optical and near-infrared brightness increased by 1--2 magnitudes on a time scale of 50 days, the X-rays varied 
slowly, by no more than a factor of two; however, the X-ray emission clearly {\it softened} during the outburst, 
suggesting that the hot plasma component disappeared during that episode. A possible model involves the disruption 
of the outer, hot magnetospheric or coronal  magnetic loops by a disk that closed in as a consequence of the 
increased accretion rate \citep{audard05}.

\subsection{X-rays from circumstellar environments: Fluorescence}
\label{sec:fluordisk}
X-rays from TTS and protostars surrounded by circumstellar disks (and possibly molecular envelopes) inevitably
interact with these environments. For the observer, this is most evident in increasingly large X-ray attenuation
toward stars of earlier evolutionary stages. The responsible photoelectric absorption mechanism ionizes disk
surfaces or envelopes quite efficiently \citep{glassgold97}, which has important consequences for chemistry,
heating, or accretion mechanisms.

Here, we discuss direct evidence for disk ionization from fluorescent emission by ``cold'' (weakly ionized) iron.
In brief, energetic photons emitted by a hot coronal source eventually hitting weakly ionized or neutral surfaces 
may eject an 1$s$ electron from Fe atoms or ions. This process is followed by fluorescent transitions, the most efficient of
which is the 2$s$--1$p$ Fe~K$\alpha$ doublet at 6.4~keV, requiring irradiating photons with energies above the 
corresponding photoionization edge at 7.11~keV. The transition is known from X-ray observations of solar
flares, in which case fluorescence is attributed to X-ray irradiation of the solar photosphere \citep{bai79}.
The physics of fluorescent emission is comparatively simple although radiative transfer calculations are required
for accurate modeling (e.g., \citealt{drake08b}). 

The same radiation can also be generated by impact ionization due to a non-thermal electron beam \citep{emslie86}, a 
mechanism that would be expected in particular during the early phases of a flare when high-energy electrons are
abundant \citep{dennis85}. Excitation by high-energy electrons is a rather inefficient process, 
however \citep{emslie86, ballantyne03}.

Although fluorescent emission is prominent in accretion disks around compact objects, detection around 
normal stars is relatively recent. Observations of the 6.4~keV Fe~K$\alpha$ line require a spectral resolving power 
of $\approx$~50 in order to separate the line from the adjacent, and often prominent, 6.7--7.0~keV Fe\,{\sc xxv-xxvi} 
line complex from hot plasma. An unambiguous detection was reported by \citet{imanishi01} who found strong
6.4~keV line flux during a giant flare in the Class I protostar YLW 16A (in the $\rho$ Oph dark cloud) and attributed 
it to irradiation of the circumstellar disk by X-rays from hot coronal flaring loops.
\begin{figure}[t]
\hbox{
\includegraphics[angle=0,width=6.2cm]{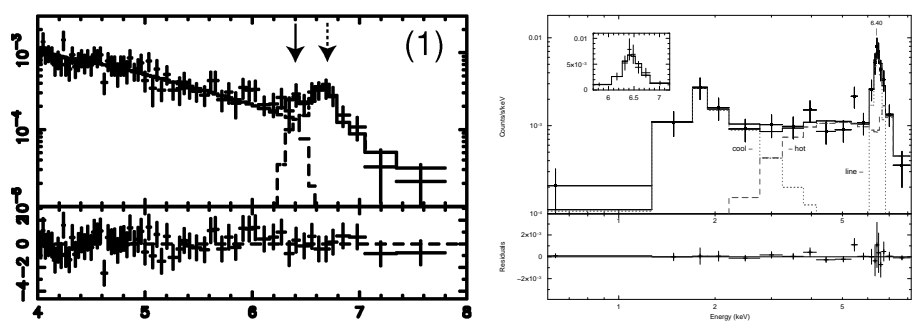} 
\hskip 0.2truecm\includegraphics[angle=-90,width=5.5cm]{ngc2071fluor.ps}
}
\vskip -4truecm
\caption{Left (a): Evidence for an Fe fluorescence line at 6.4~keV (left arrow) next to the 6.7~keV feature of highly ionized Fe 
(right arrow); best-fit models including a narrow line at 6.4~keV, as well as the residuals (bottom panel) are
also shown. (From \citealt{tsujimoto05}, reproduced by permission of the AAS.) --
Right (b): Spectrum of the deeply embedded infrared source  NGC~2071 IRS-1; note the flat spectrum. 
The 6--7~keV region is entirely dominated by the 6.4~keV feature.
 (From \citealt{skinner07}, reproduced by permission of the AAS.)  }
\label{fig:fluor} 
\end{figure}

Detectable Fe fluorescence has remained an exception among CTTS or protostars, but a meaningful sample is 
now available (\citealt{tsujimoto05}; Fig~\ref{fig:fluor}a), with several unexpected features deserving further study. Specifically,
strong fluorescence in the Class I protostar Elias 29 (in the $\rho$ Oph region) appears to be quasi-steady rather than related to 
strong X-ray flares \citep{favata05}; even more perplexing, the 6.4~keV line is modulated on time scales of
days regardless of the nearly constant stellar X-ray emission; it also does not react to the occurrence of appreciable
X-ray flares in the light curve \citep{giardino07b}. The latter authors suggested that here we see 6.4~keV
emission due to non-thermal electron impact in relatively dense, accreting magnetic loops. Given the high density,
the electrons may not reach the stellar surface to produce evaporation and hence, an X-ray flare.
The long time scale of this radiation is, however, still challenging to explain.

Occasionally, the fluorescence line dominates the 6--7~keV region entirely. An example is NGC~2071 IRS-1 in which 
a very broad and bright 6.4~keV feature (Fig.~\ref{fig:fluor}b) requires an absorbing column density of 
$\approx 10^{24}$~cm$^{-2}$ for fluorescence, larger than the column density absorbing the rest of the spectrum \citep{skinner07}.

\begin{figure}[t]
\hbox{
\includegraphics[angle=0,width=6cm]{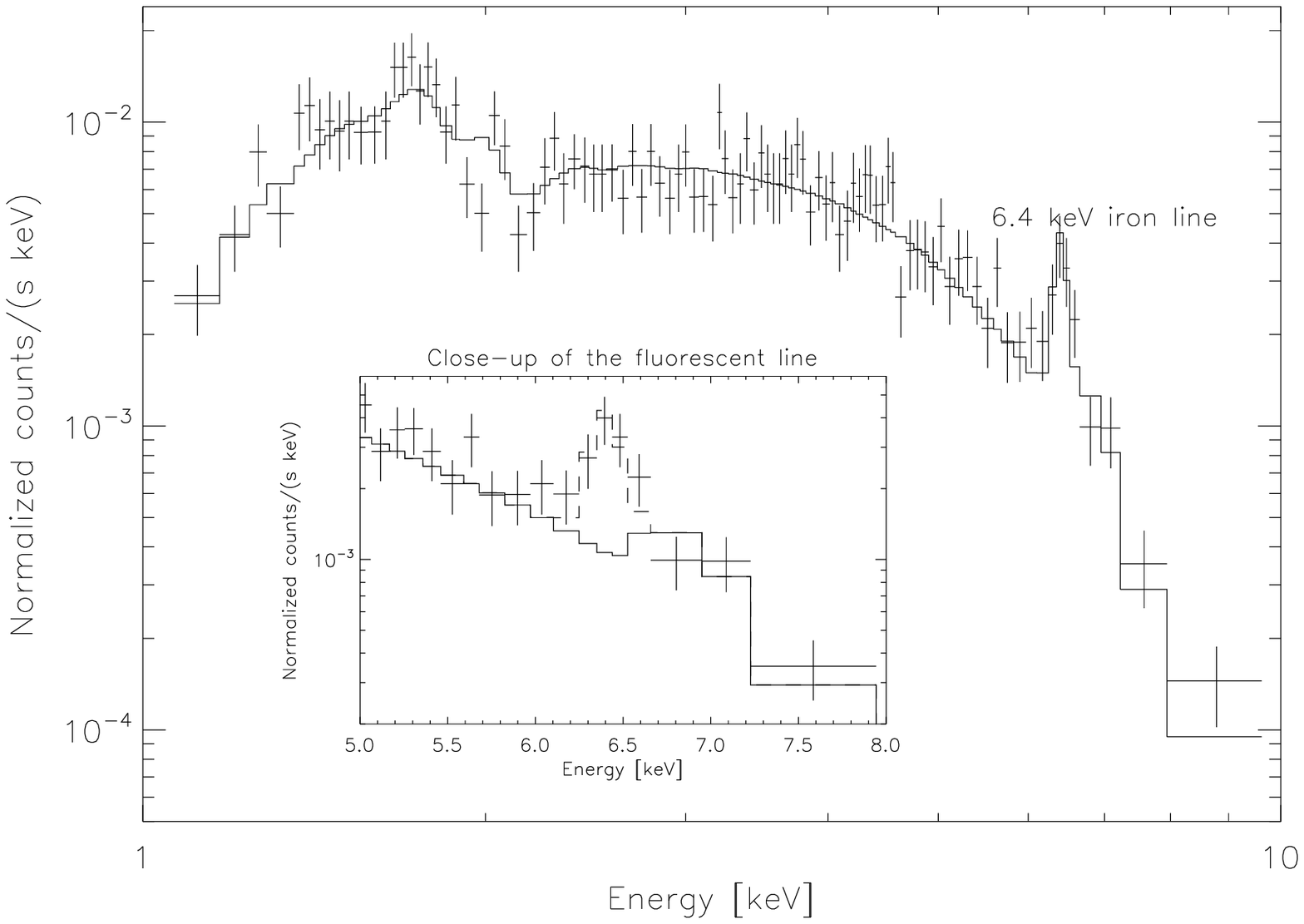}
\includegraphics[angle=0,width=6cm]{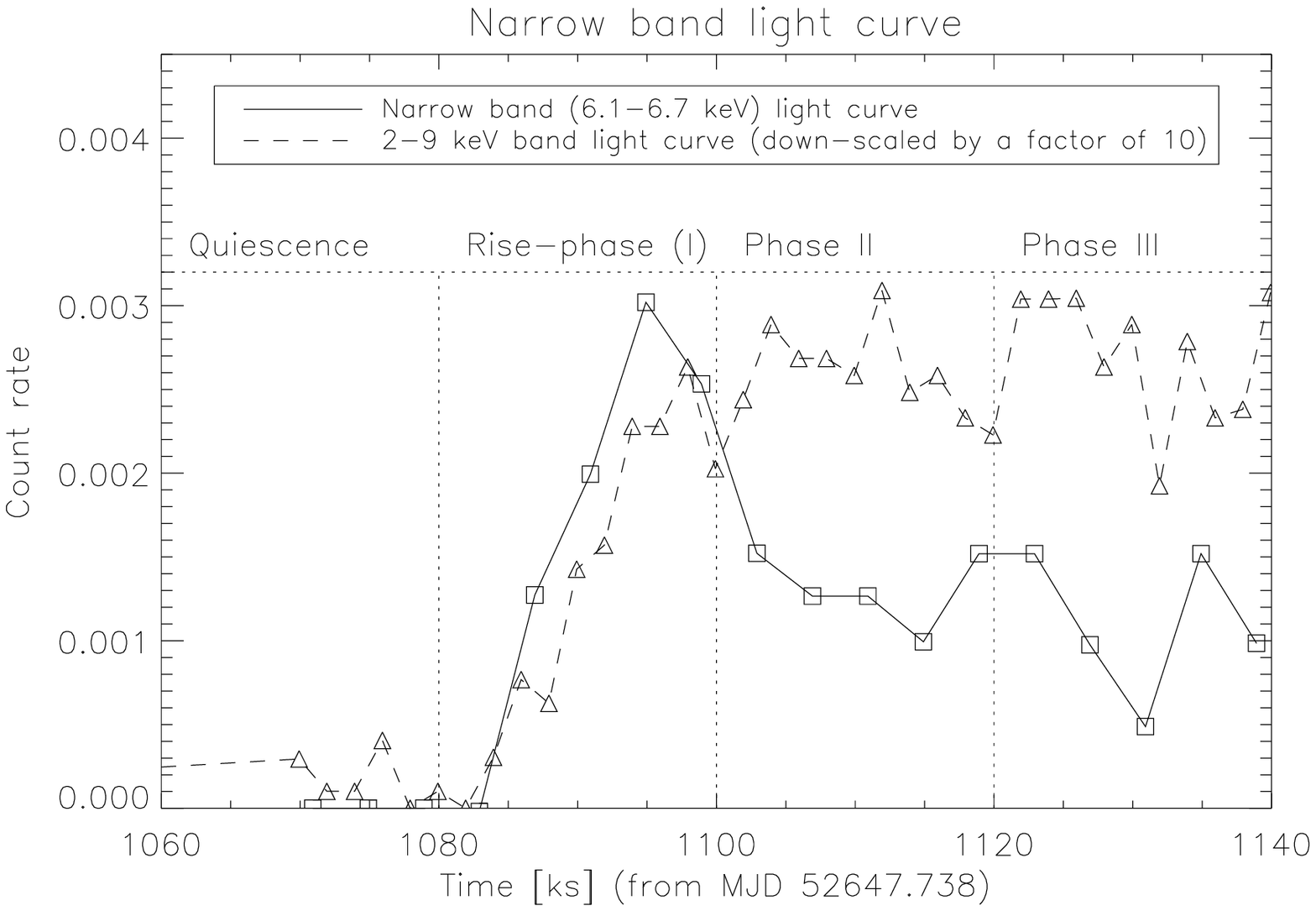}
}
\caption{Left (a): Medium-resolution spectrum of the flaring protostar V1486 Ori showing a 6.4~keV iron line with very little
contribution from the 6.7~keV Fe\,{\sc xxv} feature (see also inset).  --
Right (b): The solid light curve refers to the 6.1--6.7~keV spectral range during the same flare, while the dashed line
shows the integrated 2--9~keV light curve. Note that the $\approx$6.4~keV emission peaks early in the flare.
(From \citealt{czesla07}.)}
\label{fig:coup331} 
\end{figure}

A connection with the initial energy release in a protostellar flare is seen in the case of the Class I star
V1486 Ori in which strong 6.4~keV line flux is seen during the rise phase of a large soft X-ray flare (\citealt{czesla07};
Fig.~\ref{fig:coup331}; the average quiescent + flare spectrum of this object is shown in Fig~\ref{fig:fluor}a).
A relation with the electron beams usually accelerated in the initial ``impulsive'' phase of a (solar) flare is suggestive, 
although this same flare phase commonly also produces the hardest thermal X-ray spectra. In any case, the extremely 
large equivalent width requires special conditions; detailed calculations by \citet{drake08b} suggest fluorescence 
by an X-ray flare partially obscured by the stellar limb.

\section{X-rays from hot stars}
\label{sec:intro}
In the Hertzprung-Russell diagram, the top of the main sequence harbours stars with spectral types O and early B, which are the hottest objects of the stellar population ($>$20\,kK). They also are the most luminous (10$^5$\,L$_{\odot}$) and the most massive (M$>$10\,M$_{\odot}$) stellar objects. Early-type stars emit copious amounts of UV radiation which, through resonance scattering of these photons by metal ions, produce a strong outflow of material. Typically, the mass-loss rate of this stellar wind is about 10$^{-6}$\,\msol\,yr$^{-1}$ (i.e. 10$^8$ times the average solar value!) while the terminal (i.e. final) wind velocity is about 2000\,\kms. Though rare and short-lived ($<$10\,Myr), hot stars have a profound impact on their host galaxies. Not only are they the main contributors to the ionizing flux, mechanical input, and chemical enrichment but they can also be the precursors/progenitors of supernovae, neutron stars, stellar black holes and possibly gamma-ray bursts. Therefore, an improved understanding of these objects, notably through the analysis of their high-energy properties, can lead to a better knowledge of several galactic and extragalactic phenomena.

X-ray emission from hot stars was serendipitously discovered with the Einstein satellite thirty years ago \citep{har79}. One month after Einstein's launch, in mid-December 1978, calibration observations of Cyg X-3 revealed the presence of five additional sources to the North of that source. Dithering exposures showed that these moderately bright sources followed the motion of Cyg X-3, indicating that they were not due to contamination of the detector, as was first envisaged. A comparison with optical images finally unveiled the nature of these sources: bright O-type stars from the Cyg OB2 association. This discovery was soon confirmed by similar observations in the Orion and Carina nebulae \citep{ku79,sew79}. The detection of  high-energy emission associated with hot stars confirmed the expectations published in the same year by \citet{cas79}, who showed that the ``superionization'' species (e.g. N\,{\sc v}, O\,{\sc vi}) observed in the UV spectra could be easily explained by the Auger effect in the presence of a significant X-ray flux.

Since then, hot stars have been regularly observed by the successive X-ray facilities: indeed, only X-rays can probe the energetic phenomena at work in these objects. However, the most critical information was only gained in the past ten years thanks to the advent of new observatories, \xmm\ and \ch, which provided not only enhanced sensitivity but also the capability of high-resolution spectroscopy. 

\subsection{Global properties}
\label{sec:global}
\subsubsection{Nature of the emission}
\label{sec:nature}

\begin{figure}
\hspace{-0.2cm}
   \includegraphics[width=6.3cm, bb=20 150 570 650, clip]{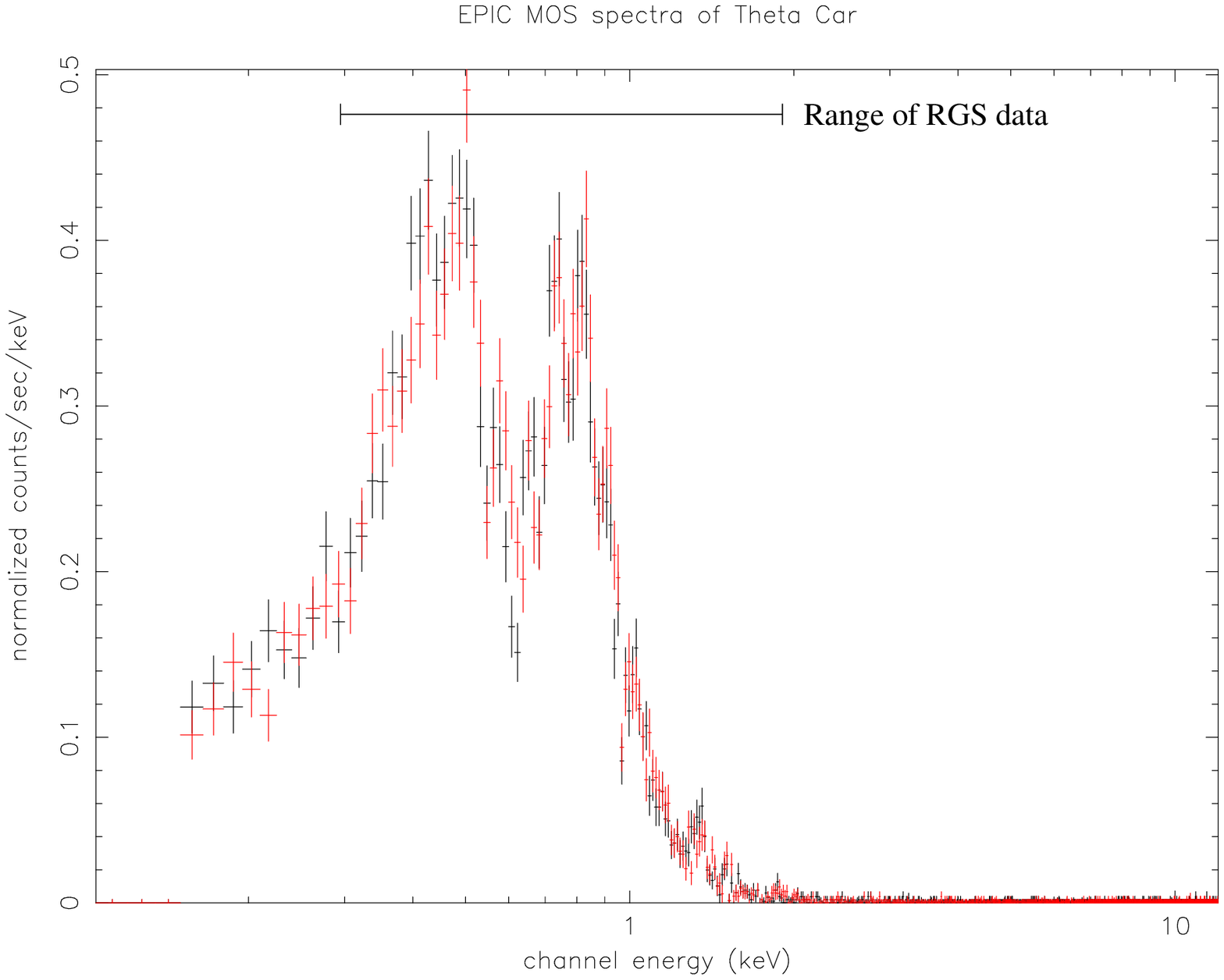}
   \includegraphics[width=6.3cm]{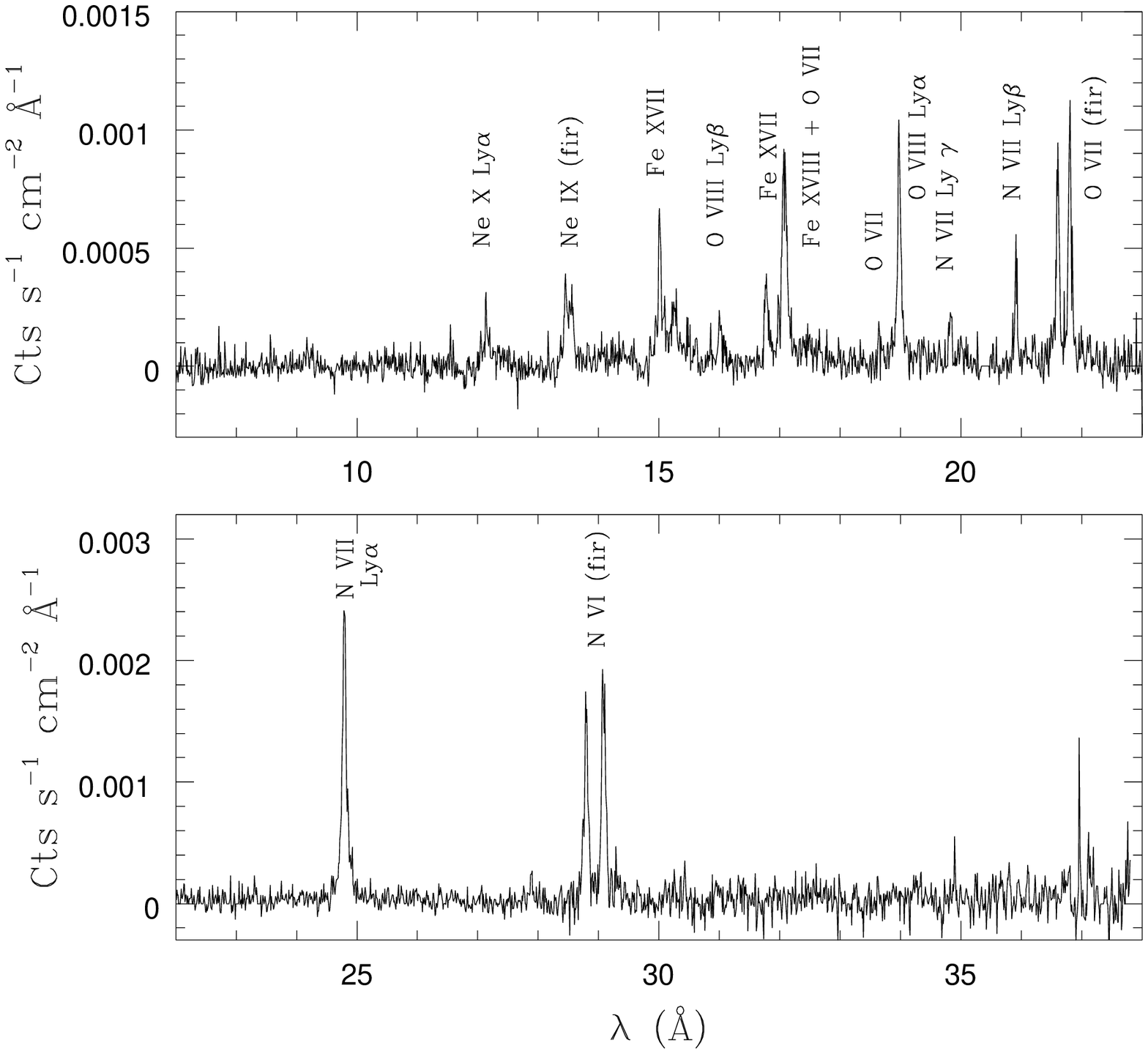}
  \caption{Medium and high-resolution X-ray spectra of the massive system $\theta$\,Car (B0.2V+ late B:), as observed by \xmm. The high-resolution spectrum distinctly shows lines from H-like and He-like ions, as well as lines associated with L shell transitions in iron ions (from \citealt{naz08c}).}
  \label{spec}
  \end{figure}

\begin{figure}
  \begin{center}
  \includegraphics[width=12cm]{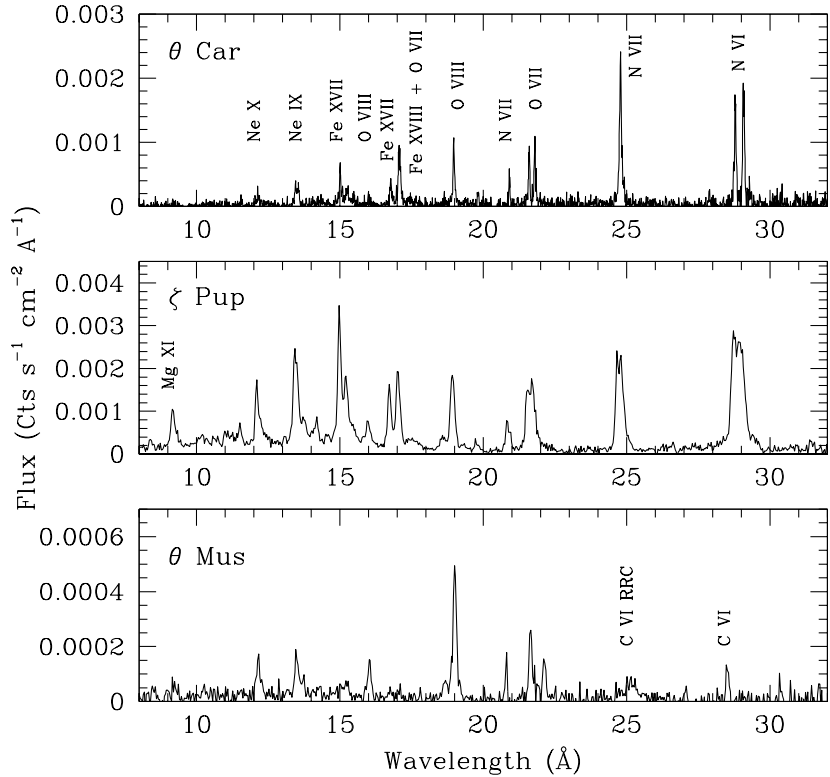}
  \caption{High-resolution X-ray spectra of a B star ($\theta$\,Car, top, adapted from \citealt{naz08c}), an O-star ($\zeta$\,Pup, middle), and a WR star ($\theta$\,Mus, bottom). (For the last two: data taken from the 2XMM database) }
  \label{highresobwr}
  \end{center}
  \end{figure}

Medium and high-resolution spectroscopy (see Figs. \ref{spec} and \ref{highresobwr}) have now ascertained that the X-ray spectrum of hot stars, single as well as binaries, is composed of discrete lines from metals whose ionization stages correspond to a (relatively) narrow range of temperatures. These lines can be superimposed on a (weak) continuum bremsstrahlung emission. The X-ray spectrum thus appears mainly thermal in nature. 

However, some hot, long-period binaries display non-thermal radio emission, which clearly reveals the presence of a population of relativistic electrons in the winds of these objects. Such relativistic electrons are likely accelerated by diffuse shock acceleration processes taking place near hydrodynamic shocks inside or (most probably) between stellar winds (for a review, see \citealt{deb07}). The possibility of a high-energy counterpart to this radio emission was investigated by \citet{che91}. In this domain, inverse Compton scattering is expected to boost a small fraction of the ample supply of stellar UV photons to X-ray energies. To be detectable, such non-thermal X-ray emission needs to be sufficiently strong compared to the thermal X-ray emission. This preferentially happens in short-period binaries, where the wind-wind interaction, and hence particle acceleration, takes place close to the stars and thus inside the radio photosphere, rather than in long-period binaries, where the interaction is located further away, outside the radio photosphere: the list of binary systems with non-thermal X-rays is thus expected to differ from that of non-thermal radio emitters \citep{deb07}. However, no convincing evidence has yet been found for the presence of non-thermal X-ray emission: only two candidates (HD~159176, \citealt{deb04}, and FO15, \citealt{alb05}) present some weak indication of a possible non-thermal component in their spectra. In addition, the observed non-thermal emission of some binary systems still appears problematic, like e.g. for Cyg\,OB2\,\#8A \citep{rau08}. Further work is thus needed to settle this question, and the answer may have to wait for the advent of more sensitive instruments, especially in the hard X-ray/soft gamma-ray range. 

\subsubsection{Temperatures}
\label{sec:temperature}

Except in the case of some peculiar objects (see Sects. \ref{sec:cwb}, \ref{sec:magnetic}), the X-ray spectrum of single hot stars appears overall quite soft. Fits to medium-resolution spectra with optically-thin thermal plasma models in collisional ionization equilibrium (e.g. {\it mekal, apec}) always favor main components with temperatures less than 1\,keV. For O stars, the best fits to good quality data can usually be achieved by the sum of two thermal components at about 0.3 and 0.7--1\,keV (e.g. the O stars in the NGC~6231 cluster, \citealt{san06}, and in the Carina OB1 association, \citealt{ant08}).

High-resolution spectroscopic data of 15 OB stars were analyzed in a homogeneous way by \citet{zhe07} for the case of X-ray emitting regions distributed throughout the wind (see Sect. \ref{sec:origin} for details). The distribution of the differential emission measure (DEM) as a function of temperature displays a broad peak centered on 0.1--0.4\,keV (i.e. a few MK, see also a similar result obtained for 9 OB stars by \citealt{woj05}), confirming the soft nature of the spectrum. A tail at high temperatures is sometimes present but is always of reduced strength compared to the low-temperature component. The only exceptions to this scheme are two magnetic objects (see Sect. \ref{sec:magnetic}) and the otherwise normal O-type star $\zeta$\,Oph ($kT\sim$0.5--0.6\,keV). In general, no absorbing column in addition to the interstellar contribution was needed in the fits, suggesting the absorption by the stellar wind to be quite small (\citealt{zhe07}, see also \citealt{san06}). 

\subsubsection{\lxlb\ relation}
\label{sec:lxlbolsection}

Already at the time of the discovery, \citet{har79} suggested that a correlation exists between the X-ray and optical luminosities. Such a relation was soon confirmed and slightly revised to a scaling law between unabsorbed X-ray and bolometric luminosities\footnote{Indeed, for stars of similar spectral type and interstellar absorption, both relations are equivalent.} of the form $L_{\rm X}^{\rm unabs}\sim 10^{-7} \times L_{\rm bol}$ (see e.g. \citealt{pal81}). \citet{sci90} examined the possibility of correlations with other properties of these hot stars but could not find any evidence for relations with the rotation rate (contrary to late-type stars), the terminal wind velocity nor the mass-loss rate (though combinations of velocity and mass-loss rates gave good results, see below). More recently, \citet{ber97} re-evaluated this scaling law using data from the \ros\ All-Sky Survey (RASS). Their relation (see Table \ref{tab:lxlbol}) presents a large scatter ($\sigma$ of 0.4 in logarithmic scale, or a factor 2.5) and flattens considerably below $L_{\rm bol}\sim10^{38}$ \ergs, i.e. for mid and late B stars.

  \begin{figure}
\hspace{-0cm}
  \includegraphics[width=4.0cm]{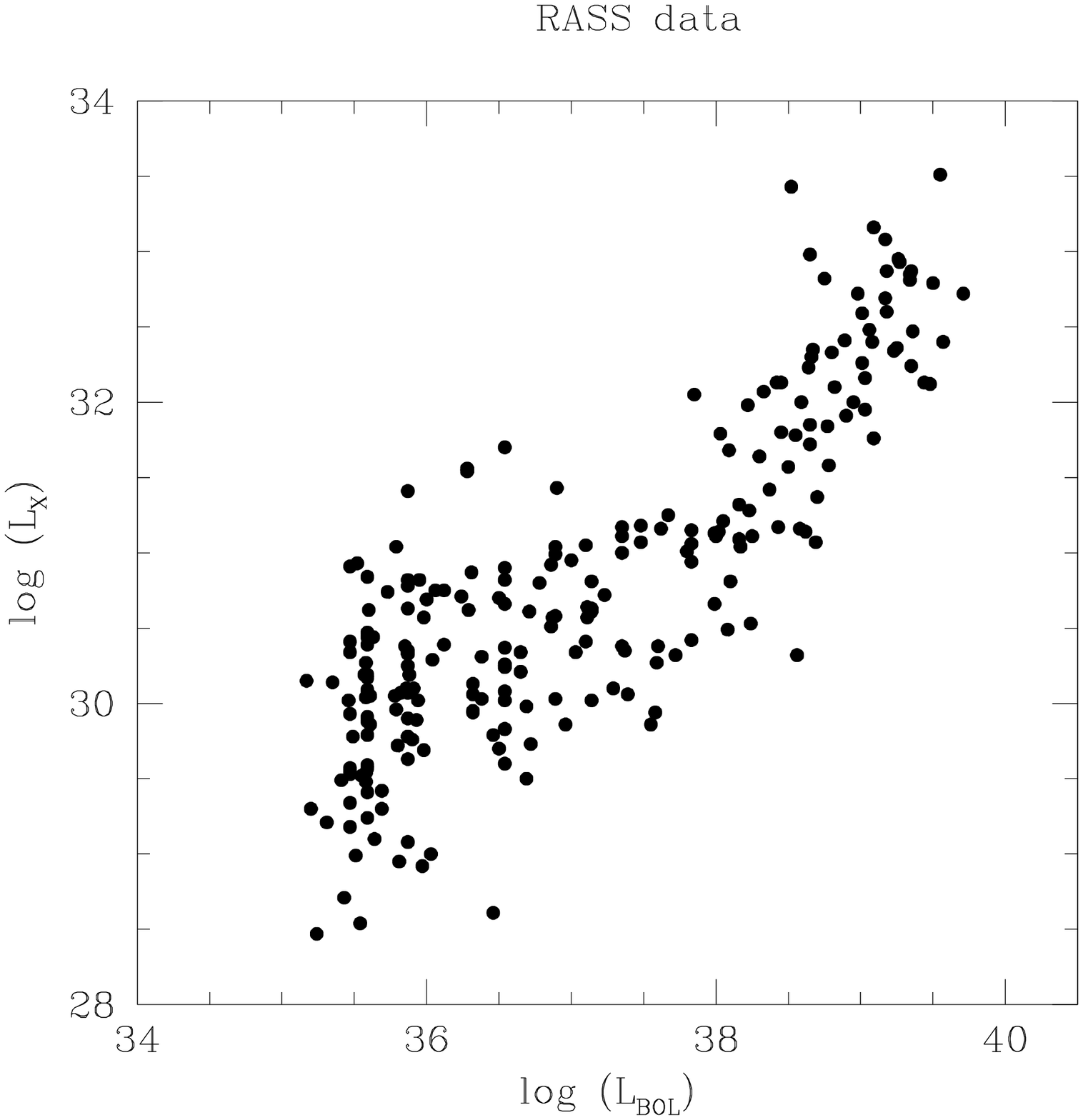}
  \includegraphics[width=4.0cm]{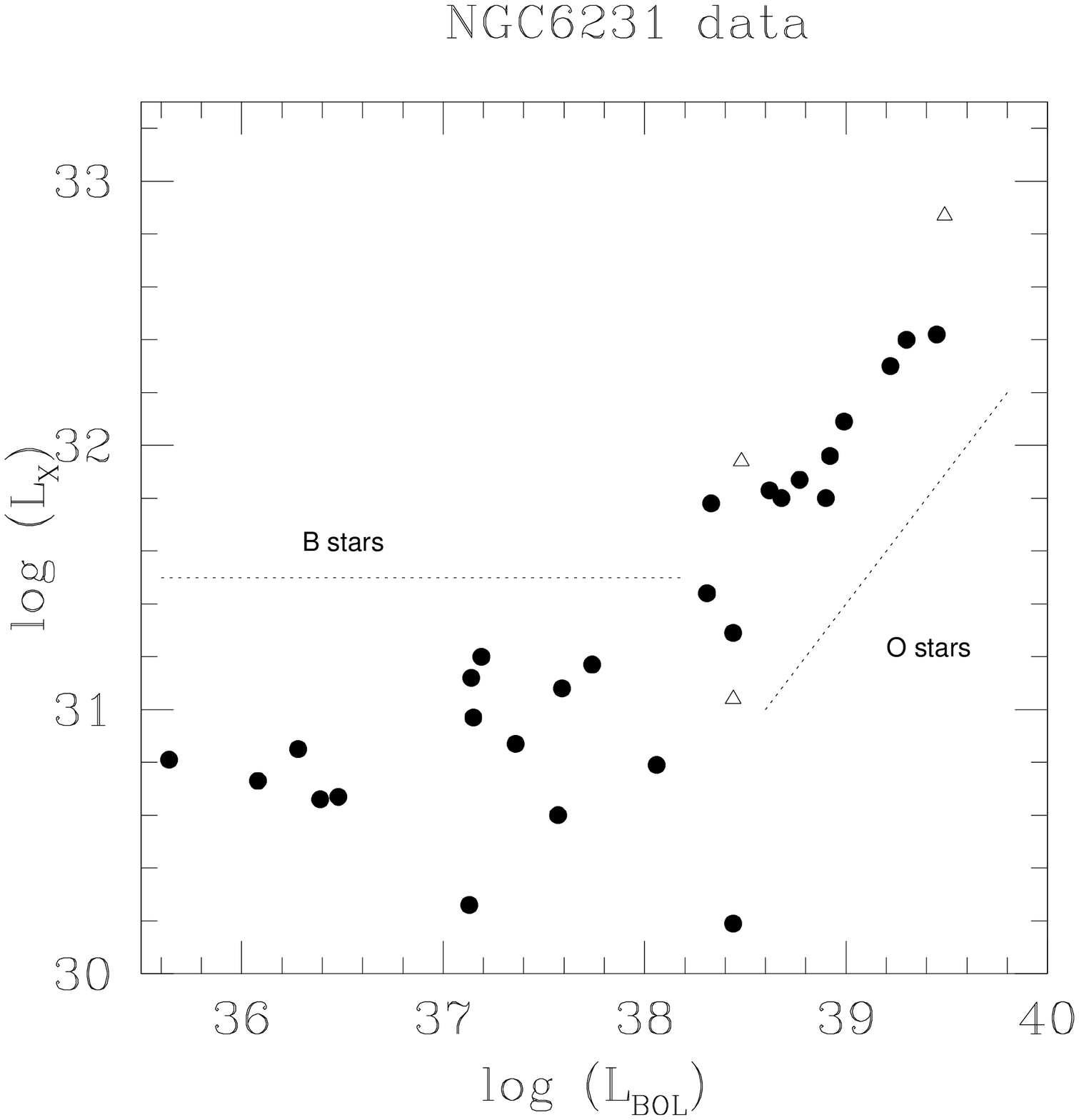}
  \includegraphics[width=4.0cm]{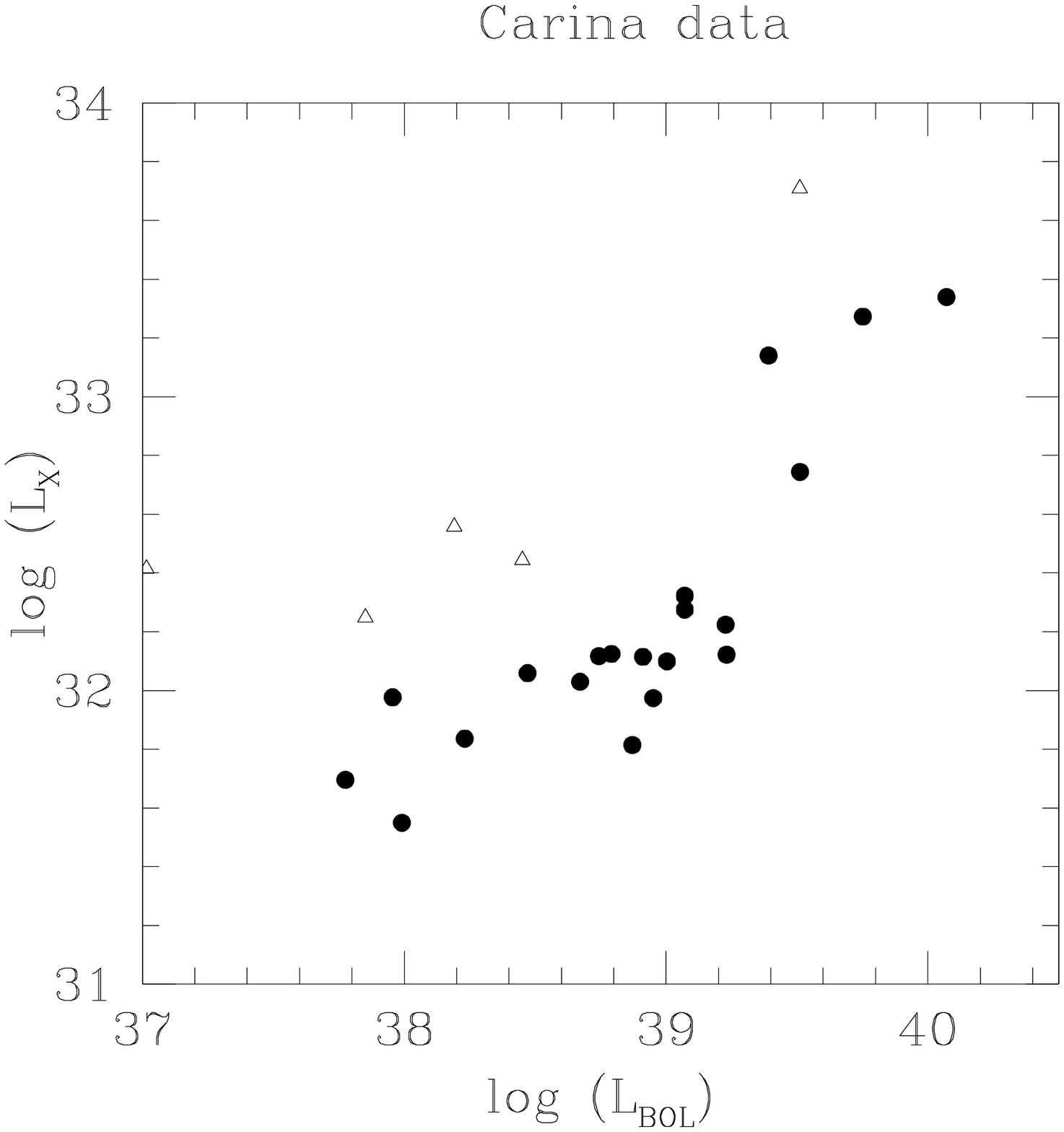}
  \caption{$L_{\rm X}-L_{\rm bol}$ relations for OB stars in RASS, for OB stars in NGC~6231 and for O stars in Carina OB1. Peculiar objects or problematic objects are shown as open symbols. (Figures based on data from \citealt{ber97,san06,ant08})}
  \label{fig:lxlbol}
  \end{figure}

With the advent of \xmm\ and \ch, a more detailed investigation of this so-called ``canonical'' relation became possible. Rather than conducting a survey, the new studies rely on detailed observations of rich open clusters and associations, notably NGC~6231 (\xmm, \citealt{san06}), Carina OB1 (\xmm, \citealt{ant08}), Westerlund 2 (\ch, \citealt{naz08a}), Cyg OB2 (\ch, \citealt{alb07}). The derived $L_{\rm X}-L_{\rm bol}$ relations are listed in Table \ref{tab:lxlbol}. For NGC~6231 and Carina OB1, the stellar content and reddening are well known, enabling a precise evaluation of both the bolometric luminosity and the reddening correction to be applied to the observed X-ray flux. In both cases, the $L_{\rm X}-L_{\rm bol}$ relation appears much tighter (dispersion of only 40\%) than in the RASS analysis. For Westerlund 2 and Cyg OB2, the optical properties are less constrained, naturally leading to a larger scatter. In the latter case, hints of a steeper relation were found for giant and supergiant stars compared to main-sequence objects \citep{alb07} but this needs to be confirmed.

The question arises about the actual origin of the difference between the RASS and \xmm\ results. Part of the answer lies in the differences between the datasets, and differences in data handling. On the one hand, the RASS provides a large sample of hot stars, covering all spectral types, but generally with only approximate knowledge of the optical properties (spectral types taken from general catalogs, bolometric luminosities taken from typical values for the considered spectral types). The X-ray luminosities were derived from converting count rates using a single absorbing column and a single temperature matching the observed hardness ratios best. On the other hand, the chosen clusters provide a smaller sample but have undergone an intensive optical spectroscopic monitoring, leading to a better knowledge of their stellar content. In addition, the X-ray luminosities were derived from fits to the medium-resolution spectra. However, it remains to be seen whether the distinct data analyses are sufficient to explain all the differences in the scatter. One additional distinction between the datasets concerns the  nature of the sample itself: clusters represent a homogeneous stellar population with respect to age, metallicity, and chemical composition, while the RASS analysis mixed stars from different clusters as well as field stars. The RASS scatter of the $L_{\rm X}-L_{\rm bol}$ relation could thus be partly real, and additional investigations, comparing more clusters, are needed. Notably, the study of high- or low-metallicity stellar populations would be of interest to better understand the X-ray emission as a whole. A first step in this direction was recently done by Chu et al. (in preparation). Their 300\,ks observation of the LMC giant H{\sc ii} region N11 revealed an $L_{\rm X}-L_{\rm bol}$ relation some 0.5\,dex higher than in NGC~6231. However, it is not yet clear if the observed OB stars are ``normal'', single objects or overluminous peculiar systems (see Sects. \ref{sec:cwb}, \ref{sec:magnetic}).

It might also be worth to note that the $L_{\rm X}-L_{\rm bol}$ relation was investigated in different energy bands \citep{san06,ant08}. The tight correlation exists in both soft (0.5--1\,keV) and medium (1--2.5\,keV) energy bands, but breaks down in the hard band (2.5--10\,keV). While the scatter due to a poorer statistics (lower number of counts) cannot be excluded, it should be emphasized that the hard band is also more sensitive to additional, peculiar phenomena whose emission is expected to be harder, such as non-thermal emission, emission from magnetically-channeled winds (see Sect. \ref{sec:magnetic}), etc. Indeed, even if such phenomena negligibly contribute to the overall level of X-ray emission, they could significantly affect the hard band. 

\begin{table}
\caption{ $L_{\rm X}-L_{\rm bol}$ relations for O-type stars from the literature
\label{tab:lxlbol}} 
\small
\begin{center} 
\begin{tabular}{l l l l}
\hline 
Region & Relation & Energy band & Reference\\
\hline
RASS & $\log L_{\rm X}$=$1.08 \log L_{\rm bol}$--11.89 & 0.1--2.4\,keV & \citet{ber97}\\
NGC~6231 & $\log L_{\rm X}$=$-6.912 \log L_{\rm bol}$ & 0.5--10.0\,keV & \citet{san06}\\
NGC~6231 & $\log L_{\rm X}$=$-7.011 \log L_{\rm bol}$ & 0.5--1.0\,keV & \citet{san06}\\
NGC~6231 & $\log L_{\rm X}$=$-7.578 \log L_{\rm bol}$ & 1.0--2.5\,keV & \citet{san06}\\
Cyg OB2 & $\log L_{\rm X}$$\sim-7 \log L_{\rm bol}$ & 0.5--8.0\,keV & \citet{alb07}\\
Carina OB1 & $\log L_{\rm X}$=$-6.58 \log L_{\rm bol}$ & 0.4--10.0\,keV & \citet{ant08}\\
Carina OB1 & $\log L_{\rm X}$=$-6.82 \log L_{\rm bol}$ & 0.4--1.0\,keV & \citet{ant08}\\
Carina OB1 & $\log L_{\rm X}$=$-7.18 \log L_{\rm bol}$ & 1.0--2.5\,keV & \citet{ant08}\\
\hline 
\end{tabular} 
\end{center} 
\end{table} 

Although a well established observational fact, the $L_{\rm X}-L_{\rm bol}$ relation is still not fully explained from the theoretical point-of-view. Qualitatively, it can of course be expected that the X-ray luminosity scales with the stellar wind properties (density, momentum or kinetic luminosity) if the X-ray emission arises in the wind (see Sect. \ref{sec:origin}). Since the wind is actually radiatively driven, these quantities should depend on the global level of light emission, i.e. on $L_{\rm bol}$. However, the exact form of the scaling relation is difficult to predict. \citet{owo99} made an attempt to explain the scaling law in the context of a distributed X-ray source. Considering that the volume filling factor varies as $r^s$, they found that $L_{\rm X}$ scales as $(\dot M/v_{\infty})^{1+s}$ if the onset radius of the X-ray emission is below the radius of optical depth unity (optically thick case), and $(\dot M/v_{\infty})^2$ otherwise. Adding the dependence of the wind parameters to the bolometric luminosity, one finally gets the observed scaling for $s$ values of $-$0.25 or $-$0.40 in the optically thick case. For less dense winds (optically thin case), $L_{\rm X}$ should be proportional to $L_{\rm bol}^{2.7\,{\rm or}\,3.5}$, i.e. the power law becomes steeper, as possibly observed for early B stars \citep[see however below]{owo99}. In this context, it may be interesting to note that recent high-resolution observations of $\zeta$\,Ori seem to favor constant filling factors (i.e. $s=0$ or $f\propto r^0$, \citealt{coh06}). 

Finally, recent studies have confirmed that the ``canonical'' $L_{\rm X}-L_{\rm bol}$ relation only applies for $L_{\rm bol}>10^{38}$ \ergs, i.e. down to B0--1 stars. It then breaks down at lower luminosities, with $L_{\rm X}/L_{\rm bol}$ ratios two orders of magnitude smaller at B3 than at B1 \citep{cas94, coh97}. Some authors therefore questioned whether B stars should emit at all in the X-ray domain. Indeed, these stars lack both putative origins of X-ray emission (see Sect. \ref{sec:origin}): strong stellar winds, except maybe for the earliest subtypes, and developed coronae (because of the absence of a subsurface convective zone). However, when detected, the X-ray emission from B stars appears to be related to their bolometric luminosity, although with a shallower scaling law (see Fig. \ref{fig:lxlbol}). This strongly suggests an intrinsic origin for the high-energy emission, and would request a revision of the current models. On the other hand, only a small fraction of these objects are detected: in the RASS, the detection fraction was less than 10\% for B2 and later objects \citep{ber97}; in NGC~6231, it is 24\% for B0--B4 stars but only 7\% for B5 and later objects \citep{san06}. Moreover, when detected, their spectra are often quite hard ($kT_2>$1\,keV) and many X-ray sources associated with B stars present flares: these properties are more typical of pre main sequence (PMS) objects, suggesting that the detected X-ray emission from B stars actually corresponds to that of a less massive, otherwise hidden PMS companion.

\subsubsection{A relation with spectral types?}
\label{sec:systematics}

High-resolution X-ray spectroscopy allowed \citet{wal08} to examine the metal lines in search of systematic effects related to the spectral classification, as are well known in the UV and optical domains. At first sight, the overall X-ray spectra seem to shift towards longer wavelengths for later types, and ratios of neighbouring lines from the same element but at different ionization stages (e.g. Ne\,{\sc x}/Ne\,{\sc ix}, Mg\,{\sc xii}/Mg\,{\sc xi}, Si\,{\sc xiv}/Si\,{\sc xiii}) are seen to evolve or even reverse. These trends clearly call for confirmation using a larger sample but, if real, they would have to be explained by theoretical models.

\subsection{Origin of the X-ray emission: insights from high-resolution spectroscopy}
\label{sec:origin}

Different origins for the high-energy emission of hot stars were proposed in the first decades after its discovery. Observational tests are indeed crucial to distinguish between the models. Not surprisingly, the strongest constraints were derived only recently with the advent of high-resolution spectroscopy. The new results lead to modifications of the original models, which also affect other wavelength domains.

\subsubsection{Proposed models and a priori predictions}
\label{sec:oldmodel}

The first proposed model to explain the X-ray emission from hot stars was a corona at the base of the wind, analogous to what exists in low-mass stars (see e.g. \citealt{cas79}). However, several observational objections against such models were soon raised. First, while neutron stars in close high-mass X-ray binaries showed the presence of significant attenuation by the stellar wind of their companion, no strong absorption was found for the intrinsic X-ray emission of massive stars \citep{cas83}. This suggests that the source of the X-ray emission lies significantly above the  photosphere, at several stellar radii. Second, the coronal [Fe\,{\sc xiv}]\,$\lambda$\,5303\AA\ line was never detected in hot stars spectra \citep{nor81,baa87}. Finally, the line profiles from ``superionized'' species were incompatible with coronal models, unless the mass-loss rate was drastically reduced \citep{macf93}.  

An alternative scenario relies on the intrinsic instability of the line-driving mechanism (responsible for propulsing the stellar wind). Indeed, the velocity in an unstable wind is not the same everywhere: fast-moving parcels of material will overcome the slow-moving material, generating shocks between them. This causes the formation of dense shells which should be distributed throughout the whole wind. At first, strong forward shocks between the fast cloudlets and the ambient, slower (``shadowed'') material were considered as the most probable cause of the X-ray emission \citep{luc80}. Subsequent  hydrodynamical simulations rather showed the presence of strong {\it reverse} shocks which decelerate the fast, low density material \citep{owo88}. However, the resulting X-ray emission from such rarefied material is expected to be quite low and cannot account for the level of X-ray flux observed in hot stars. Instead, \citet{fel97} proposed that mutual collisions of dense, shock-compressed shells could lead to substantial X-ray emission (simulations yield values only a factor of 2--3 below the observations). In such a model, the X-ray emission arises from a few (always $<$5, generally 1 or 2) shocks present at a few stellar radii ($<$10\,$R_*$). Since these shocks fade and grow on short time scales, such models predict significant short-term variability, which is not observed. To reconcile the observations with their model, \citet{fel97} further suggested wind fragmentation, so that individual X-ray fluctuations are smoothed out over the whole emitting volume, leading to a rather constant X-ray output. At the end of the 20th century, this wind-shock scenario was the most favored model amongst X-ray astronomers. 

  \begin{figure}
\begin{center}
  \includegraphics[width=9cm]{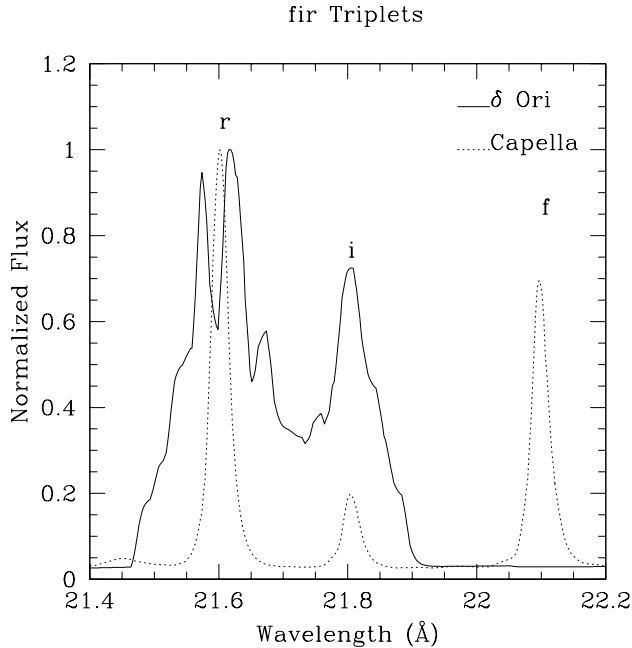}
  \caption{
Examples of observed He-like triplets (data taken from the Xatlas database and smoothed by a box of 4 pixels 
half width). Due to the presence of a strong UV flux, the forbidden line is nearly entirely suppressed for the hot star $\delta$\,Ori while it appears with full strength in the spectrum of the late-type star Capella.}
  \label{fircompa}
  \end{center}
  \end{figure}

To get a better insight into the properties of the X-ray emission region, high-resolution data were badly needed. Indeed, they provide crucial information. For example, the location of the X-ray emitting plasma can be derived from the analysis of line ratios. The most interesting lines to do so are triplets from He-like ions, which arise from transitions between the ground state and the first excited levels (see Fig. \ref{fir}). The line associated with the $^1P_1\rightarrow {^1S}_0$ transition is a resonance line ({\it r}), the one linked with the $^3S_1\rightarrow {^1S}_0$ transition is forbidden ({\it f}), and those associated with the $^3P_{1,2}\rightarrow {^1S}_0$ transition are called intercombination lines ({\it i}), hence the name {\it fir} triplets. It is important to note that the metastable $^3S_1$ level can also be depleted by transitions to the $^3P$ level. Such transitions are either collisionally or radiatively excited (see  Sects.  \ref{sec:densities} and \ref{sec:herbigs}). The {\it f/i} ratio (often noted $\cal R$) therefore decreases when the density $n_{\rm e}$ is high or the photoionization rate $\phi$ is significant: as already mentioned in Sect. \ref{sec:herbigs}, a useful approximation is 
\begin{equation}
{\cal R}=\frac{{\cal R}_0}{1+\phi/\phi_c+n_{\rm e}/N_c}
\end{equation}
where the critical values of the photoexcitation rate and the density, noted $\phi_c$ and $N_c$ respectively, are entirely determined by atomic parameters \citep[see also Table \ref{lineratios}]{blu72}. For hot stars, the density remains quite low, even close to the photosphere, and its impact is therefore often neglected\footnote{Only the Si\,{\sc xiii} triplet is marginally affected at very small radii ($<$0.05\,$R_*$) above the photosphere (see Fig. 2 in \citealt{wal01}).}; however, the UV flux responsible for the $^3S_1\rightarrow {^3P}_{1,2}$ transition is not negligible in these stars (see Fig. \ref{fircompa}). The UV radiation is diluted, as the distance from the star increases, by a factor 
\begin{equation}
w(r)=0.5\times\left(1- \sqrt{1-(R_*/r)^2}\right).
\end{equation}
Hence, its influence on the {\it f/i} ratio decreases with radius\footnote{It should be noted that the reasoning above applies only to single objects. For some hot binaries, strong forbidden lines have been observed, while intercombination lines remain negligible \citep{pol05}. This is sometimes taken as a typical signature of X-rays arising in wind-wind interactions (see also Sect. \ref{sec:cwb}). The proposed explanation is twofold: first, the wind-wind interaction occurs far from the stellar surfaces, thereby decreasing the impact of the UV flux; second, the forbidden line could be boosted by inner-shell ionization of Li-like ions.}. For hot stars, the observation of a given {\it f/i} ratio will thus yield an estimate of the dilution factor and thereby of the position in the wind of the X-ray emitting plasma. \citet{leu06} have further shown that the {\it f/i} ratios could also be interpreted in the framework of a wind-shock model (i.e. with X-ray emission regions distributed throughout the wind). In this case, the integrated {\it f/i} ratio increases faster with radius than in the localized case. Furthermore, they predict slight differences in the profiles of the {\it fir} lines: the {\it i} line should be stronger where the UV flux is maximum, i.e. close to the star, and it should thus have weaker wings than the {\it f} line, which is produced further away and should appear more flat-topped. Those subtle differences might however not be easily detectable with current instrumentation. Finally, the {\it (f+i)/r} ratio (often noted $\cal G$) can also pinpoint the nature of the X-ray plasma. For $\cal G$ values close to 4, the plasma is dominated by photoionization; if close to 1, the collisions are preponderant. In addition, the {\it (f+i)/r} ratio is a sensitive indicator of the temperature, as is the ratio of H-like to He-like lines, although at very high densities the {\it r} line is affected by the collisional transition of $^1S_1\rightarrow {^1P}_1$. To help in the interpretation of high-resolution spectra, \citet{porquet01} performed extensive calculations of $\cal R$ and $\cal G$ values for a large range of densities, plasma temperatures and stellar temperatures in the case of the most abundant He-like ions (C, N, O, Ne, Mg, Si). These results, available at CDS, do not rely on the approximate formula quoted above for the {\it f/i} ratio but take into account all atomic processes, notably the blending of {\it fir} lines with dielectronic satellite lines (whose exact contribution depends on the assumed temperature and spectral resolution). On the other hand, they were calculated assuming the stellar flux to be a blackbody, whereas most other authors rather use the approximate formula but rely on detailed atmosphere models (such as Kurucz or TLUSTY). 

\begin{figure}
  \begin{center}
  \includegraphics[width=9.cm]{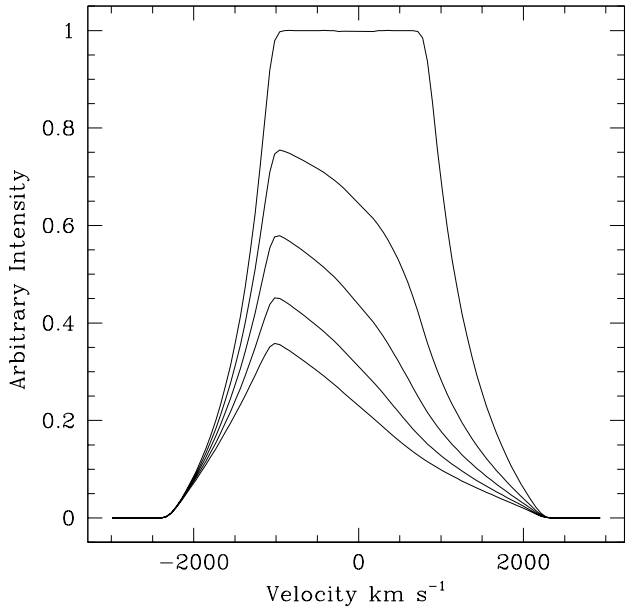}
  \caption{Theoretical X-ray line profiles expected for an exospheric model with $v_{\infty}=2300$\,\kms, $q=0$, $\beta=1$, $R_0=1.8\,R_*$ and increasing opacities $\tau_*$ (from 0 to 2 in steps of 0.5). 
  (Figure courtesy of G. Rauw).}
  \label{lineprof}
  \end{center}
  \end{figure}

To go further in the interpretation of the new high-resolution data, \citet[see also the precursor works of \citealt{macf91}]{owo01} presented detailed calculations of the X-ray line profiles expected for hot stars, when assuming that the X-ray emitting material follows the bulk motion of the wind. Such so-called exospheric models depend on four parameters. First, the exponent $\beta$ of the velocity law 
\begin{equation}
v(r)=v_{\infty} \left[ 1-\frac{R_*}{r} \right]^{\beta}.
\end{equation}
Second, the exponent $q$ involved in the radial dependence of the volume filling factor ($f[r]\propto r^{-q}$), which reflects variations with radius of the temperature or of the emission measure of the X-ray emitting plasma. Third, a typical optical depth $\tau_*$, which depends on the wind properties: 
\begin{equation}
\tau_*=\frac{\kappa \dot M}{4 \pi v_{\infty}R_*}.
\end{equation}
Finally, the onset radius $R_0$, below which there is no production of X-rays. The last two parameters were found to be the most critical ones: the blueshifts and line widths of the line profiles clearly increase with increasing $\tau_*$ and/or $R_0$, while the profiles are much less sensitive to changing values of $q$ or $\beta$. In this framework, a coronal model can be reproduced by setting $R_0=R_*$ and choosing high values of $q$, while a wind-shock model rather requires lower values of $q$ and larger values of $R_0$. The former always produces narrow, symmetric profiles, with little or no blueshift; the latter generates broad lines (generally FWHM$\sim v_{\infty}$) whose exact shape depends on the opacity. In the optically thin case ($\tau_*\sim0$), the line profile expected for a wind-shock model is flat-topped, unshifted and symmetric. For more absorbed winds, the redshifted part of the line, arising from the back side of the star's atmosphere, suffers more absorption and is therefore suppressed: the line profile appears blueshifted and skewed (see Fig. \ref{lineprof}). In addition, there should be some wavelength dependence of the line profiles since the absorption is strongly wavelength-dependent: the high-energy lines should appear narrower (lower values of $\tau_*$ and $R_0$), although this effect is often mitigated by the larger instrumental broadening at these energies. 

  \begin{figure}
  \begin{center}
  \includegraphics[width=12.5cm]{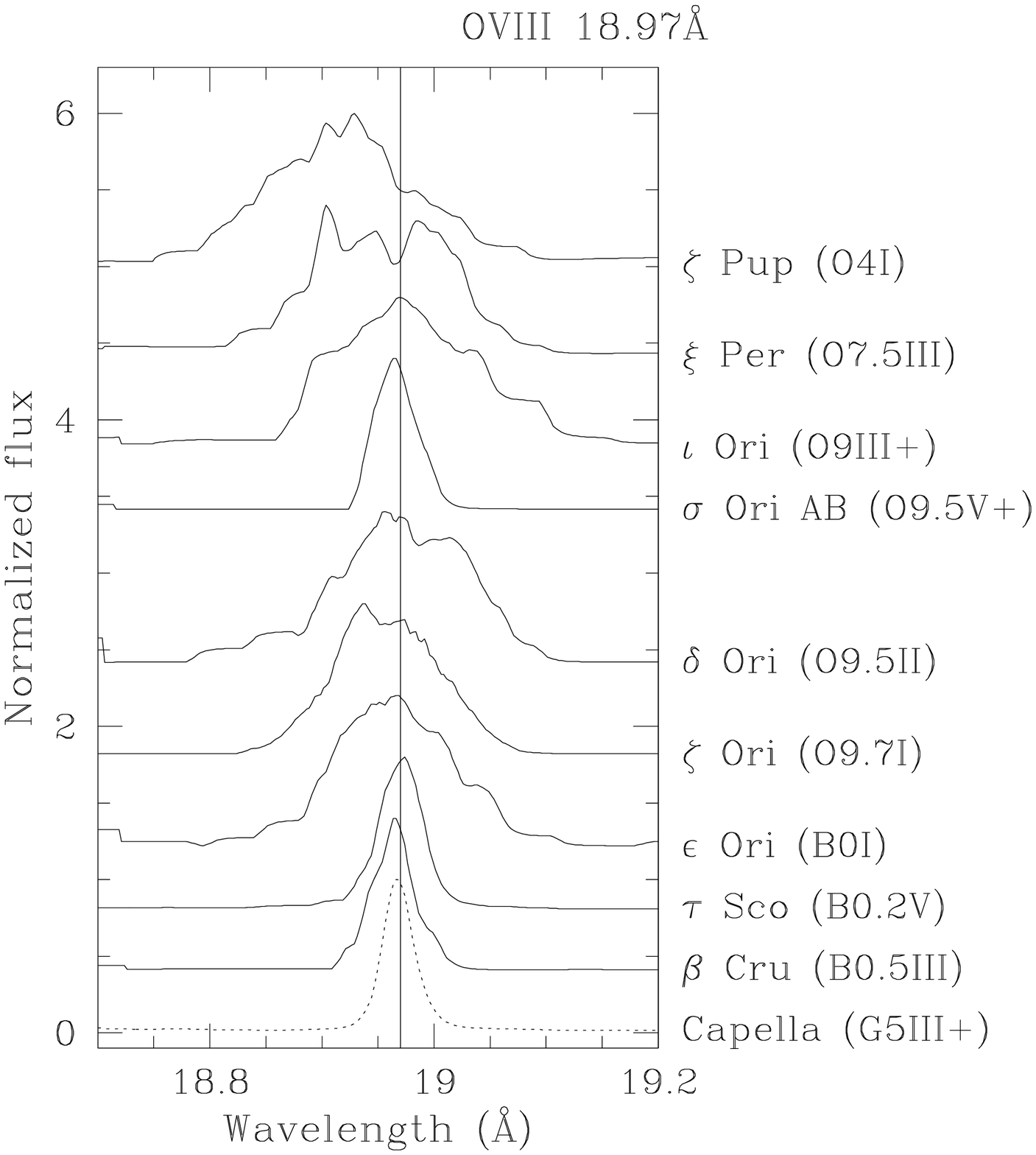}
  \caption{Observed line profiles for O\,{\sc viii}\,$\lambda$18.97\AA\ in a sample of bright OB objects and in the coronally active late-type star Capella (where the line width is mostly instrumental). The vertical line corresponds to the rest wavelength of the line; the normalized profiles have been shifted upwards by an arbitrary amount for clarity (data taken from the Xatlas database and smoothed by a box of 4 pixels 
half width).}
  \label{obsprof}
  \end{center}
  \end{figure}

\subsubsection{Results from high-resolution spectra}
\label{sec:highres}

Before the advent of \xmm\ and \ch, only low or medium-resolution spectra were available to X-ray astronomers. These data led to the determination of four main observables: the temperature(s) of the plasma, the degree of absorption, the temporal variability, and the overall flux level. The current X-ray facilities have indeed given some better insights into these properties (see Sect. \ref{sec:global}), but they have also provided high-resolution spectra which permit to study the line ratios and the detailed shape of the line profiles for the first time. Such observations (see example in Fig. \ref{obsprof}) have resulted in new, strong constraints on the models. 

One of the first hot stars observed at high-resolution was $\zeta$\,Pup (O4Ief), an early supergiant which possesses a strong wind. \citet{kah01} and \citet{cas01} report on \xmm\ and \ch\ observations, respectively. They found that the line profiles are blueward skewed and broad (HWHM$\sim$1000\,\kms), as expected for a wind-shock model. By estimating the emission measures (EMs) for each line \newline(EM $=4 \pi d^2 F_{\rm line}/{\rm Emissivity}(T_e)$, with $d$ being the distance and $F_{\rm line}$ the line flux), \citet{kah01} further showed that $\zeta$\,Pup displays a large overabundance in nitrogen and that only a small fraction of the wind emits X-rays (since the measured EMs are much smaller than the EM available from the whole wind). The {\it fir} triplets indicate line formation regions at a few stellar radii, compatible with the radii of optical depth unity at the wavelengths considered. These radii appear smaller for high nuclear charge ions (such as Si and S) than for lighter elements (such as O, Ne, Mg, \citealt{cas01}). Finally, the N\,{\sc vii}\,Ly$\alpha$\,24.79\,\AA\ line seems broader and either flat-topped \citep{cas01} or structured \citep{kah01}, suggesting a formation in more remote regions of the wind. However, this feature could rather be explained by a blend of this line with N\,{\sc vi}\,He$\beta$\,24.90\,\AA\ \citep{pol07}. Overall, although a qualitative agreement with the wind-shock model was readily found, a more detailed, quantitative comparison awaited the work of \citet{kra03}. These authors found that profiles of the kind derived by \citet{owo01} yield rather good fits, but they discovered two intriguing problems. First, the good fits by a spherically symmetric wind model were found for a wind attenuation about five times smaller than the one expected from the known mass-loss rate ($\tau^{\rm exp}_*=4-30$ vs. $\tau^{\rm obs}_*=0-5$, see also \citealt{osk06}). Second, the derived $\tau_*$ appeared rather similar for all lines, i.e. it seemed independent of wavelength \citep{kra03}, although the error bars were rather large. \citet{coh09} recently re-investigated the \ch\ spectrum of $\zeta$\,Pup, by considering more lines and by using the HEG data in addition to the MEG data. They now found a significant variation of the optical depth with wavelength, compatible with the atomic opacity expected for a mass-loss rate of $\sim$3$\times10^{-6}$\,M$_{\odot}$\,yr$^{-1}$ (about three times lower than in the literature).

The situation at first appeared quite different for the O9.7Ib star $\zeta$\,Ori\,A. In \citet{wal01}, the line profiles were reported as broad (HWHM $=850\pm40$~\kms) but symmetric, unshifted and Gaussian rather than flat-topped. Again, as for $\zeta$\,Pup, emission from high-energy ions seems to be produced closer to the stellar surface than for low-energy ions (at 1.2$\pm$0.5\,$R_*$ for Si\,{\sc xiii}, but at a few stellar radii for other elements). These results were confirmed by \citet{raa08}, but challenged by \citet{leu06} and \citet{coh06} - the latter two using the same \ch\ data as \citet{wal01} - as well as by \citet{pol07}, who relies on the same \xmm\ data as \citet{raa08}. Although they used different techniques, both \citet{coh06} and \citet{pol07} detected slight asymmetries and blueshifts (up to $-$300\,\kms) in the line profiles of $\zeta$\,Ori\,A. 
\citet{coh06} also showed that wind profiles with $\tau_*=0.25-0.5$, $R_0\sim 1.5\,R_*$ and $q\sim 0$ yield significantly better fits than simple Gaussians: some wind attenuation is thus needed. However, as first found for $\zeta$\,Pup, the observed mean opacity is smaller than expected \citep{coh06} and no clear trend of the opacity with wavelength was detected. \citet{coh06} argue that, taking into account a reduced mass-loss rate and the presence of numerous ionization edges, the opacity variations over the considered wavelength range might be smaller than first thought albeit compatible with the data, within uncertainties. As was done for $\zeta$\,Pup, Wollman et al. (2009, in preparation) re-analyzed the \ch\ data of $\zeta$\,Ori: a significant, although small, trend in the optical depths is now detected, which is compatible with the wind opacity if the mass-loss rate from literature is decreased by a factor of 10.

The late-O stars $\delta$\,Ori\,A (O9.5II+B0.5III) and $\zeta$\,Oph (O9.5Ve) both display symmetric and broad X-ray lines, although they are much narrower than those of $\zeta$\,Pup and $\zeta$\,Ori\,A (HWHM $= 430\pm 60$\,\kms, \citealt{mil02}, and HWHM $=400\pm 87$\,\kms, \citealt{wal05}, respectively). On the basis of its mass-loss rate, $\delta$\,Ori\,A should be an intermediate case between $\zeta$\,Pup and $\zeta$\,Ori\,A: narrower profiles are therefore quite surprising. In addition, the lines of $\delta$\,Ori\,A are unshifted (centroid at 0$\pm$50\,\kms) and the {\it fir} ratios indicate a formation region one stellar radius above the photosphere (which is more in agreement with a wind-shock model than a coronal model), at a position similar to that of optical depth unity. However, $\delta$\,Ori\,A is a binary and one may wonder whether peculiar effects such as interacting winds could have an impact on its X-ray emission. This does not seem to be the case since no phase-locked variations of the X-ray flux were ever reported for this object \citep{mil02} - it must be noted, though, that the changes might be quite subtle for such late-type O+B systems, see e.g. the case of CPD$-$41$^{\circ}$7742 in Sect. \ref{sec:cwb}. In contrast, the lines of $\zeta$\,Oph appear slightly blueshifted, their width might decrease for low-energy lines, and the formation radii derived from the {\it fir} ratios differ from the radius of optical depth unity. This star is a known variable, in the optical as well as in X-rays \citep[see e.g.][]{osk01}, and \citet{wal05} reports a change in the formation radius derived from the Mg\,{\sc xi} triplet between two \ch\ observations separated by two days. 

The narrowest profiles amongst O-type stars were found for $\sigma$\,Ori\,AB (O9.5V+B0.5V, \citealt{ski08} and Fig. \ref{obsprof}): HWHM=264$\pm$80\,\kms, or 20\% of the terminal velocity. For this star, the X-ray lines again show no significant shift and the most constraining {\it fir} ratios point to a formation region below 1.8\,$R_*$. 

Since more observations are now available, some authors have undertaken a consistent, homogeneous analysis of several objects. This indeed eases the comparison between different objects but, as already seen above, the results sometimes disagree between different teams. Using their ``distributed'' {\it fir} model, \citet{leu06} found onset radii of 1.25--1.67\,$R_*$ for $\zeta$\,Pup, $\zeta$\,Ori, $\iota$\,Ori, and $\delta$\,Ori\,A, without any large differences for high-mass species such as Si\,{\sc xiii}. Such a radius rules out a coronal model but can still be reconciled with a wind-shock model. \citet{leu06} further explain their differences with \citet{wal01} by (1) calibration problems at the beginning of the \ch\ mission, and (2) different atmosphere models (TLUSTY vs Kurucz) which give fluxes different by a factor of 2--3 shortward of the Lyman edge, at wavelengths responsible for exciting the Si\,{\sc xiii} transitions.

\citet[see errata in \citealt{wald08}]{wal07} examined the high-resolution spectra of 17 OB stars. The sample contains 2 known magnetic objects (see Sect. \ref{sec:magnetic}) and 15 stars considered as ``normal'' by the authors - however, we note that 3 of these, 9\,Sgr and the stars in Cyg\,OB2, are well known binaries displaying wind-wind interactions detected at X-ray energies \citep{rau02,rau05,deb06}; this likely explains the peculiar, deviant results sometimes found by the authors for these stars. Overall, when the lines are analyzed by simple Gaussian profiles, 80\% of their peaks appear within $\pm$250\,\kms\ of the rest wavelength. The distribution of the line positions peaks at 0\,\kms, but is heavily skewed (many more blueshifts than redshifts) and this asymmetry increases with the star's luminosity. The line HWHMs are distributed between 0 and 1800\,\kms, i.e. they are always less than the terminal velocity and most of the time they are less than half that value; there is a slight trend towards narrower lines (expressed as a fraction of the terminal velocity) for main-sequence objects. These widths appear also larger than the velocity expected at the position derived from the {\it fir} triplets (using a localized approach and Kurucz atmosphere models). Both width and shift seem to be independent of wavelength (see however the results of new analyses mentioned above). Finally, the formation radii found from triplets of heavier ions are smaller than those of lighter ions and the lower temperatures, derived from the ratios of lines from H and He-like ions, are only found far from the photosphere.

Despite some discrepancies in the interpretations, as shown above, several conclusions are now emerging from the analysis of high-resolution spectra:
\begin{enumerate}
\item Only a small fraction of the wind emits, as expected. 
\item The X-ray lines are broad, although not as broad as predicted (HWHM $= 0.2-0.8\,v_{\infty}$). 
\item The profiles are more symmetric than expected, without evidence for flat-topped shapes. 
\item Except for $\zeta$\,Pup, the blueshifts are small or non-existent. 
\item The opacities derived from fits to the profiles are lower than expected.
\item For a given star, the line profiles and fitted opacities are generally quite similar regardless of the wavelength, although error bars are large and shallow trends cannot be excluded (indeed, some were found in a few cases).
\item From the analysis of the {\it fir} ratios, the (minimum) radius of the X-ray emitting region appears rather close to the photosphere (often $<$2\,$R_*$ for the most stringent constraints), although there is disagreement about the possibility of a formation at or very close ($r<$1.1\,$R_*$) to the photosphere. The formation radius is generally compatible with the radius of optical depth unity, which can be easily understood: the emissivity scales with the square of the density, therefore the flux will always be dominated by the densest regions that are observable; as the density decreases with radius, these regions are the closest to the star, i.e. they are just above the ``X-ray photosphere'' (radius of optical depth unity).
\end{enumerate}

\subsubsection{A new paradigm?}
\label{sec:solution}

The characteristics of the X-ray lines enumerated above were not fully in agreement with the expectations of the wind-shock model. Therefore, some adjustments of the models have been proposed. They fall in five broad classes: resonance scattering, decrease of mass-loss rate, porosity, combination corona+wind, and a complete change of concept.

In most cases, X-ray plasmas associated with hot stars are considered to be optically thin, hence the use of optically-thin thermal plasma models. \citet{ign02} rather suggested that resonant scattering could play a significant role, especially for the strongest lines. In fact, their detailed line profile modeling showed that an optically-thick line appears much more symmetric and less blueshifted than in the optically-thin case. This arises from the fact that the Sobolev line interaction region is radially elongated and that X-rays can thus escape more easily laterally than radially, reducing the strength of the blue/red wings. Their calculations of line shift and width agreed well with those of $\zeta$\,Pup, but the line profiles of $\zeta$\,Ori and $\theta^1$\,Ori\,C were more difficult to reproduce. An evaluation of the impact of resonance scattering for Fe\,{\sc xvii} in $\delta$\,Ori\,A showed that some weak opacity effect might affect the strongest X-ray lines but the optically-thin case was still well within the error bars \citep{mil02}. \citet{leu07} incorporated resonance scattering into the models of \citet{owo01}. They showed that for $\zeta$\,Pup, including resonance scattering provides a better fit, but that some discrepancies remain. In addition, they showed that resonance scattering only has a significant effect for large filling factors ($\gg 10^{-3}$), which could be a problem for winds of massive stars (see below). Finally, it should be noted that in the above calculations including resonance scattering, \citet{leu07} were able to fit the line profiles without any large reduction of the mass-loss rate.

\begin{figure}
  \begin{center}
  \includegraphics[width=9cm,angle=90]{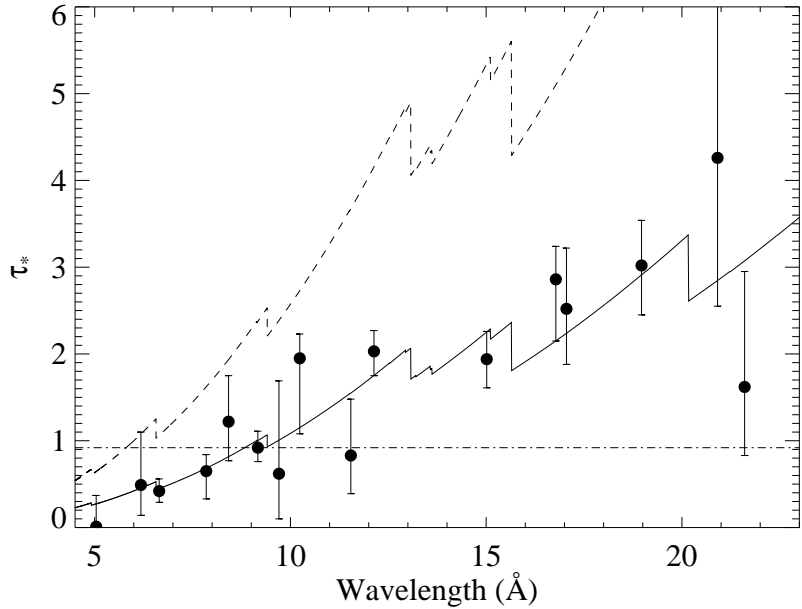}
  \caption{Variation with wavelength of the opacity $\tau_*$ derived from line-profile models of \citet{owo01} for the star $\zeta$\,Pup. The observations correspond to the dots, the dash-dot line represents a constant opacity, the dashed line the opacity variation expected from the literature mass-loss rate and the solid line the opacity variation for a reduced mass-loss rate. (Figure courtesy of D. Cohen)}
  \label{zpuptau}
  \end{center}
  \end{figure}

Reducing the mass-loss rates can naturally solve many of the discrepancies mentioned above since it leads to more symmetric and less blueshifted line profiles (see \citealt{owo01}) as well as reduced and less wavelength-dependent opacities. \citet{coh06} strongly advocate in favor of that solution following their analysis of $\zeta$\,Ori. The trends detected by \citet{coh09} and Wollman et al. (2009, in preparation) again favor this scenario. Such a reduction of the mass-loss rate was also envisaged from results at other wavelengths (UV, optical), some authors even proposing a decrease by one or two orders of magnitude \citep[e.g. ][see however remarks in \citealt{osk07}]{ful06}. However, such a drastic reduction of the mass-loss rates might be in conflict with the absorption values measured for high-mass X-ray binaries. Notably, to explain the  X-ray observations, \citet{wat06} need a mass-loss rate of 1.5--2.0$\times$10$^{-6}$\msol\,yr$^{-1}$ for the B0.5Ib primary of Vela\,X-1. More reasonable decreases by a factor of a few (2--10) are now commonly envisaged, especially in the context of (micro)clumping (see below and the outcome of the 2007 Potsdam meeting), and they are compatible with fits to the X-ray lines - at least for $\zeta$\,Pup \citep[ and Fig. \ref{zpuptau}]{coh09} and $\zeta$\,Ori (Wollman et al. 2009, in preparation).

\begin{figure}
\hspace{-0cm}
  \includegraphics[width=6.3cm]{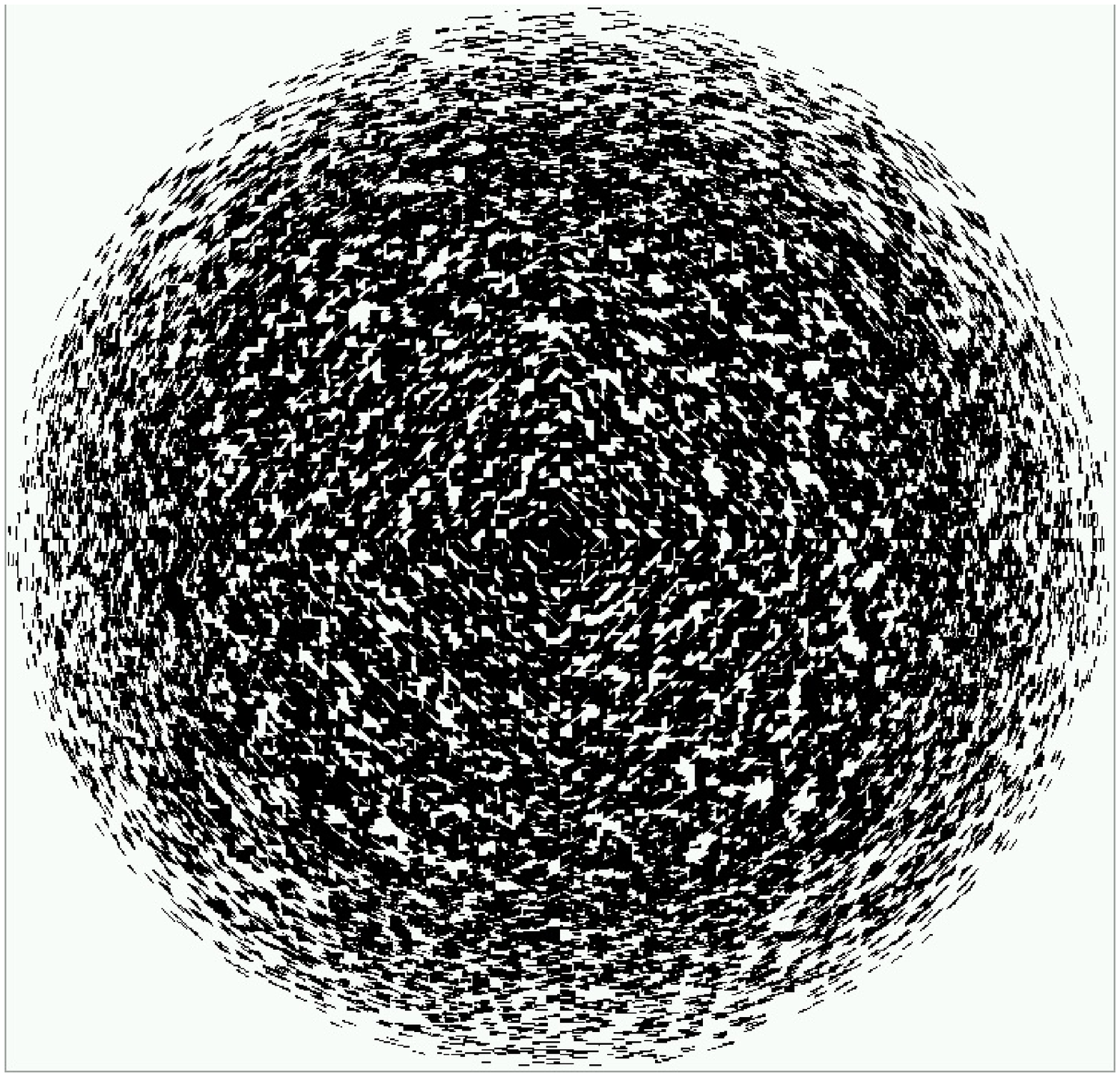}
  \includegraphics[width=6.3cm]{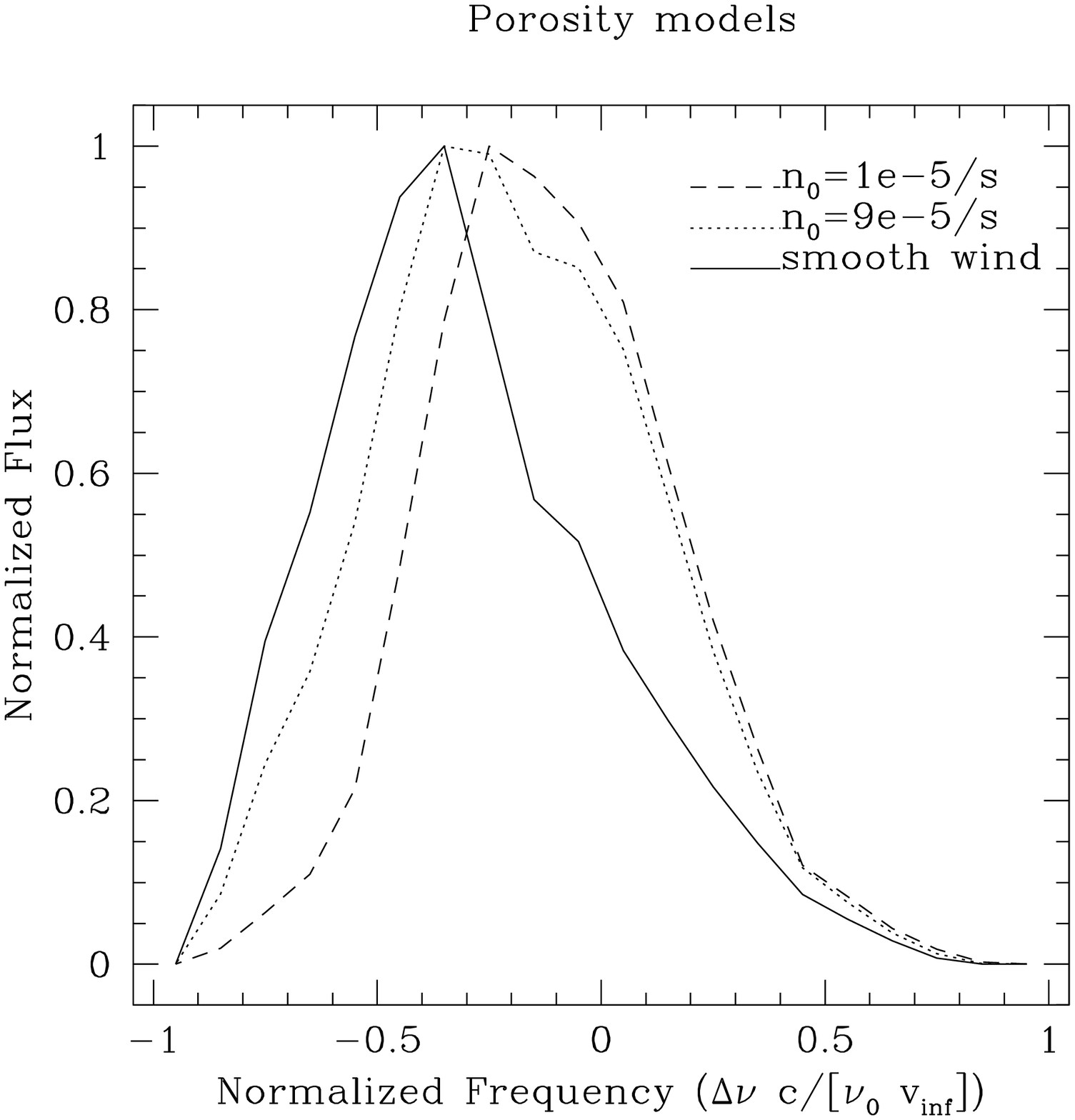}
  \caption{Left: Sketch of a porous wind (see \citealt{fel03}). Right: X-ray lines expected in a homogeneous wind and winds with optically-thick clumps (the fragmentation frequency $n_0$ represents the number of clumps passing through some reference radius per unit time - data courtesy L. Oskinova).}
  \label{porosity}
  \end{figure}

Because of the unstable line-driving mechanism propelling the wind, it is expected that the wind is not smooth. Indeed, evidence for clumping has been found observationally: stochastic variability of optical lines associated with the wind (e.g. \citealt{eve98,mar05}), model atmosphere fits of UV lines \citep{bou05}, temporal and/or spectral properties of some high-mass X-ray binaries \citep{sak03,van05}. A wind composed of dense clumps in a rarefied gas has two interesting consequences. First, a reduction of the mass-loss rate by a factor $\sqrt{f}$, where $f$ is the volume filling factor of these clumps. This reduction amounts to less than 10 and is thus not as drastic as those mentioned above. Second, the possibility of leakage, resulting in reduced opacity and more symmetric profiles, in the case of optically-thick clumps (Fig. \ref{porosity}). In this case, X-rays can avoid absorption by the dense clumps if the radiation passes in the (nearly) empty space between them. The wind is then said to be porous and the opacity is no more atomic in nature, i.e. determined by the plasma physical properties, but is rather {\it geometric}, i.e. it depends only on the clump size and interclump distance. \citet{fel03} and \citet{osk06} have envisaged the consequences of such a structured wind. Their model uses a wind composed of hot parcels emitting X-rays and cool dense fragments compressed radially and responsible for the absorption. The line profile is severely affected, being less broad and less blueshifted as well as more symmetric than in the homogeneous wind case of equivalent mass-loss rate (Fig. \ref{porosity}). For optically thick clumps, the resulting opacity is effectively grey, and the line profile should thus be independent of wavelength. As the clumps become more and more optically thin, the differences with a homogeneous wind decrease and finally disappear. For $\zeta$\,Pup, \citet{osk06} showed that using the strongly reduced mass-loss rate suggested by \citet{ful06} causes the blueshift to be too small compared to observations. However, a porous wind with only moderate reduction of the mass-loss provides good fits. Porosity can thus greatly help in reducing the opacity and getting wavelength-independent properties without a drastic reduction of the mass-loss rate which could potentially be problematic. However, the origin and consequence of porosity have been questioned by \citet{owo06}. These authors showed that, for a clumped wind, the opacity depends on the so-called porosity length $l/f$ where $l$ is the size of the clumps and $f$ the volume filling factor. For optically-thick clumps, the most favorable case, getting symmetric lines requests $l/f \gtrsim r$: either the clumps have large scales, or the filling factor is small (hence the clump compression is high), or a combination of both is needed. However, hydrodynamic simulations show that the line-driven instability leads to small-scale (typically $l\sim0.01$\,$R_*$) and moderately compressed ($f\sim 0.1$) clumps. A porosity in agreement with the observations would thus have another origin than the intrinsic instabilities of the wind. In addition, large clumps separated by large distances might lead to significant temporal variability in the X-ray light curve, which is not detected. Finally, the grey opacity contradicts the shallow but non-zero wavelength-dependence observed at least in $\zeta$\,Pup and $\zeta$\,Ori (\citealt{coh09} and Wollman et al. 2009, in preparation). Therefore, the presence of large, optically-thick clumps can most probably be discarded. However, we cannot rule out that smaller, less opaque clumps exist in the stellar winds.

\citet{wal07} and \citet{wal09} rather advocated for a partial return of the magnetic corona hypothesis. More precisely, they assumed a wind-shock model in the outer wind regions, where lines associated with low-mass ions arise, combined with a coronal model close to the star, which would produce the lines from the highly-ionized high-mass species. As support for their combined idea, \citet{wal07} emphasized three annoying facts. First, the formation radii derived from {\it fir} triplets agree quite well with radii of optical depth unity, as could have been expected. If a decrease of the mass-loss rate is considered, then one would have to explain why the formation region lies much above the X-ray photosphere. Second, the radii derived from {\it fir} triplets indicate, at least for Si\,{\sc xiii}, formation regions very close to the photosphere. In some cases, the small estimated radii might even disagree with the typical onset radius, 0.5\,$R_*$ above photosphere, of the wind-shock model. Third, the high-mass ions systematically display smaller formation radii than their lower-mass counterparts. At these small radii, the wind velocity is too low to get the observed broad lines and high post-shock temperatures. \citet{wal09} proposed a theoretical model where plasmoids produced by magnetic reconnection events are rapidly accelerated; they emphasized that these plasmoids are not clumps but ``isolated magnetic rarefactions". However, as mentioned above, \citet{leu06}, \citet{coh06} and \citet{osk06} have challenged their radius calculations. Another caveat can be noted: \citet{wal07} rely on the wind opacity calculated as in \citet{wal84}; in that model, the wind is supposed to have a temperature of 0.8$\times T_{\rm eff}$ at any position in the wind, which leads to a high ionization throughout the wind, thereby strongly reducing the opacity at low energies. However, such high temperatures at any radii might not be entirely physical (see e.g. \citealt{macf93}). In a different context, we will come back to the impact of magnetic fields on stellar winds in Sect. \ref{sec:magnetic}.

A more radical shift in thought was proposed by \citet{pol07}. Up to now, it was considered that particle interactions in winds take place through long-range Coulomb interactions. This explained the redistribution of momentum from a minority of UV-driven ions to the rest of the flow. However, it seems that the mean free path for ion-ion Coulomb collisions is not small \citep{pol07}: for $\zeta$\,Ori, it is 0.1\,$R_*$ at 3\,$R_*$ above the photosphere and 1\,$R_*$ at 10\,$R_*$. Therefore, \citet{pol07} proposes the shocks to be collisionless, like in SNRs. Any dissipation of energy would occur through other phenomena, probably linked to magnetic fields. In addition, the time scale for post-shock equilibrium appears very long (corresponding to a flow distance of a few stellar radii for electrons), thereby suggesting that no full equilibrium is ever established. In this model, the X-ray lines would arise far from the star, through ion-ion interactions such as ionization and charge exchange. \citet{pol07} claims that this would explain the faint continuum emission  at X-ray wavelengths (the electrons being too cold for a significant bremsstrahlung in this domain) and the observed line profiles. As the author mentions in his paper, a more quantitative assessment of this idea must await a detailed modeling. It is however worth noting that collisionless shocks and non-equilibrium phenomena were also proposed, although in a different context, by \citet[see below]{zhe07b}.

At present, it is still difficult to assert which solution applies best, or which combination of the above is closer to reality. Certainly, this will be a task for the next generation of X-ray facilities.

\subsection{The case of early B-stars}
\label{sec:bstars}

The previous section focused on results derived from observations of O-type stars. However, a sample of early B-type stars was also observed at high resolution with the current facilities. In their case too, surprises are common and the overall picture appears quite contrasted.

While the X-ray lines of the O-type stars often appear broad, especially for luminous objects with strong winds, the X-ray lines of B-type stars were found to be particularly narrow, except for the B0I supergiant $\epsilon$\,Ori (see Fig. \ref{obsprof} and Table \ref{tab:Bstars}).
These narrow lines were reported to be symmetric and unshifted, with the exception of a possible asymmetry in $ \beta$\,Cen \citep{raa05}. The {\it fir} triplets usually favored formation radii at distances of 1.5--5\,$R_*$ from the star.

The hardness of the spectrum and the overall emission level were found to vary from star to star (see Table \ref{tab:Bstars}). On the one hand, $\tau$\,Sco emits hard X-rays 
and displays a clear overluminosity \citep{mew03,coh03}. 
This seems also to be the case of $\theta^1$\,Ori\,A and E \citep{sch03}.
Note however that the presence of a low-mass companion to $\theta^1$\,Ori\,E \citep{her06} might lead to a revision of the above results for this star.

On the other hand, the X-ray emission from the B-type stars $\epsilon$\,Ori, $\theta$\,Car, $\beta$\,Cru A, Spica, $\beta$\,Cen, and $\beta$\,Cep appears much softer. For the first four stars, the distributions of the differential emission measure (DEM) as a function of temperature display peaks centered on 0.2--0.3\,keV, with FWHMs of 0.1--0.3\,keV, and without hard tails \citep{zhe07,naz08c}. Hottest thermal components (0.6\,keV) were detected for $\beta$\,Cen, $\beta$\,Cep, and $\beta$\,Cru A but they are clearly not dominant (the main component is at about 0.2\,keV for these objects, \citealt{raa05,fav08,coh08}). The X-ray emission levels are also much more modest, with $L_{\rm X}-L_{\rm bol}$ ratios of about $-7$~dex (see Table \ref{tab:Bstars}). 

\begin{table}
\caption{Observed properties of the high-resolution X-ray spectra of early B-type stars. An asterisk indicates unresolved lines.}
\label{tab:Bstars}       
\begin{tabular}{llcccl}
\hline\noalign{\smallskip}
Name & Sp. Type & HWHM & $kT$ & $\log(L_{\rm X}$ & References \\
 &  &  (\kms)&  (keV) & $/L_{\rm bol})$&  \\
\noalign{\smallskip}\hline\noalign{\smallskip}
$\epsilon$\,Ori   & B0I   & $\sim$1000 & 0.2--0.3 &        &\citet{zhe07} \\
$\tau$\,Sco       & B0.2V & 200--450   & 1.7      & $-6.3$ &\citet{mew03}\\
                  &       &            &          &        & \citet{coh03} \\
$\theta$\,Car     & B0.2V+late & $<$400$^*$ & 0.2 & $-7.0$ &\citet{naz08c} \\
$\beta$\,Cru\,A   & B0.5III+B2V& 140        & 0.2 & $-7.7$ &\citet{coh08} \\
$\theta^1$\,Ori\,A& B0.5V+A0 & $<$160$^*$ & $>$1.3$^a$& $-6^b$ &\citet{sch03} \\
                  & +lateB&            &          &        & \citet{ste05} \\
$\beta$\,Cep      & B1V & $<$450$^*$ & 0.2 & $-7.2$ &\citet{fav08} \\
$\beta$\,Cen      & B1III & $<$440$^*$ & 0.2 & $-7.2$ &\citet{raa08} \\
Spica             & B1III-IV &  & 0.2--0.3& &\citet{zhe07} \\
\noalign{\smallskip}\hline
\end{tabular}
\\
{\scriptsize $^a$ 70\% of the X-ray flux come from hot plasma (i.e. with $T>$1.5$\times10^7$\,K \citep{sch03}.\\
$^b$ Note that the flux values of \citet{ste05} do not agree with those of \citet{sch03}.}\\
\end{table}

Amongst the investigated B-stars were three known $\beta$ Cephei pulsators: $\beta$\,Cru\,A, $\beta$\,Cen, and $\beta$\,Cep \citep{coh08,raa05,fav08}. While no significant variation of the flux was detected for the latter two objects, $\beta$\,Cru\,A displays some variability of its hard emission at the primary and tertiary pulsation periods. However, as the maxima of the optical and X-ray light curves appear phase-shifted by a quarter of a period and as the variability is of very modest amplitude, these changes need to be confirmed. The blue straggler $\theta$\,Car, which is not a known pulsator, displays some variability but only on long-term ranges (between \ros, Einstein and \xmm\ observations, \citealt{naz08c}).

While the soft character of the X-ray emission and the values of the formation radii are compatible with the wind-whock model described in Sect. \ref{sec:origin}, the narrow lines clearly constitute a challenge for this scenario. Hard X-ray emission is also a problem in such objects since they possess much weaker winds than O stars. The exact origin of the X-ray emission of these B-type stars therefore remains a mystery up to now (see, e.g., the extensive discussion in \citealt{coh08}) but some proposed explanations rely on the possible impact of a magnetic field (see Sect. \ref{sec:magnetic}). 

To conclude this section, a last comment on the peculiar case of Be stars should be made. One such star was studied in detail with \ch: $\gamma$\,Cas \citep[B0.5IV, ][]{smi04}. $\gamma$\,Cas appears particularly bright in X-rays, although it is less luminous than X-ray binaries. Its X-ray emission consists of a strong continuum (a bremsstrahlung with $kT=11-12$\,keV), some broad (HWHM$\sim$600\,\kms) and symmetric  X-ray lines from ionized metals (an optically thin plasma with 4 components, with temperatures ranging from 0.15 to 12\,keV), and fluorescence K features from iron and silicon. For the photoionized plasma, different iron abundances were found for the hot and the warm component, which might indicate the presence of the FIP effect and at least point towards different emission sites. In addition, while the warm and most of the hot components display similar absorption, part of the hot component appears more strongly absorbed. \citet{smi04} suggest that these hard X-rays are seen through the dense regions of the Be disk, while most of the X-ray emission (warm+most of the hot plasma) is only absorbed by the stellar wind or the outer, less dense regions of the disk. The presence of fluorescence features indeed suggests that some cold gas is present close to the X-ray emission regions. \citet{li08} further presented models for generating X-rays in Be stars. According to these authors, the collision of the magnetically channeled wind (see Sect. \ref{sec:magnetic}) with Be disks could explain the emission of B-stars down to types B8 without the need for a companion.

\subsection{Do Wolf-Rayet stars emit X-rays?}
\label{sec:wr}

Wolf-Rayet stars (WRs) are the evolved descendants of O-type stars. They come in three ``flavors'', WN, WC, and WO, which refer to spectra dominated by nitrogen, carbon, or oxygen lines, respectively. In the evolutionary sequence of hot stars, WC and WO objects are expected to correspond to later stages than WN stars, explaining their metal-rich composition. Both types display mass-loss rates on average ten times larger than for their O-type progenitors. These winds, denser and enriched in metals, are much more opaque to X-rays than those of O stars, especially in the WC case. One may thus naively expect WRs stars to be less X-ray bright than O stars. 

The first X-ray observations of WRs indeed provided results quite in contrast with those of O-type stars. The detection fraction was much smaller for WRs and no $L_{\rm X}-L_{\rm bol}$ relation was found for the detected objects \citep{wes96}, although the latter fact could be explained theoretically by assuming a particular relation of the X-ray flux on the wind properties, and by taking into account the absorption by the dense, enriched winds \citep{ign99}. An overall view of the situation ten years ago was presented by \citet{pol95a} and \citet{pol95b}: only 20 WRs were clearly detected by \ros\ (i.e. detection statistic $\lambda>$10), which corresponds to about 10\% of the WRs listed in the VIIth catalog of Galactic WRs \citep{van01}. Using this catalog, it appears that 7 of these stars belonged to the WC category, 12 were WN and 1 WN/WC; 6 of these were supposedly single (4WN+1WC\footnote{This object is WR111, whose clear detection was subsequently challenged \citep{osk03}. It now appears that this old detection was a false alarm.}+1WN/WC\footnote{This object is WR20a, subsequently found to be a very massive binary of type WN6ha+WN6ha \citep{rau04}.}), 4 were candidate binaries (3WN+1WC), and 10 were known binaries (5WN+5WC) at that time. As for O-type stars, WR binaries appear brighter than single objects, most probably because of wind-wind interactions (see Sect. \ref{sec:cwb}). In addition, WN stars were found to be on average four times brighter than WC stars.

\begin{table}
\caption{X-ray properties of suspected single Wolf-Rayet stars in our Galaxy. Quoted luminosities correspond to observed values, i.e. without dereddening, for detected objects and to upper limits for undetected stars. The letters H and P in the remarks column indicate the presence of a hard tail (or hot component) and of periodicities (either in optical, X-rays, or both), respectively. Note that the quoted limits correspond to different significance levels (e.g. conservative estimate from \citealt{gos05} vs 1-$\sigma$ in \citealt{osk03}).}
\label{tab:wr}       
\begin{tabular}{llllll}
\hline\noalign{\smallskip}
Name & Sp. Type & $L_{\rm X}$ (erg\,s$^{-1}$) &  $\log(L_{\rm X}/L_{\rm bol})$ & Remarks & References \\
\noalign{\smallskip}\hline\noalign{\smallskip}
\multicolumn{5}{l}{\it Detected Wolf-Rayet stars (\xmm\ and \ch)} \\
WR1  & WN4 & 2.0$\times 10^{32}$ & $-6.6$ & P &\citet{ign03} \\
WR3  & WN3 & 2.5$\times 10^{32}$ & $-6.8$ &   &\citet{osk05} \\
WR6  & WN4 & 1.7$\times 10^{32}$ & $-6.6$ & HP&\citet{ski02a} \\
WR20b& WN6 & 1.6$\times 10^{33}$ & $-6.2$ & H &\citet{naz08a} \\
WR42d& WN5 & 4.0$\times 10^{32}$ & $-6.7$ &   &\citet{osk05} \\
WR44a& WN5 & 1.6$\times 10^{31}$ & $-8.1$ &   &\citet{osk05} \\
WR46 & WN3 & 2.4$\times 10^{32}$ & $-7.0$ & P &Gosset et al., in prep. \\
WR110& WN5 & 7.2$\times 10^{31}$ & $-7.1$ & HP&\citet{ski02b} \\
WR136& WN6 & 7.9$\times 10^{30}$ & $-8.8$ &   &\citet{osk05} \\
WR152& WN3 & 3.2$\times 10^{31}$ & $-7.3$ &   &\citet{osk05} \\
\multicolumn{5}{l}{\it Other detected Wolf-Rayet stars (\ros)} \\
WR2$^a$& WN2 & 7.9$\times 10^{31}$ & $-7.3$ &   &\citet{osk05} \\
WR7  & WN4 & 2.0$\times 10^{32}$ & $-6.6$ &   &\citet{osk05} \\
WR18 & WN5 & 2.5$\times 10^{32}$ & $-6.9$ &   &\citet{osk05} \\
WR78 & WN7 & 1.3$\times 10^{31}$ & $-8.7$ &   &\citet{osk05} \\
WR79a& WN9 & 7.9$\times 10^{31}$ & $-7.7$ &   &\citet{osk05} \\
\multicolumn{5}{l}{\it Undetected Wolf-Rayet stars} \\
WR5  & WC6 &  $<9.3\times 10^{30}$ & $<-7.7$& &\citet{ski06} \\
WR16 & WN8 &  $<1.0\times 10^{30}$ & $<-8.9$& &\citet{osk05} \\
WR40 & WN8 &  $<4.0\times 10^{31}$ & $<-7.6$& &\citet{gos05} \\
WR57 & WC8 &  $<8.9\times 10^{30}$ & $<-8.1$& &\citet{ski06} \\
WR60 & WC8 &                       & $<-8.2$& &\citet{osk03} \\
WR61 & WN5 &  $<5.0\times 10^{30}$ & $<-8.1$& &\citet{osk05} \\
WR90 & WC7 &  $<1.9\times 10^{30}$ & $<-8.7$& &\citet{ski06} \\
WR111& WC5 &                       & $<-7.8$& &\citet{osk03} \\
WR114& WC5 &                       & $<-9.2$& &\citet{osk03} \\
WR118& WC9 &                       & $<-7.4$& &\citet{osk03} \\
WR121& WC9 &                       & $<-7.4$& &\citet{osk03} \\
WR124& WN8 &  $<2.0\times 10^{32}$ & $<-6.6$&P&\citet{osk05} \\
WR135& WC8 &  $<6.6\times 10^{29}$ & $<-9.1$& &\citet{ski06} \\
WR144& WC4 &                       & $<-7.4$& &\citet{osk03} \\
WR157& WN5 &  $<4.0\times 10^{31}$ & $<-7.9$& &\citet{osk05} \\
\noalign{\smallskip}\hline
\end{tabular}\\
{\scriptsize $^a$ In a recent poster, \citet{ski08b} presented the first results from \ch\ observations of WR2 ($L_{\rm X}$=1.3$\times 10^{32}$\,erg\,s$^{-1}$, $\log[L_{\rm X}/L_{\rm bol}]$=$-6.9$), but also of WR24 (WN6, $L_{\rm X}$=9.5$\times 10^{32}$\,erg\,s$^{-1}$, $\log[L_{\rm X}/L_{\rm bol}]$=$-7.0$), and WR134 (WN6, $L_{\rm X}$=2.7$\times 10^{32}$\,erg\,s$^{-1}$, $\log[L_{\rm X}/L_{\rm bol}]$=$-6.8$). As these authors consider their analysis preliminary, we prefer to include it only in a note.}\\
\end{table}

With so few WRs detected, small number statistics could bias the results; investigations with higher sensitivity were thus crucially needed in order to settle the question about the intrinsic strength of the WR X-ray emission. About 20\% of the presumably single WCs were observed using ASCA, \xmm, or \ch:  WR60, 111, 114, 118, 121, 144 \citep{osk03} and WR5, 57, 90, 135 \citep{ski06}. These studies now place stringent constraints on the intrinsic emission of WC objects, with limits on $\log[L_{\rm X}/L_{\rm bol}]$ between $-$7.4 and $-9.1$ (Table \ref{tab:wr}). This suggests that either there is no intrinsic X-ray emission in WRs of WC type, which would be surprising because of the presence of radiatively-driven winds similar to O-stars (see Sect. \ref{sec:origin}), or that the emission takes place deep in the wind and is then completely absorbed.

The case of single WN stars is less clear-cut (Table \ref{tab:wr}). On the one hand, a few putatively single WN-type objects were found to be clear X-ray emitters in recent observations. WR6 and WR110 are rather bright ($L_{\rm X}^{\rm unabs}\sim5\times10^{32}$~\ergs) and display two main thermal components at 0.5 and $>$3\,keV \citep{ski02a,ski02b}. For WR46, 3 temperatures provide the best fit: 0.15, 0.6 and 2--3\,keV (Gosset et al. 2009, in preparation). The high temperature is quite unusual for single objects and is not predicted by shocked wind models - it is rather reminiscent of X-ray emission from interacting winds. The observed variations in the optical (WR6 and WR46) or in X-rays (WR46 and WR110) also suggest a possible binary nature for these stars. However, WR147 displays hints of a non-zero X-ray emission associated with the WR component \citep{pit02} and WR1 presents a soft spectrum, with little emission above 4\,keV, in contrast to the three previous objects \citep{ign03}. In the latter case, the possible presence of an absorption edge due to ionized sulfur further indicates absorption unrelated to the neutral ISM and rather linked to the hot wind: the X-ray emission would thus occur at some depth inside the wind. \citet{ign03} therefore proposed that the X-ray emission of WR1 would be typical of single WN objects and could explain the soft part of the spectra of WR6 and 110. On the other hand, WR40 remained undetected in a 20ks \xmm\ observation \citep{gos05}. This was quite surprising since this object is strongly variable in the optical domain and the changes are attributed to the unstable stellar wind: on the basis of the wind-shock model, WR40 was thus expected to be a moderate X-ray emitter. However, its non-detection yields a (very) conservative upper limit  on its $\log[L_{\rm X}/L_{\rm bol}]$ of $-7.6$, about one order of magnitude below that of O-type stars and similar to the limits found for WCs. To reconcile the detection of WR1 and the invisibility of WR40, one might have to consider the structure of their winds in detail. The ratio $\dot M/v_{\infty}$, which intervenes in the estimation of the wind density, is 13 times larger for WR40; this difference is however slightly attenuated by the larger radius of WR40. In addition, WR1 presents an earlier spectral type, and its wind is consequently more ionized, i.e. less opaque. Indeed, WR1 and WR40 are not similar objects, as far as their wind and stellar properties are concerned, and their X-ray emission could thus well differ. Finally, differences in wind porosity/clumping might also impact the detection level of WRs. Contrary to WCs, whose winds invariably present large optical depths, the WN stars display a wider variety, in the visible as well as in X-rays: individual modeling is thus probably necessary for each case.

The first detection of a WO star, WR142, revealed a rather faint source with an inferred $L_{\rm X}/L_{\rm bol}$ ratio of only 10$^{-8}$ \citep[e.g. similar to the upper limits on non-detections of other WRs, see above and][]{osk09}. The faintness of the source prevented any detailed spectral analysis but the hardness ratio clearly indicates a high temperature for the X-ray source which cannot be explained by the wind-shock model. This puzzling result might require to consider magnetic activity as a source of X-rays in WRs \citep{osk09}.

While most of the studies have been performed for galactic objects, \citet{gue08a,gue08b} as well as \citet{gue08c} presented a first analysis of the X-ray emission of WRs belonging to the LMC. From the available \ros, \ch, and \xmm\ observations, which cover more than 90\% of the known WRs in the LMC, only 32 objects (out of 125) were detected. They are mostly binaries: about half of the known WR binaries were observed as X-ray sources whereas the detection rate is only 10\% for supposedly single objects. Many similarities with the Galactic case were found: non-detection of single WC stars, preferential detection of binaries. However, some clear differences were also discovered: larger values for the X-ray luminosities and $L_{\rm X}/L_{\rm bol}$ ratios. While more sensitive data are certainly needed to confirm these trends, it remains to be seen whether these observations could be linked to the lower opacity of the winds in the low-metallicity environment of the LMC.

\subsection{Interacting winds in hot binaries}
\label{sec:cwb}

\begin{figure}
  \begin{center}
  \includegraphics[width=8cm, bb=140 220 575 685, clip]{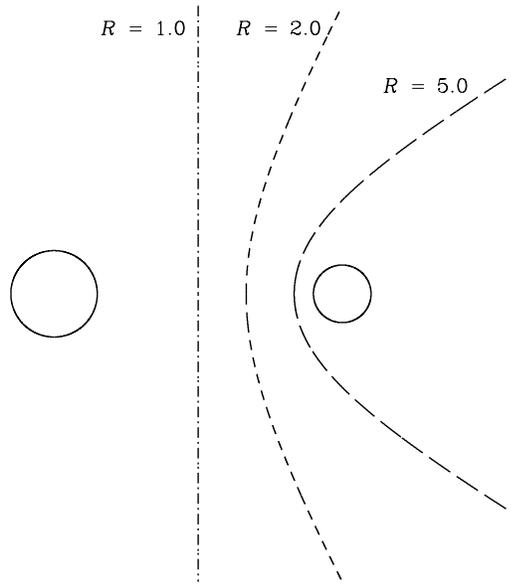}
  \caption{Shape of the wind-wind interaction region derived from pressure equilibrium, for different wind momentum ratios $R$. For two winds of equal strengths ($R = 1$), the equilibrium occurs in a plane in-between the two stars; when the winds differ, the interaction region takes a more conical shape and begins to fold around the star with the weaker wind.}
  \label{cwb}
  \end{center}
  \end{figure}

It is well established that hot massive stars possess strong stellar winds. Therefore, if two such stars form a close pair, an interaction between the winds can be expected. The interaction region is likely to be planar if the winds are of equal strength or it will rather appear cone-like if the winds are different (with the weaker wind inside the cone, Fig.~\ref{cwb}). As the winds flow at tremendous speeds before colliding, the gas is heated to high temperatures: $kT=(3/16) mv^2$, which is about 4.7\,keV assuming a wind of solar composition and a typical wind velocity of 2000\,\kms\ \citep{ste92}. Plasma with such temperatures can only be observed in the high-energy domain, and X-rays thus constitute the best means to study this phenomenon. It must be noted that before the advent of sensitive X-ray facilities, only a handful of wind-wind interactions had been investigated in depth. 

Theoretically, the wind-wind interactions introduced above can be separated in two classes, depending on the importance of cooling in the post-shock gas. To quantify this, one introduces the cooling parameter $\chi$, defined as the ratio between the cooling time of the post-shock gas and the typical escape time from the shock region. It can be expressed as $\chi=v^4D/\dot M$ where the wind velocity $v$ is in units of 1000\,\kms, the distance $D$ from the star to the shock\footnote{This distance can be found by considering the ram pressure equilibrium between the two flows. Considering the wind properties of the two stars (noted 1 and 2 in the following for the primary and secondary, respectively), the ratio of the separations between the shock and the star, measured along the binary axis, is $\frac{D_1}{D_2}=\sqrt{\frac{\dot M_1v_1}{\dot M_2v_2}}=R$ \citep[the wind momentum ratio, see Fig. \ref{cwb}, ][]{ste92}.} in 10$^7$~km, and the mass-loss rate $\dot M$ in 10$^{-7}$~\msol\,yr$^{-1}$. For $\chi \ll 1$, the gas cools rapidly and the collision is to be considered as radiative. This situation generally occurs in short-period binaries. In this case, hydrodynamic models predict that instabilities arise in the interaction region, making the collision quite turbulent, and the X-ray luminosity then follows a relation of the form $L_{\rm X}\propto\dot M v^2$ \citep{ste92}. For $\chi\ge1$, the collision is adiabatic, the interaction thus appears smoother and the X-ray luminosity rather scales as $L_{\rm X}\propto\dot M^2 v^{-3/2} D^{-1}$. This behaviour is preferentially expected for long-period binaries.

According to the above considerations, several observational hints for the presence of a wind-wind interaction can be expected. First, this phenomenon provides an additional source of X-rays, on top of the intrinsic stellar contributions: such systems should thus appear overluminous. Indeed, the first X-ray observations showed that hot binaries are on average more luminous than single stars (e.g. \citealt{pol87,chl91}). There are indications that this overluminosity might increase with the combined bolometric luminosity of the system \citep{lin06}. This could be readily explained if one considers that the winds are radiatively-driven: larger luminosities mean stronger winds, hence stronger wind-wind interactions. Second, because of the high speeds of these winds which are colliding face-on, the X-ray emission should appear harder than the typical emission from O-type stars ($kT$ of 1--10\,keV vs 0.3--0.7\,keV). However, it should be noted that these first two criteria are not sufficient for ascertaining the presence of a wind-wind interaction in a given system, as other phenomena also produce additional hard X-ray emission (e.g. magnetically channeled winds, see Sect. \ref{sec:magnetic}). Finally, further evidence for the presence of a wind-wind interaction comes from the detection of modulations of the high-energy emission with orbital phase. These phase-locked variations are produced by changing absorption along the line-of-sight and/or changing separation in eccentric binaries. 

Absorption changes occur as the interaction region is alternatively observed through the wind of one or the other star. These variations are indeed best seen in systems containing two very different winds, e.g. WR+O binaries. The most famous system in this category is definitely $\gamma^2$Vel (WC8+O7.5III, $P$=78.5\,d). \citet{wil95} reported on multiple \ros\ observations of the system: they observed a recurrent strong increase of the observed X-ray emission when the O-type companion was in front of the WR star. This increase was detected in the ``hard'' \ros\ band, i.e. 0.5--2.5\,keV\footnote{Quite surprisingly, very soft (0.1--0.5\,keV) emission was also detected in $\gamma^2$Vel but, as it seems stable with phase, it must be produced quite far in the winds, where the absorption is low. Since no single WC star was ever detected in X-rays (see Sect. \ref{sec:wr}), this soft flux can probably not be considered as intrinsic to the WR star.}, and was interpreted as due to reduced absorption when the interaction region is seen through the less dense and less metal-rich (thus less opaque) O-star wind. The width of the light-curve peak was further related to an opening angle of about 50$^{\circ}$ for the shock cone. Short-term variations, detected by ASCA \citep{ste96}, provide evidence for instabilities linked to the interaction region. Using high-resolution spectra, \citet{sch04} and \citet{hen05} showed that the X-ray emission was produced far from the UV sources (broad lines, large {\it f/i} ratio). The spectrum of the hot plasma was fitted with 3 temperatures, interpreted as coming from different regions \citep{sch04}: 0.25\,keV (with constant absorption, this flux should be emitted far in the winds), 0.65 and 1.8\,keV (both along the shock cone, the observed absorption decreasing when the line-of-sight lies inside the shock cone). For the latter components, reproducing exactly the change in the absorbing column proved rather difficult: either a decrease of a factor of four in the mass-loss rate is needed (microclumping, porosity?), or a change in the abundances (hints of a neon enrichment, suggesting a mix of O and WR material for the emitting plasma). Regarding the overall properties of the spectra, it must be noted that different metals sometimes yield different plasma temperatures, suggesting that non-equilibrium ionization could be present in $\gamma^2$Vel \citep{hen05}.

Similar effects are seen in a few other systems. In the eccentric system WR22 (WN+O7-9, $P$=80.3\,d), no emission is present at soft energies, but the flux at medium energies (0.7--2\,keV) varies strongly due to changes in absorption (Gosset et al. 2009, submitted). The minimum absorption occurs when the O-type star is in front of the WR: as for $\gamma^2$Vel, the interaction region is then seen through the less opaque wind of the companion. The behaviour of the absorption at other phases can be qualitatively understood if one considers the interaction region to dive deeper and deeper inside the WR wind, as seen from Earth. Outside our Galaxy, a similar phenomenon might occur in the peculiar system HD~5980 (WR+WR, \citealt{naz07}). Binaries composed of two O stars can also undergo similar absorption changes, although of smaller amplitude. The small decrease of the soft X-ray flux of Plaskett's star (O7.5I+O6I, $P=14.4$\,d) and HD~93403 (O5.5I+O7V, $P=15.1$\,d) can be explained by the optically thicker wind of the primary star obscuring our view to the interaction region (\citealt{lin06,rau02b}, respectively). Absorption effects have also been proposed for HD~165052 \citep[O6.5V+O6.5V, ][]{cor96} and 29 CMa (O8.5I+O9.7V, \citealt{ber95} - although in this case it should concern the intrinsic emission of the stars, since there is apparently no overluminosity typical of an interacting region).

  \begin{figure}
  \begin{center}
  \includegraphics[width=12.5cm, bb=40 260 570 550, clip]{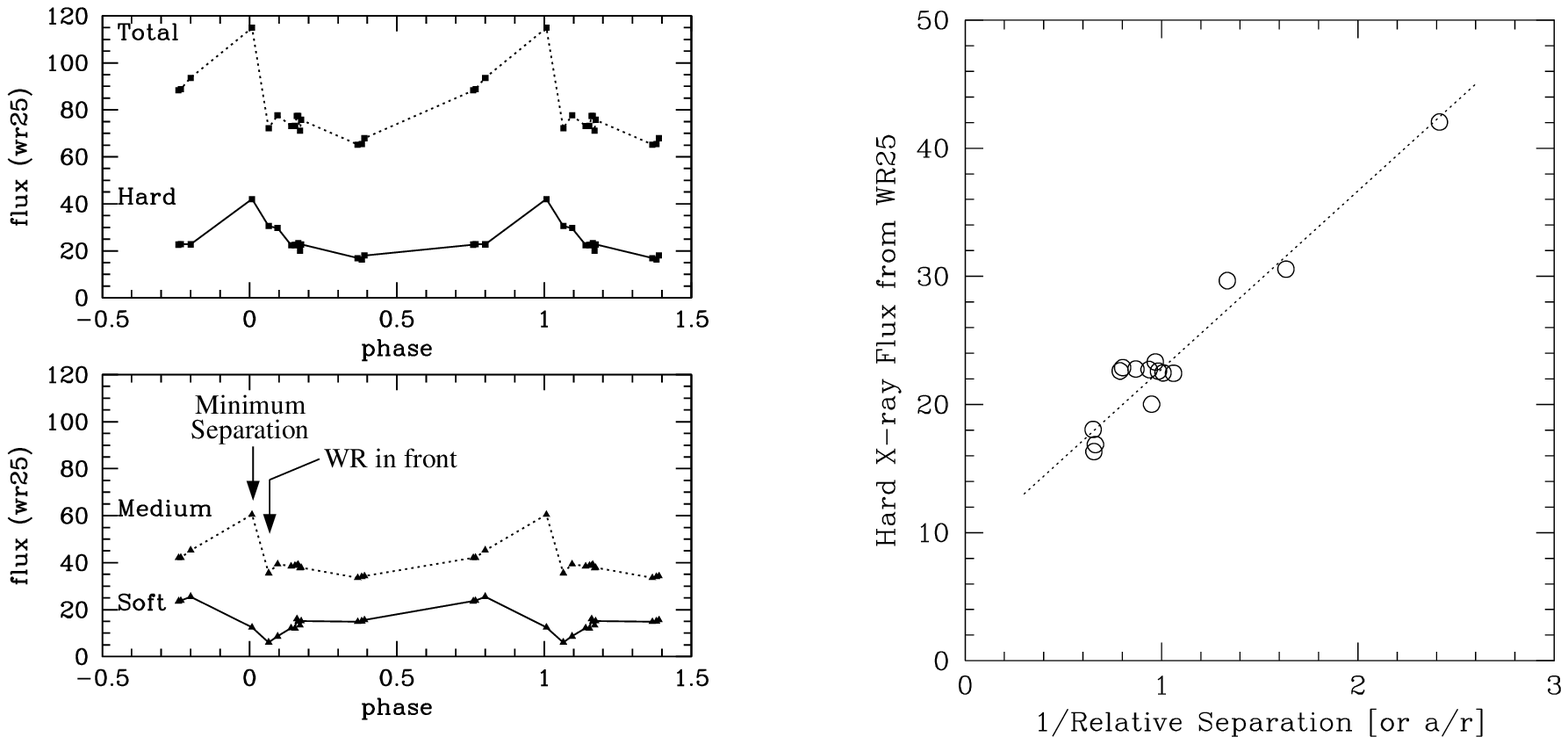}
  \caption{Left: X-ray light curves of WR\,25 in different energy bands. At minimum separation ($\phi$=0) the hard and medium fluxes are maximum, while the soft flux is minimum (or the absorption is maximum) when the Wolf-Rayet star is in front. Right: The observed hard X-ray flux as a function of the inverse of the relative separation between the stars ($r$ divided by the semi-major axis $a$). Figures adapted from \citet{gos07}, which rely on \xmm\ data.}
  \label{cwbabs}
  \end{center}
  \end{figure}

A changing separation in eccentric systems can also be responsible for a phase-locked modulation of the wind-wind X-ray emission. In adiabatic situations (where $L_{\rm X}\propto\dot M^2 v^{-3/2} D^{-1}$), the stars are rather distant from one another and the stellar winds have plenty of time to reach their terminal velocity before interacting. Therefore, no variation in the wind speed is expected and changes in the X-ray luminosity can be attributed to the varying distance between the stars and the shock zone. As theory predicts a $1/D$ effect (see equation above), the intrinsic X-ray flux is expected to be maximum at periastron, where the plasma density is higher. Such a variation has now been observed in several systems: WR25 (WN+O4, $P=208$\,d, see Fig. \ref{cwbabs}, \citealt{gos07}), HD~93205 (O3.5V+O8V, $P=6.1$\,d, \citealt{ant03}), HD~93403 (O5.5I+O7V, $P=15.1$\,d, \citealt{rau02b}). However, it must be noted that in other eccentric systems, like $\gamma^2$Vel and WR22, the hard X-ray flux is found to be constant, without evidence for a $1/D$ variation (\citealt{rau00}, Gosset et al. 2009, submitted). The flux of the long-period binary WR140 (WC7+O4--5, $P=7.94$\,yr, $e=0.88$) rises towards periastron but then decreases suddenly as the WR star passes in front of the O-type star and its dense wind occults both the companion and the wind-wind interaction \citep{cor03}. However, the flux does not follow perfectly the expected $1/D$ variation \citep{pol02}, maybe because the collision becomes radiative near pariastron or because of the energy lost to accelerate non-thermal particles at that time \citep{pit06}. Interestingly, this system is one of the few where changes of the profiles of the X-ray lines have been detected. \citet{pol05} report broad and blueshifted ($-$600\,\kms) lines before periastron and even wider but slightly redshifted lines just after periastron. Such line profile changes clearly deserve an in-depth study: future X-ray facilities might help detect them for a wide range of interacting wind systems and the observations might then be compared to model predictions, see e.g. those of \citet{hen03}. It is worth noting that hints of a collisionless nature of the wind-wind shock as well as evidence for non-equilibrium ionization were reported for WR140 by \citet{pol05} and \citet{zhe00}. Similar clues were also found for WR147 \citep{zhe07b}.

In the case of a radiative interaction ($L_{\rm X}\propto\dot M v^2$), the stars are closer to each other and the stellar winds are still in the acceleration zone when they interact. Therefore, the interaction occurs at lower speeds at periastron than at apastron: following the equation above, the X-ray flux should thus be minimum at periastron. A good example of this case is Cyg OB2 \#8A (O6If+O5.5III(f), $P=21.9$\,d, \citealt{deb06}) where the flux and plasma temperature present a minimum close to periastron. This is probably also the case of HD~152248 (O7.5III(f)+O7III(f), $P=5.8$\,d, \citealt{san04}). Note that this idea can be applied on a more global scale, with longer-period systems permitting more acceleration for the wind thereby explaining their larger X-ray overluminosities \citep{lin06} - of course, this is valid only in the radiative regime (i.e. up to $P\sim$~10--20~d): for longer periods, the $1/D$ effect dominates.

A last case of wind interaction occurs when one of the star has little, if any, mass-loss rate. In this case, the wind of one star crashes onto the photosphere of its companion (or close to it): the X-ray emitting region thus roughly corresponds to the hemisphere facing the primary star. Such a possibility was envisaged to explain the peculiar shape of the light curve of CPD$-$41$^{\circ}$7742 (O9V+B1-1.5V, $P=2.4$\,d, \citealt{san05}). From a non-zero value, the count rate smoothly increases during a quarter of the orbit; this is followed by an equally smooth decrease during the next quarter of the orbit; when the maximum flux should be reached, a narrow dip makes the light curve come back to its original level. This level corresponds to the intrinsic emission of the two stars; the broad peak occurs when the X-ray emitting hemisphere comes slowly into view as the system rotates (an effect analogous to the lunar phases); the narrow drop corresponds to an X-ray eclipse when the primary occults the secondary (with an inclination of 77$^{\circ}$, the system is seen nearly edge-on). Such a phenomenon can be expected for other O+B systems, but has not yet been specifically searched for.

  \begin{figure}
  \begin{center}
  \includegraphics[width=9.cm, bb=143 158 460 560, clip]{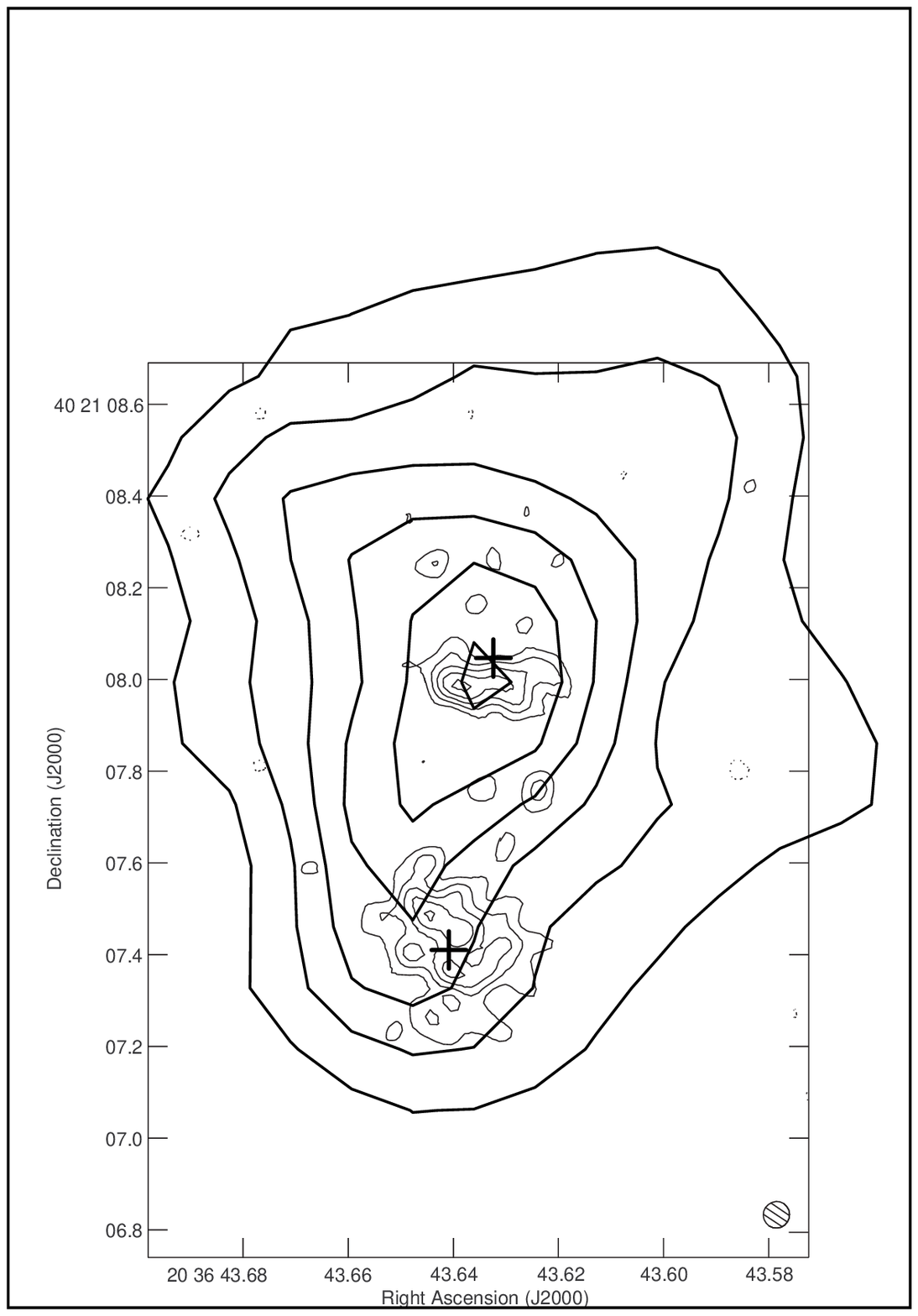}
  \caption{The emission from WR147, with X-ray contours (thick lines) superimposed on radio contours (thin lines, figure reprinted from \citealt{pit02}). The two crosses give the positions of the stars: the WR star is to the south, the O star to the north. The WR star is associated with the southern radio source, while the elongated radio source to the north is non-thermal radio emission from the wind-wind collision. The X-ray emission, which is extended, is not co-spatial with the WR star but is likely associated with the wind-wind interaction. }
  \label{wr147}
  \end{center}
  \end{figure}

Further evidence for X-rays associated with a wind-wind interaction is the spatial extension of the high-energy source. Indeed, the X-ray emission is not point-like but occurs in an extended region close to the stagnation point (the intersection of the shock with the binary axis). Unfortunately, spatial resolution is still limited and one did not expect any direct evidence of this extension to be found with either \xmm\ or \ch. However, an indirect signature for the source extension was detected in a few systems. For example, in WR22, it is necessary to take extension into account to closely model the absorption at the most absorbed phase (Gosset et al. 2009, submitted). Such an effect could also help for a better modeling of absorption in $\gamma^2$Vel \citep{sch04}. In the very massive eclipsing binary WR20a (WN6ha+WN6ha, $P$=3.7\,d, \citealt{naz08a}), the X-ray emission brightens during the optical eclipse: the lack of an X-ray eclipse is directly related to a non-zero extension of the X-ray emission; furthermore, the brightening requires the collision zone to be rather opaque when seen edge-on (which happens a quarter of phase after/before the eclipse). Finally, {\it direct} evidence for source extension was found in \ch\ observations of WR147 (WN8+OB, see \citealt{pit02} and Fig. \ref{wr147}). This second closest WR system (the closest one being $\gamma^2$Vel) has been resolved at both radio and IR wavelengths: the non-thermal radio source corresponding to the wind-wind interaction lies 0.58$^{\prime\prime}$ north of the WR star and the OB companion is located 0.06'' further away. The size of the observed X-ray emission is about 70\% larger than the \ch\ point spread function and its position does not correspond to that of the WR star. No other possibility than wind-wind collision can explain such observations and these data thus constitute important evidence showing the reality of the phenomenon. It should be noted that the best fit to the observations considers a main component associated with the extended interaction zone plus weaker contributions from the point-like stars: in this case, the WR star should thus have intrinsic, non-zero X-ray emission (see Sect. \ref{sec:wr} for further discussion on this subject).

A strange feature discovered in two interacting systems of type WC+O ($\gamma^2$\,Vel, \citealt{sch04}, and $\theta$\,Mus, \citealt{sug08}) must also be mentioned: the presence of narrow radiative recombination continua (RRC) associated with highly ionized carbon. It indicates that cool gas ($kT=$ a few eV) is present not far from the X-ray sources. Its origin is not yet clear: because of the lack of phase-locked variations, \citet{sch04} favor recombination occurring far in the winds or far out in the post-collision flow while \citet{sug08} suggest that the RRC feature displays a similar shift as the X-ray lines arising from the hot plasma of the wind-wind interaction. 

Once that clear evidence was found for wind-wind interactions, a better modeling was attempted. Indeed, when well understood, such interactions can help probe the stellar wind parameters, which are notoriously difficult to estimate (e.g. \citealt{ant04,deb06}). However, one should not forget that there is more than just one interaction in the system: there are the stars. This has two consequences. First, for a meaningful comparison, the modeled wind-wind contribution should always be added to the stars' intrinsic high-energy emissions. Indeed, the predicted changes are generally larger than the observed ones since the stellar X-rays dilute the variations of the interaction emission (e.g. HD~152248, \citealt{san04}). Second, wind acceleration needs to be cautiously modeled. Indeed, in hot stars, the winds are accelerated by UV radiation; in hot binaries, there are two sources of UV photons: along the binary axis, the stellar winds might hence not be accelerated to their full strength (a process called radiative inhibition) and, in asymmetric systems (e.g. WR+O), the strongest wind might even suddenly decelerate as it approaches the companion (a process called sudden radiative braking). Such braking effects alter the wind velocity, hence the X-ray emission: the flux and plasma temperature will be lower than expected. Observational evidence for that process is still quite elusive in the X-ray domain. For $\gamma^2$Vel, \citet{hen05} used a geometrical model (with X-ray emission along a cone revolving with the orbital period) to reproduce the observed unshifted line profiles: they found that the latter could only be fitted with a half-opening angle of 85$^{\circ}$, i.e. much more than the 25$^{\circ}$ expected on the basis of the shape of the X-ray light curve. This result was interpreted as a possible consequence of sudden radiative braking close to the O star. 

Finally, a last remark must be made: not all hot binaries display X-ray bright interactions and this is not solely due to the type of performed observations (a short snapshot prohibiting the production of light curves, contrary to a dedicated monitoring). In NGC~6231, only HD~152248 and CPD$-$41$^{\circ}$7742 present a clear overluminosity in the $L_{\rm X}-L_{\rm bol}$ diagram, although several other binary systems exist in the cluster \citep{san06}. In the case of $\iota$ Orionis \citep{pit00}, no significant variation of the X-ray emission was detected between periastron and apastron observations, although they were eagerly expected. The reasons for these differences are unclear, and definitely require more investigation.  

\subsection{Hot magnetic objects}
\label{sec:magnetic}

Up to recent years, magnetic fields were quite an elusive subject for hot stars. Indeed, hot stars lack outer convection zones, and there seemed to be no correlation between X-ray emission and rotation rate (as seen in cool stars). However, magnetism had been proposed to explain the variability of some peculiar objects, e.g. $\theta^1$\,Ori\,C, and it was also needed for producing the non-thermal radio emission observed in a few hot systems. Direct evidence for its presence is difficult to find as the broadening of the stellar lines in the spectra of hot stars prevents the direct observation of Zeeman splitting. However, the Zeeman effect also induces polarization of light and it was finally through spectropolarimetric studies that magnetic fields were first detected in 2002 ($\theta^1$\,Ori\,C, \citealt{don02}). More efforts to search for magnetic fields are currently under way \citep{bou08,hub08,pet08,sch08}. Table \ref{tab:mag} summarizes the current status of magnetic field detections in hot stars.

\begin{sidewaystable}
\caption{Properties of magnetic stars earlier than B1 (a '+' in the spectral type indicates the presence of companions). The dipolar field strength is quoted when available; if absent, the values of $\eta$ were calculated by assuming the observed field strength to be the dipolar one. Quoted X-ray temperatures correspond to the main component. }
\label{tab:mag}       
\begin{tabular}{llrrccccccccl}
\hline\noalign{\smallskip}
Name & Sp. Type & $B_{\rm obs}$ & $B_{\rm dip}$ & $v \sin i$ & $v_{\infty}$ & $R_*$ & $\dot M$ & $kT$ & HWHM & $\log(L_{\rm X}$ & $\eta$ & Ref. \\
 & & (G) & (G) & \multicolumn{2}{c}{(\kms)} & (R$_{\odot}$) & (M$_{\odot}$\,yr$^{-1}$) & (keV) & (\kms) & $/L_{\rm bol})$& & \\
\noalign{\smallskip}\hline\noalign{\smallskip}
9\,Sgr            & O4V+        & 211$\pm$57&                        & 128& 2950& 16.0& 2.4$\times10^{-6}$& 0.26& 500--1600& $-6.4$& 0.3& 1,2\\
HD~148937          & O5.5f?p     & $-276\pm88$&                       &  45& 2600& 15.0& $\lesssim10^{-7}$& 0.2& 873& $-6.0$& 13& 2,3\\
$\theta^1$\,Ori\,C\,$\dagger$& O5.5V+      & & 1060$\pm$90                      &  24& 2980&  8.3& 1.4$\times10^{-6}$& 2.5--3& 300& $-6.0$& 3.6& 4\\
                           & O7V+        & &                                  &     & 2760&  9.1& 5.5$\times10^{-7}$&   &   &   & 12& \\
HD~191612          &O6.5f?p--    & & 1500                             &  45& 2700& 14.5& 1.6$\times10^{-6}$& 0.2--0.3& 900& $-6.1..2$& 21& 5\\
                  &--O8fp+      & &                                  &    &     &     &     & & & & \\
HD~36879           & O7V         & 180$\pm$52&                        & 163& 2170--2400& 10& 4$\times10^{-7}$& & & & 0.6--0.7& 6\\
HD~155806          & O7.5Ve      & $-115\pm$37&                       &  91& 2390& 8.9& 2$\times10^{-7}$& soft & & $-6.7^*$ & 0.4& 7\\
HD~152408          & O8I         & $-89\pm$29&                        &  85&  955--2200& 31& 5--12$\times10^{-6}$& & & $<-7.3^*$& 0.05--0.3& 8\\
$\zeta$\,Ori\,A   & O9.7I       & & 61$\pm$10                        & 110& 2100& 25.0& 1.4--1.9$\times10^{-6}$& 0.2& 850 & $-7.1$& 0.11--0.15& 9\\
$\tau$\,Sco       & B0.2V       & \multicolumn{2}{c}{300 everywhere} &   5& 2000&  5.2& 2$\times10^{-8}$& 0.6& 200--450& $-6.5$& 12& 10\\
NU\,Ori           & B0.5V+      & & 650$\pm$200                      & 225& & & & 0.2 & & $-6.7$& 55\,$\ddagger$& 11\\
$\xi^1$\,CMa      & B0.5-1III   & 306&                               &  20& 1518& 7.1& 1.9$\times10^{-8}$& 0.3$^*$& & $-6.6^*$& 31& 12\\
$\beta$\,Cep      & B1V+        & & 360$\pm$40                       &  27& 800--1500& 6.9& $10^{-9}$& 0.2& $<$450 & $-7.2$& 790--1481& 13\\
\noalign{\smallskip}\hline
\end{tabular}\\
$^*$ \ros\ observations (from \citealt{ber96}). \\
$\dagger$ The two lines correspond respectively to the hot and cool models of \citet{gag05}.\\
$\ddagger$ Assuming the same values of radius, mass-loss rate and terminal velocity as for $\tau$\,Sco.\\
References: (1) \citet{rau02},  (2) \citet{hub08}, (3) \citet{naz08d,naz08b}, (4) \citet{don02}, \citet{gag05}, (5) \citet{don06a}, \citet{naz07b,naz08d}, \citet{how07}, (6) \citet{hub08}, \citet{pri90}, \citet{vil92}, (7) \citet{hub08}, \citet{pri90}, \citet{chl91}, \citet{ber96}, \citet{fmar05}, (8) \citet{hub08}, \citet{pri90}, \citet{vil92}, \citet{chl91}, \citet{ber96}, (9) \citet{raa08}, \citet{bou08}, (10) \citet{mew03}, \citet{coh03}, \citet{don06b}, (11) \citet{ste05}, \citet{pet08}, (12) \citet{cas94}, \citet{hub06}, (13) \citet{hen00}, \citet{fav08}\\
\end{sidewaystable}

To study the impact of a magnetic field on the stellar winds, hydrodynamical modeling was undertaken \citep{udd02}. It revealed that the crucial factor is the importance of magnetic energy relative to the kinetic energy of the wind, better evaluated through the use of a wind confinement parameter \begin{equation}
\eta=\frac{B^2_{\rm eq}R_*^2}{\dot M v_{\infty}}.
\end{equation}
If $\eta\ll$1, the field is weak and the outflowing wind remains rather unaltered; if $\eta>$1, the field is strong enough to channel the wind towards the stellar equator. In the latter case, it must be noted that no stable disk forms. The head-on collision of the two channeled wind streams at the equator heats the gas to high temperatures: as for interacting wind binaries, hard X-ray emission should thus be produced. In addition, since the phenomenon occurs quite close to the photosphere, in shocked plasma with little radial velocity, the lines of the emitting hot plasma are expected to be narrow ($\sigma<$250\,\kms) and only slightly blueshifted ($-$100 to 0\,\kms, \citealt{gag05}). Finally, if the magnetic and rotational axes differ, a periodic modulation of the X-ray emission should be observed, as the region of magnetically confined wind is alternatively observed edge-on/face-on (though this modulation can be of limited amplitude depending on the characteristics of the system, see e.g. \citealt{fav08}).

$\theta^1$\,Ori\,C was observed at four phases with \ch\ gratings and also at medium resolution in the framework of the COUP campaign (\citealt{gag05}, see errata in \citealt{gag05b}). The observed spectrum is indeed quite hard, with a dominant temperature at 2.5--3\,keV, and a second, much weaker component at 0.7\,keV. Flux variations of about 30\% are detected, with a simultaneous maximum of X-ray and H$\alpha$ emissions when the ``disk'' is seen face-on; a slightly larger absorption column is also observed when the ``disk'' is seen edge-on. Most X-ray lines are narrow (350\,\kms, slightly larger than predicted) but the broader lines (e.g. O\,{\sc viii}) correspond to a cooler plasma and their formation could thus be explained by the ``usual'' wind-shock model. Moreover, the line centroids are on average close to 0\,\kms, though slightly variable (from a small blueshift of $-$75\,\kms\ when seen face-on to a redshift of $+$125\,\kms when edge-on). Finally, the comparison of the line triplets from He-like ions suggests a formation region very close to the stellar surface, at radial distances $r=1.6-2.1\,R_*$. All in all, the case of $\theta^1$\,Ori\,C is well explained by the hydrodynamical models. 

This is also the case for $\tau$\,Sco, whose magnetic field was detected by \citet{don06b} shortly after high-resolution X-ray spectroscopy had revealed its striking similarities to $\theta^1$\,Ori\,C: narrow (HWHM of 200--450\,\kms) and unshifted lines, hard emission (strong component at $kT=1.7$\,keV), clear overluminosity, X-ray formation region at $r$=1--3\,$R_*$ \citep{mew03,coh03}. The magnetic field geometry is however much more complex than a simple dipole, and MHD simulations still need to be performed to check if the agreement is also quantitative.

One can now easily imagine a whole continuum of magnetic effects in massive stars. On the one hand, strong magnetic confinement ($\eta\gg 1$) will produce hard X-rays, narrow lines, a large overluminosity, and a clear periodic modulation, as exemplified by $\theta^1$\,Ori\,C. On the other hand, small magnetic confinement will not affect the X-ray emission, which keeps its usual properties from the wind-shock model (soft X-rays, broader lines, constant emission, and $\log[L_{\rm X}/L_{\rm bol}]\sim-7$, see Sect. \ref{sec:origin}). This other extreme would be represented by $\zeta$\,Ori\,A, whose magnetic field is very weak \citep{bou08} and which displays a ``normal'' X-ray emission (see Sect. \ref{sec:highres}). In between those two extreme cases, a whole range of possibilities opens up, which still needs to be tested. 

In view of the surprising spectra of early B-type stars, which cannot be explained by the current wind-shock model (see Sect. \ref{sec:bstars}), magnetic fields are often seen as a potential solution. Indeed, B-type stars possess weaker stellar winds than O-stars and even small magnetic fields can produce large magnetic confinements, as exemplified by $\beta$\,Cep (Table \ref{tab:mag} and \citealt{fav08}). Magnetically channeled winds naturally produce the observed narrow X-ray lines and the observed formation radii are compatible with this scenario. However, detailed MHD simulations are still missing, as well as sensitive polarimetric observations of problematic B-stars ($\theta$\,Car, $\beta$\,Cru). It remains to be seen if the variety of temperatures and luminosities can also be reproduced by such models, e.g. hard and bright X-rays for $\tau$\,Sco (with $\eta\sim10$) vs. soft emission without overluminosity for $\beta$\,Cep (which has a larger $\eta\sim1000$). 

The magnetic models were, however, less successful in some cases. For example, the non-thermal radio emitter 9\,Sgr, discussed above, displays a high-resolution X-ray spectrum quite typical of ``normal'' O-type stars (low $kT$, broad and slightly blueshifted lines, see \citealt{rau02}) although its magnetic field is significant \citep{hub08}. Indeed, wind-wind interactions most probably play a role in this system, and a full modeling taking into account both magnetic fields and colliding winds should help understand the high-energy characteristics of 9\,Sgr. The problems are more critical for the Of?p stars. This category gathers at least three peculiar Galactic stars: HD~108, HD~191612, and HD~148937 \citep{wal72,wal73}. To explain their cyclic variability in the optical domain, a magnetic oblique rotator model was proposed. Indeed, a magnetic field has been detected for HD~191612 \citep{don06a} and HD~148937 \citep{hub08}, and the derived confinement parameters are large, with values similar to or larger than for $\theta^1$\,Ori\,C (Table \ref{tab:mag}). In the X-ray domain, all three stars present similar spectra but their properties do not fully agree with the predictions of the magnetic models \citep{naz04,naz07,naz08d,naz08b}. On the positive side, two similarities with $\theta^1$\,Ori\,C can be underlined: a large overluminosity (an order of magnitude compared to the ``canonical'' relation) and the flux variations of HD~191612, in phase with those of the optical emissions. In addition, although the overall flux is soft, a second thermal component at 1--3\,keV is present but it is of reduced intensity (it only accounts for 30\% of the intrinsic flux) and it can certainly not explain the overluminosity. On the negative side, these stars display soft spectra with a dominant component at 0.2--0.3\,keV, as expected for ``typical'' O-type stars (see Sect. \ref{sec:temperature}), and broad X-ray lines (HWHM $\sim 900$\,\kms). For these Of?p stars, it thus seems that an additional phenomenon must be at work to explain the characteristics observed at X-ray energies.

As spectropolarimetric analyses are still on-going, it is difficult to assess at the present time the overall validity of the magnetically-channeled wind model at high-energies. Sensitive magnetic detections and high-resolution X-ray spectroscopy of OB-stars need first to be accumulated. Once definitive values for the physical properties of the magnetic systems will become available, statistical analyses will be performed to test whether significant differences between magnetic and non-magnetic systems exist and detailed modeling will be able to check if their X-ray emission follows the expectations or not. Until then, hard X-ray emission with narrow unshifted lines might certainly be considered as a typical signature of strong magnetic fields. This result, at least valid for $\tau$\,Sco and $\theta^1$\,Ori\,C, was also used to explain the observations of numerous X-ray bright O-type stars in very young clusters. Following some authors \citep[e.g. ][]{sch03}, the number of bright, hard sources associated with O-type objects increases as the cluster age decreases. This result was linked to observational hints that magnetic field strength decreases with age \citep{don06a}. However, the impact of binarity (and associated interacting wind emission) was not fully assessed. In this context, it should be remembered that young clusters still harbor the most massive stars: these objects, displaying the strongest winds, are the most susceptible to produce strong wind-wind interactions. Decreasing magnetic field might thus not be the only cause for the smaller number of X-ray bright sources observed in older clusters.

\section{Conclusions}
\label{sec:conclusion}
\xmm\ and \ch\ have provided entirely new views of stellar X-rays, in particular through high-resolution grating
spectroscopy offering a resolving power of several hundred. Stellar X-ray sources are rich in emission lines from which
information on physical conditions and processes can be extracted.

Coronae of cool stars reveal standard thermal spectra that have, nevertheless, offered a number of surprises, such as
previously unknown systematics of element abundances. Subtle effects related to line broadening, optical depths,
resonant scattering, and fluorescence have been crucial to develop models of the emitting sources or irradiated surfaces.
Much attention has also been paid to the unique methodology of measuring average coronal densities that are pivotal
for the construction of coronal models.

Density measurements were at the origin of new hypotheses related to pre-main sequence stars. Soft X-ray emission
revealing very high densities has been interpreted as a signature of accretion shocks near the stellar surface. Outflows or
jets are also X-ray sources, perhaps related to the same physical mechanisms also operating in winds of massive stars.

The current X-ray observatories have also clearly provided better insight into the properties of hot, 
massive stars. In particular, high-resolution spectra have proved a crucial tool to test the proposed 
models in detail: the new observations do not fully agree with expectations and require a change of paradigm. 

Continuing investigations in the domain of high-resolution spectroscopy are clearly needed. The higher-sensitivity facilities 
of the next generation should notably be able to resolve the profile variations of X-ray lines, which are 
expected in various contexts (wind-wind interaction in binaries, channeled wind in magnetic objects, and even 
in winds of single stars, rotating coronal plasma, plasma flows in magnetic fields, etc). Studies of X-ray emission 
in different metallicity environments, e.g. the Magellanic 
Clouds, should also provide crucial information since the stellar wind strength and absorption depend on the 
metal content. In this context, it must be noted that  high sensitivity and high spectral resolution must be 
complemented by a high spatial resolution - it is of little interest to get a well-exposed spectrum of a whole 
cluster!

\begin{acknowledgements}
Both authors warmly thank the Editorial Board of {\it The Astronomy and Astrophysics Review} and 
in particular its Editor-in-Chief, Thierry Courvoisier, for the invitation to write this review article. 
They further thank Marc Audard for critically reviewing their paper.
The authors are indebted to several colleagues who gave permission to use original figure material, namely
Costanza Argiroffi,
Markus Aschwanden,
David Cohen,
Stefan Czesla,
Eric Gosset,
Hans Moritz G\"unther,
Mihalis Mathioudakis,
Raanan Nordon,
Lida Oskinova,
Rachel Osten,
Julian Pittard,
Gregor Rauw,
Luigi Scelsi,
Stephen Skinner, and
Masahiro Tsujimoto.
 YN acknowledges support from the Fonds National de la Recherche Scientifique (Belgium), the PRODEX XMM and Integral contracts, and the `Action de Recherche Concert\'ee' (CFWB-Acad\'emie Wallonie Europe). YN also wishes to thank G. Rauw for his constant help and for providing useful suggestions on the manuscript; 
J.-P. Swings, J.-M. Vreux, and E. Gosset for a careful reading of the text; D. Cohen, E. Wollman, and L. Oskinova for 
interesting discussions. The X-Atlas, 2XMM, and ADS/CDS databases were used for preparing this document.
\end{acknowledgements}

\bibliographystyle{spbasic}      
\bibliography{mergebib}   

%
%

\end{document}